\documentclass[twoside,12pt]{article}
\usepackage{epsfig}
\usepackage{amssymb}
\usepackage{cite}
\def\Journal#1#2#3#4{{#1} {#2} (#4) #3}

\def\NPA{{\em Nucl. Phys.} A}

\def\NPB{{\em Nucl. Phys.} B}

\def\PLB{{\em Phys. Lett.} B}

\def\PRL{\em Phys. Rev. Lett.}     
\def\PREV{\em Phys. Rev.}
\def\PREP{\em Phys. Rep.}

\def\PRD{{\em Phys. Rev.} D}
\def\PRC{{\em Phys. Rev.} C}

\def\ZPC{{\em Z. Phys.} C}
\def\ZPA{{\em Z. Phys.} A}
\def\ANNP{\em Ann. Phys. (N.Y.)}

\def\INT{{\em Int. J. Mod. Phys.} E}
\def\INTA{{\em Int. J. Mod. Phys.} A}
\def\JHEP{{\em JHEP}}
\def\EPJC{{\em Eur. Phys. J.} C}
\def\EPJA{{\em Eur. Phys. J.} A}
\def\JPG{{\em J. Phys.} G}
\def\PPNP{{\em Prog. Part. Nucl. Phys.}}

\newcommand{\be}{\begin{equation}}
\newcommand{\ee}{\end{equation}}
\newcommand{\bea}{\begin{eqnarray}}
\newcommand{\eea}{\end{eqnarray}}

\begin{document}

\title{ \vspace{1cm} Chiral Perturbation Theory and Baryon Properties}
\author{V\'eronique Bernard\\
\\
Lab. de Physique Th\'eorique, Universit\'e Louis Pasteur, 
F67084, Strasbourg, France \\
E-mail: bernard@lpt6.u-strasbg.fr}
\maketitle
\begin{abstract} 


Theoretical as well as  experimental 
progress has been  made in the last
decade in describing the properties of baryons. In this review I will mostly
report on the theoretical issues. Two non-perturbative methods are privileged
frameworks for studying these properties in the low energy domain: chiral 
perturbation theory, the 
effective field theory of the Standard Model at energies below 1 GeV and
lattice QCD. I will mainly concentrate here on the first one but I will
also discuss the complementarity of the two methods. Chiral
extrapolations for lattice simulations of  some nucleon properties will be 
investigated. I will then 
concentrate on processes involving at most two nucleons, describing for example
pion-nucleon and pion-deuteron scattering, 
pion photo- and electroproduction off the nucleon and the deuteron and doubly 
virtual Compton scattering. Three flavor calculations will also be reviewed.
\end{abstract}

\tableofcontents

\section{Introduction}
\label{intro}
Quantum chromodynamics (QCD) has emerged at the 
end of the sixties as the theory of strong interaction, for a
brief history see \cite{eck0604}. 
At high energies it is
perturbative, i.e. observables can be expanded in terms of the strong coupling
constant $\alpha_s$. However the
theory becomes highly non-perturbative  at low energies since $\alpha_s$ 
becomes large. At those low energies
a central role is played by the spontaneously and explicitly broken
chiral symmetry. Taking advantage of that one can formulate an effective
field theory equivalent to QCD at low energy. This is the 
model-independent framework, Chiral Perturbation 
Theory (CHPT), which will be the topic of this review.    
Another model-independent way to solve QCD is the lattice approach. As we will 
see lots of progress has been made in that 
framework, however, most calculations
 involve pion masses which are still far from 
their physical value. It is also limited in the lattice spacing and in 
the volume used in the calculations due to computational costs.  
There has thus
been 
in the last decade a whole development of chiral extrapolations  within
CHPT as well as techniques to try to get a handle on the systematic errors
inherent to the lattice calculations. There is thus nowadays a nice interplay
between lattice and CHPT which will be reported here. 

Testing QCD in the low energy domain has been  and still is the subject of 
lots 
of work from the theoretical as well as experimental side. I will 
report here on some of these 
in the framework of CHPT restricting  myself to the case when nucleons 
are present and 
especially to processes with one nucleon in the initial and  mostly one nucleon
in the final states. The ideas underlying CHPT have
been generalized to the few nucleon sector where one has to deal with a 
nonperturbative problem. I will touched upon it while discussing processes
involving two bound nucleons, namely the deuteron. For more details on 
chiral EFTs in the few-nucleon sector see \cite{e06}.

In Section \ref{eft} 
after  briefly reviewing   what is an effective field theory
and defining chiral symmetry, I will give some
general properties of  CHPT. 
I will then concentrate on Baryon Chiral Perturbation 
theory
without and with explicit $\Delta$ degrees of freedom, Sections \ref{bchpt} 
and 
\ref{eftdel} respectively, discussing  different regularizations of the 
theory. I will then show the
complementarity of the two frameworks, lattice QCD and CHPT. In Section 
\ref{chextrapol}
I will report on chiral extrapolations of some baryon properties. 
In the two next sections I
will discuss a few applications within $SU(2)$, namely $\pi N$ and $\pi d$
scattering  and different electromagnetic properties of the nucleon. 
The last section will be devoted to the case of three flavors. 

Baryon CHPT is a very active field. A great amount of studies have been done
with and without the $\Delta$ degrees of freedom. I will unfortunately not
be able to review all of what has been done and I apologize to those whose
work I did not include. 
 
\section{Effective Field Theory}
\label{eft}

Many reviews and lectures  concerning effective field theories and ChPT are at 
present available \cite{do90}-\cite{bm06}.
I will thus be rather brief here and refer
the interested reader to these reviews for more details.
 
Effective field theories (EFTs) date back to the 70's and have become
a popular tool ever since. Wilson, Fisher and Kadanoff first 
studied critical phenomena in condensed matter systems within
an EFT. These were then developed in 
this context and enlarged to other fields.   
EFTs are powerful techniques
when the relevant degrees of freedom depend on the typical energy of the
problem. For energies beyond a certain scale $\Lambda$ the fundamental theory
is applicable while for energies below it can be more practical to use an
effective theory. One has to keep in mind that a fundamental theory can
in turn become an effective theory of some other fundamental theory  
at higher energies. 
There are different types of EFTs depending on the 
structure of the transition
between the fundamental and the effective level. 

$\bullet$ decoupling 

In that case one integrates out the heavy degrees of freedom (heavy with 
respect to $\Lambda$). One is left with the light degrees of freedom
described by a Lagrangian which contains a renormalizable part and 
non-renormalizable couplings suppressed by inverse powers of $\Lambda$.
The coupling constants of the low-energy EFT can be obtained by perturbative
matching to the underlying theory.
The best known example is the Standard Model. 

$\bullet$ non-decoupling

In that case the relevant degrees of freedom are different from the
one of the fundamental theory due to a phase transition. 
This is the type we are interested in here, the phase transition being, as we
will see, the
spontaneous breakdown of chiral symmetry. Renormalizability
is not a  meaningful concept anymore and in general EFTs in this category are
nonrenormalizable.   
Thus an infinite numbers
of counterterms have to be added to make it finite. However in an EFT
where the range of applicability is below the scale $\Lambda$,
the number of counterterms needed at each level of the expansion is finite
though at increasing expansion powers, it increases drastically. 

\subsection{\it Chiral Symmetry \label{chisym} }

In the Standard Model the  degrees of freedom are the gluons 
described by the non-abelian field strength tensor $G_{\mu \nu}$ and
$N_f$ flavors of quarks $q$. The QCD Lagrangian is given by:
\be 
{\cal L}_{QCD} = -\frac{1} {4} G_{\mu \nu}^a  G^{\mu \nu \, a}
+ \sum_{N_f} \bar q (i \gamma^\mu D_\mu -m_q) q \, \, .
\ee
In what follows we will consider only the light 
quark sector $(u,d,s)$ for which the masses $m_q$ are much smaller than the 
hadronic scale
of the order of 1 GeV. Typically the up and down mass are a few MeV and
the strange mass is $\sim 100$ MeV. 
Therefore ${\cal L}_{QCD}$ can very well be approximated 
by a Lagrangian ${\cal L}^0_{QCD}$ with no mass term.
In that case left and right handed quarks live their own lives and the  
symmetry group of the Lagrangian, apart from the discrete
symmetries of parity, charge conjugation and time reversal, is 
$SU(N_f)_L \times SU(N_f)_R \times U(1)_V$. $U(1)_V$ generates conserved
baryon number while the remaining group $SU(N_f)_L \times SU(N_f)_R$ is 
the chiral symmetry group $G$ we are mostly interested in this review. 
We will not discuss here the chiral anomalies and CP violation. A classical
theory can be realized in quantum field theory in 
two different ways, the so-called Wigner-Weyl  mode and the Nambu-Goldstone 
one, depending on how the vacuum responds
to the symmetry transformation.
In fact
there are indications from experiment and theory that chiral symmetry is 
spontaneously broken in nature.  
Chiral symmetry is a symmetry of ${\cal L}^0_{QCD}$ but not of the vacuum 
which 
is invariant only under $SU(3)_V \times U(1)_V$, where $SU(3)_V$ is the
subgroup of vector rotations. Evidences are for example the non-existence 
of degenerate parity 
doublets in the hadron spectrum, the existence of the triplet of unnaturally
light pseudoscalar particles and the very different behaviour of the
correlators of 
axial and vector currents. It is also supported by lattice simulations
(for a recent study, see \cite{ne05} and references therein). 
To a spontaneously broken symmetry
corresponds non-vanishing order parameters and    
the appearance of  Goldstone bosons. I will come back to that below.
A typical model which describes spontaneous symmetry breaking is the 
linear sigma-model. However it has a few drawbacks. First it involves
a $\sigma$ particle, the chiral partner of the pion which is not observed.
Second  its renormalizability requires specific relations between various
couplings which are not in agreement with experiment \cite{gl84}.   

As we have seen the masses are in fact not exactly zero. The quark mass
term leads to the so-called explicit chiral symmetry breaking: 
the vector and axial-vector currents are no longer conserved.  
In the limit of equal quark masses, QCD  possesses an additional SU(3)
flavor symmetry. Restricting to the up and down quarks this is the well
known isospin symmetry. However this symmetry is again broken, slightly
in SU(2) and more strongly within SU(3).  
The difference in the quark masses as well as electromagnetic effects explains
for example
the splittings in mass for the hadrons appearing in 
the same multiplet. In the following I will mostly restrict to the 
equal mass case. Isospin violations will be discussed in Section \ref{isoviol}.

\subsection{\it Effective Chiral Meson Lagrangian} \label{eftm}
 
Chiral symmetry and its spontaneous symmetry breaking thus
leads to: 

\begin{itemize}
\item the existence of $N_f^2-1$ Goldstone particles, massless excitation of
the vacuum
corresponding to the $N_f^2-1$ axial generators of the chiral group as stated
by the Goldstone theorem \cite{go61}: 
these are identified with 
the three (eight) lightest hadronic states, the pseudoscalar mesons $\pi$ in $SU(2)$
($\pi, K, \eta$ in $SU(3)$).

\item the  Goldstone theorem also predicts that the interaction 
between these particles is weak. Thus a Lagrangian describing
them must be of the derivative type.  This allows for  
an {\it expansion in  the external momentum p} as long as one
stays below a certain scale. This scale corresponds in fact to the 
occurrence of other hadrons in the spectrum,  $\Lambda_\chi \sim 1$ GeV, usually
called the scale of chiral symmetry breaking.

\item the quark masses being small one does not expect much changes while
going from  ${\cal L}^0_{QCD}$ to  ${\cal L}_{QCD}$. 
An {\it expansion in $m_q$} around the chiral limit can be performed again below the scale
$\Lambda_\chi$. These small masses generate the physical values of the
pseudoscalar mesons, typically to leading order in these quark masses one has:
\be
M_\pi^2 = 2 B m_q
\label{eq:mpilo} 
\ee
where the constant of proportionality $B$ will be discussed below.
\end{itemize}

These features are best realized writing 
an effective theory in terms of the Goldstone bosons 
which will be valid up to the scale 
$\Lambda_\chi$.
This effective
field theory is known as {\it Chiral Perturbation Theory (CHPT)} and has been 
first worked out by \cite{we79} and developed by \cite{gl84,gl85}.
The idea underlying effective field theories
is to write down the {\it most general} Lagrangian in terms of the
observed asymptotic fields in agreement with the  symmetries
of the system:

\begin{equation}
{\cal L}_{\rm QCD} [ \bar q,q,G_{\mu \nu} ] \to {\cal L}_{\rm eff} [ U, \partial U,
\cdots {\cal {M}}, v_\mu, a_\mu ] \, \, .
\label{eq:lchpt}
\end{equation}
In this equation $U$ is a unitary matrix field  representing the Goldstone 
bosons and transforming under a chiral 
transformation
as $U \stackrel{G}{\to} g_R U g_L^{-1}$, $(g_R,g_L) \in  G$. In order to be 
able to generate Green 
functions of quark currents, locally chiral invariant external fields, 
that is non-propagating objects,
are introduced.
${\cal M}$ which contains scalar and pseudoscalar fields enables 
one to describe the mass 
matrix while  $ v_\mu, \,  a_\mu$,
vector and axial fields, respectively, allow to describe the couplings
to external photons and W bosons. The equivalence between the two 
Lagrangians has been shown in \cite{le94b,we94}. 

The heavy fields not included in the theory  can be 
seen through virtual effects included via low-energy constants (LECs) 
connected to the light fields, i.e the pseudo-Goldstone bosons. 
Indeed, the decoupling 
theorem \cite{apca75} states that all effects from heavy fields will reveal
themselves through renormalization of coupling constants.

\subsubsection{Power counting} \label{powc}

CHPT is thus a two-fold expansion in external momentum and 
quark masses and can be written as:
\be
{\cal L}_{eff}=\sum_{i,j} {\cal L}_{i,j}  \, , \quad \quad {\cal L}_{i,j}=
{\cal O}(p^i m_q^j) \,\,  .
\ee
It is based on chiral countings rules: the field $U$ counts as 
a quantity  ${\cal O}(1)$,
$\partial_\mu U, v_\mu$ and $a_\mu$ are thus ${\cal O}(p)$ ($v_\mu$ and 
$a_\mu$ appear in the covariant derivative at the same level as 
$\partial_\mu$). In order to 
relate the two expansions one assigns a chiral counting to the 
quark masses through Eq.~(\ref{eq:mpilo}) where $M_\pi^2$ should be
counted as a quantity of order 
$p^2$.

As we have seen the spontaneous breaking of a symmetry is also related to 
the existence of non-vanishing order parameters. The simplest 
one is the quark condensate connecting the left and right handed quarks.
It also plays a special role being related to the density of small
eigenvalues of the QCD Dirac operator \cite{bc80,ls92,st98}.  
The question is: how big is the value of
this condensate? The notion of big is related to another order parameter 
which is the pion
decay constant $F$ which will be defined below and can be expressed as
an integral:
\be
F^2= \frac{1}{6} \int d^4x \langle 0| \bar u_L (x) \gamma_\mu d_L(x)
\bar d_R (0) \gamma^\mu u_R (0) |0 \rangle \, \, .
\ee
There exists two scenarios the so-called standard one and  
Generalized CHPT \cite{mst94} depending on the answer to that question.
The standard scenario assumes that this condensate  is 
large meaning that the quantity $B/F \equiv  -\langle 0 | \bar q q | 0 \rangle/F^3 $ is much
bigger than one  while the Generalized one allows for a value comparable or
even much smaller than one. These two scenarios lead to a different chiral
counting, in the first case $B$ is  ${\cal O}(1)$ and thus $m_q$ is 
${\cal O}(p^2)$, see Eq.~(\ref{eq:mpilo}), while in the second case 
$B$ is ${\cal O}(p)$ and thus $m_q$ is 
also ${\cal O}(p)$. One thus has:
\be
{\cal L}_{eff}^{std}= \sum_{i+2j=d} {\cal L}_{ij} \, , \quad \quad
{\cal L}_{eff}^{gchpt}= \sum_{i+j+k=d} B^k {\cal L}_{ij} \,\, . 
\ee 

How can one decide between the two cases? 

\begin{itemize} 
\item the Gell-Mann Okubo relation which relates the $\pi$, $K$ 
and $\eta$ masses which is well verified experimentally and holds naturally in 
standard CHPT is modified 
in Generalized CHPT. 
 
\item lattice simulations give values for $-\langle 0 | \bar q q | 0 
\rangle^{1/3} \sim 200-370$ MeV in agreement with standard CHPT, for a 
summary of recent results see
\cite{ne05}.

\item QCD like theories are expected to undergo a number of 
phase transitions at zero temperature upon varying $N_f$ at fixed number
of colours.  
It has been discussed in \cite{mou00} that $N_f=3$ could be not
far from a chiral phase transition suggesting that in  $SU(3)$ 
the quark condensate might be much smaller than in $SU(2)$. The
different behaviour between the SU(2) and the SU(3) case could come
from large vacuum fluctuations of strange $q \bar q$ pairs related
to a large OZI-rule violation in the scalar channel \cite{dsg00}. 
I will come back to this in more detail in Section \ref{cdy}. In this
context it has been pointed out \cite{dsg02}  that it might not 
be necessary to introduce a different counting when 
the condensate is small, a non perturbative resummation of 
the vacuum fluctuations encoded in some LECs  might in fact
be sufficient. This is called Resummed Chiral Perturbation Theory.

\end{itemize}
At first sight the standard picture seems to be privileged at least
within $SU(2)$. However one has certainly to await for better lattice results
as well as more precise experimental results to really settle this
issue definitively. For example the $\pi \pi$ scattering lengths
are actually under debate since the preliminary analysis of  
$K_{e4}$ data from
NA48/2 \cite{mas06}  seems to be in conflict with the very precise 
determination \cite{cgl00,cgl01} combining CHPT and dispersion relations.

In the rest of the review I will only discuss the standard case. Note that
there has been only one attempt to apply the Generalized picture to 
the nucleons \cite{mk00}. 
 
At tree level one can reconstruct the current algebra predictions
of the 1960's as well as 
low energy theorems. 
With a hermitian Lagrangian tree amplitudes are necessarily real.
Unitarity and analyticity however lead to complex amplitudes. 
A systematic low energy expansion thus requires a loop expansion. 
To do this in a consistent fashion
one has to introduce a {\it counting scheme}.  
In fact it has been shown by Weinberg
in the meson sector \cite{we79} that there is a one to one correspondence
between loop and chiral order, that is diagrams with $L$ meson loops are
suppressed by powers of $(p^2)^L$.  The power counting formula of Weinberg 
orders the various contributions to any S-matrix 
element for the pion interaction according to the 
chiral dimension
$D$,
\be 
D=2 + \sum_d N_d(d-2)+2 L
\ee
with $N_d$ the number of vertices with dimension $d$ (derivatives and/or 
pion mass insertions). Chiral symmetry
gives a lower bound for $D$, $D \ge 2$.   
The experimental 
precision reached in the meson sector makes it necessary to include 
corrections of $O(p^6)$, 
for a review on two-loop calculations see \cite{bij07}. 
At that order one has, according to the formula to include
tree graphs (L=0) from ${\cal L}_{\pi \pi}^{(2,4,6)}$, one loop graphs (L=1) 
with insertion from $ {\cal L}_{\pi \pi}^{(2,4)}$ and finally two loop graphs
(L=2) with insertion from  $ {\cal L}_{\pi \pi}^{(2)}$.

\subsection{\it  Baryon Chiral Perturbation Theory\label{bchpt} }

I will now extend the discussion to include matter fields. 
In this section I will concentrate on nucleons and pions
coupled to external sources. 
The $\Delta$ degree of freedom will be discussed in Section \ref{eftdel}. 
Couplings to photons will be given in Section \ref{isoviol}.
As we will see some
problems arise concerning the power counting when matter fields are 
introduced. In order to solve these, 
different regularization have been introduced which will be 
discussed in Sections \ref{ir}, \ref{or}.

\subsubsection{Lagrangian} \label{lag}

The minimal relativistic effective $\pi N$ Lagrangian \cite{fms98} up to 
${\cal O}(p^4)$  is given below. 
For the heavy baryon case and the relation between the LECs in the two
framework see that reference. Only the terms which will be needed
in the review are shown explicitly:

\bea 
{\cal L}_{\rm eff} &=& {\cal L}_{\pi N}^{(1)} +  {\cal L}_{\pi N}^{(2)} +
{\cal L}_{\pi N}^{(3)} +  {\cal L}_{\pi N}^{(4)} +
{\cal L}_{\pi N}^{(5)}+ {\cal L}_{\pi N}^{(6)} +
{\cal L}_{\pi \pi}^{(2)} +  {\cal L}_{\pi \pi}^{(4)} 
+ \cdots \, \, ,\nonumber 
\\
{\cal L}_{\pi N}^{(1)}
&=&\bar{\psi} \bigl( i \not \! \! D - m_0 +\frac{1}{2} 
g_0 \not \! u \gamma_5
\bigr) \psi \, \, ,\nonumber
\\
{\cal L}_{\pi N}^{(2)}&=&\sum_{i=1}^{7} {\it{c_i}}
 \bar{\psi} {\cal O}_i^{(2)} \psi=
\bar{\psi}_N\left[{\it{c_1}} \langle\chi_+\rangle
            -{\it{c_2}}\frac{1}{8 m_0^2}\left\{\langle u_\mu u_\nu
                         \rangle \{D^\mu, D^\nu\}+ {\rm h.c.}\right\}
              +{\it{c_3}}\frac{1}{2} \langle u^2\rangle \right. 
\nonumber \\
& &\left.
 +{\it{c_4}}\frac{i}{4}[u_\mu,u_\nu] \sigma^{\mu \nu} 
+ \ldots \right]\psi_N \, \, , \nonumber \\ 
{\cal L}_{\pi N}^{(3)}&=& \sum_{i=1}^{23} {\it{d_i}} \bar{\psi} 
{\cal O}_i^{(3)} 
\psi =\cdots \frac{d_{16}}{2} \gamma_\mu \gamma_5 \langle \chi_+  \rangle
u_\mu +\frac {i d_{18}}{2} \gamma^\mu \gamma_5 [D_\mu, \chi_-] 
  +\cdots           \, \, ,          \nonumber \\
{\cal L}_{\pi N}^{(4)}&=& \sum_{i=1}^{118} {\it{e_i}} \bar{\psi} 
{\cal O}_i^{(4)} \psi \, \, ,\nonumber \\
{\cal L}_{\pi \pi}^{(2)} &=& \frac {F^2}{4} \langle \nabla_\mu U 
\nabla_\mu U^\dagger
\rangle + \langle \chi_+ \rangle \, \, ,
\nonumber \\
{\cal L}_{\pi \pi}^{(4)} &=&\sum_{i=1}^{7} {\it{l_i}}
 \tilde  {\cal O}_i^{(4)}= \cdots \frac{l_3}{16} \langle \chi_+\rangle^2
 + \frac{l_4}{16} \Bigl\{ 2[\langle \nabla_\mu U  \nabla^\mu U^+\rangle
   \langle \chi_+ \rangle \Bigr .\nonumber \\
& & + \Bigl. \langle \chi^+ U \chi^+ U +\chi U^+\chi U^+\rangle]
-4 \langle \chi^+ \chi \rangle - \langle \chi_-\rangle^2 \Bigr\} + \cdots \, \,
,
\label{eq:lag}
\eea
where the following standard notations have been used:
\bea 
U& =& u^2 \quad,  \quad \quad  u_\mu= i u^\dagger \partial_\mu U u^\dagger
\nonumber \\
\nabla_\mu U &=& \partial_\mu U -i(v_\mu+a_\mu)U
+i U(v_\mu -a_\mu) \quad,  \quad D_\mu=\partial_\mu + \frac{1}{2} [u^\dagger,
\partial_\mu u]
\nonumber \\
\chi_\pm&=&u^+ \chi u^+ \pm u \chi^+ u \quad,  \quad \quad \chi=2B(s+ip) \, \, 
.
\label{eq:deffield}
\eea

At lowest order the Lagrangian is given by the non-linear $\sigma$ model. One
recovers in ${\cal L}_{\pi \pi}^{(2)}$ the two
parameters $B$ and $F$ discussed previously. $F$ is the chiral limit value of the pion decay constant:
\be
\langle 0| A_\mu^a(0) | \pi^b(p)\rangle= i \delta^{ab} F p_\mu \,\, .
\ee
In Eq.~(\ref{eq:lag})  
$m_0$ is the chiral limit  of the nucleon mass $m$ and $g_0$ 
the one of the axial vector coupling constant
$g_A$ defined by: 
\be   
\langle N(p')| A_\mu^a |N(p)\rangle = \bar u(p') \Bigl [ \gamma_\mu 
G_A(t) + \frac{(p'-p)_\mu} {2 m} G_P(t) \Bigr ] \gamma_5 \frac {\tau^a}{2}
u(p)
\label{gaff}
\ee
with $t=(p'-p)^2$ the invariant momentum transfer and $g_A \equiv  G_A(0)$. 
The form
of Eq.~(\ref{gaff}) follows from Lorentz invariance, isospin conservation, the 
discrete symmetries C, P and T and the absence of second class currents 
\cite{we58} which is consistent with experimental information, see e.g. 
\cite{mmk94}.
$F$, $m_0$ and  $g_0$ differ from their physical values by terms proportional
to the quark masses, e.g.
\be
F_\pi=F(1+ {\cal O(M)})= F\Bigl(1+ \frac{M^2} {16 \pi^2 F^2} 
\bigl(16 \pi^2 l_4^r (\lambda) -\ln \frac{ M^2}{\lambda^2} \bigr)+
{\cal O}(M^4)\Bigr). \label{fpi}
\ee  
The next-to-next-to-leading order contributions to $F_\pi$ can be found in
\cite{cod04} (see also \cite{bcegs97} and references therein).
The expression for the nucleon mass and the axial 
vector coupling will be given in Section  \ref{chextrapol} and discussed in detail there. 

The coefficients of the monomials ${\cal O}^{(n)}$ are the so-called LECs.
The numbers of counterterms in   ${\cal L}_{\pi N}^{(3),(4)}$ 
correspond to  \cite{fms98}. 
In that reference, terms  in ${\cal L}_{\pi N}$ are included which could 
be eliminated by
using equation of motion (EOM) for the classical fields as done in
\cite{em96}. In fact the knowledge of both these physically equivalent 
Lagrangians allows for a
non-trivial test of the calculations performed. In the case 
of  ${\cal L}_{\pi \pi}^{(4)}$ we have used the expression given in 
\cite{gss88} which also differs from the original one \cite{gl84}
by EOM terms though in that case the number of terms are the same. 
Clearly, the number of counterterms increases fastly with the order
one is working. However, as we will see in the applications Sections
\ref{chextrapol}, \ref{pind} and \ref{emprop}, only
few of them appear in each process. We will be mostly concerned in this review
 with 
the order 2 and two of the order 3 counterterms for which we will give
numerical values in the next section. 

\subsubsection{Low-energy constants} \label{lec}

\vspace{0.2cm}
i) {\bf {some general comments}}

\vspace{0.2cm}

The general properties of the LECs are the following:

\begin{itemize}

\item[$\bullet$] as we have seen they describe the influence of ``heavy'' 
degrees of freedom not contained explicitly in chiral Lagrangians.

\item[$\bullet$] their values are {\it {not constrained by symmetries}}.

\item[$\bullet$] most of them are {\it {scale-dependent}},
see Eq.~(\ref{fpi}). Indeed the Goldstone 
loops are in general divergent and need to be renormalized. Since 
by construction the full effective Lagrangian contains all terms permitted
by the symmetries, the divergences can be absorbed in a renormalization of the 
coupling constants occuring in the Lagrangian. Thus one has for a generic 
LEC $b_i$ in the  modified $\overline{MS}$ scheme
\be
b_i=b_i^r(\lambda) +\kappa_i L \quad , \quad \quad L=(4 \pi)^{-2}
\lambda^{d-4} \Big \{ \frac {1}{d-4} -\frac{1}{2} (\ln 4 \pi + \Gamma'(1)+1) 
\Bigr \} \,\, . \label{kapa} 
\ee
Of course the
scale dependence thus 
introduced cancels with the one from the loops so that physical quantities
are scale-independent. Since loops starts at order 
three in the 
nucleon sector the LECs of order two are scale-independent.
The $\kappa_i$, Eq.(\ref{kapa}) have been obtained in 
\cite{gl84} for the mesons and \cite{mms00,fms98,em96} for the nucleons in 
HBCHPT
from an explicit calculation of the one-loop generating functional
(note 
that there is a misprint in \cite{em96} concerning the LEC $d_{11}$ which
was corrected in \cite{gilmr02}.). The complete divergence structure
of the one-loop generating functional corresponding to the heavy baryon
Lagrangian in the presence of virtual photons has been worked out 
in \cite{gilmr02} using a super-heat-kernel method \cite{ne98}. 
For the LECs which specifically
enter this review the $\kappa_i$ are given in Tables 
\ref{tab:lecsm} and 
\ref{tab:lecsn}. Care has to be taken to the exact definition of the 
Lagrangian, for relations between the $\kappa_i's$ for  different forms 
of the Lagrangian with or 
without EOM terms see \cite{fms98,em96,gilmr02}.  

\item[$\bullet$] A very important property is that {\it they relate different 
observables}.
This is illustrated in Table \ref{tab:lecsm} 
where   different sources 
from which  the $l_i's$ can be obtained are given 
(in Table  \ref{tab:lecsn} only the sources
from which the $c$'s have been determined are shown). 
 This severely
constrains the result of the calculation performed. 
  
\item[$\bullet$] They are assumed to fulfil the criteria of 
{\it naturalness} based on a dimensional analysis.
Comparing the Lagrangian ${\cal L}_{\pi \pi}^{(2)}$ and
${\cal L}_{\pi \pi}^{(4)}$ for example one can make an estimate
of the expected size of the couplings $l_i$ in terms of the
symmetry breaking scale $\Lambda_\chi \sim 4 \pi F_\pi$.
One has:
 
\be
|l_i| \sim \frac {16 F_\pi^2/4}{\Lambda_\chi^2} \sim \frac{4}{ (4 \pi)^2}
\sim 0.025 \,\, .
\ee 

For the nucleon the $c$'s should be typically of the order $g_A/\Lambda_\chi
\sim 1$ GeV$^{-1}$.
\end{itemize}
One can distinguish between two different classes of LECs, the so-called 
dynamical LECs and the symmetry breakers. The first ones are proportional 
to $\partial_\mu^{2n}$ while the second are proportional to quark masses
$m_q^{2n+1}, 
m_q \partial^{2n}, \cdots$.
The mixed LECs parametrizing operators with quark mass insertions and 
derivatives enter the second class since at fixed pion mass they can 
be absorbed in the values of certain dynamical LECs.  
$L_{\pi N}^{(2)}$ has, for example, two symmetry breakers and five dynamical 
LECs, see Eq.~(\ref{eq:lag}).

Determining the LECs from QCD is a difficult non-perturbative 
problem. Thus either one makes phenomenological evaluations based 
on experimental 
information at low energies or one uses additional inputs from theory
in order to pin them down. The most commonly used approach is 
resonance saturation but sum rules \cite{dha92,mouss99,abm01} as well as
matching to dispersion theory  are also considered. Example of such 
a matching will be given in Section \ref{strange}.   
These are the best candidates 
in the case of the dynamical LECs. The progress of lattice calculations in the 
last years 
has however opened the 
possibility to determine the LECs directly from QCD. These
calculations  are best suited
in the case of the symmetry breakers since these can be obtained through 
variation of the quark masses.   
There has also been a lot of activity to determine the LECs from models
and especially the Nambu--Jona-Lasinio (NJL) model, for reviews see 
for example \cite{bi96}. 
These will not be discussed  here. 
At present 
most attempts concern the meson sector. 
In the nucleon sector as we will see the LECs have been evaluated 
mostly by fitting the experiments, though some resonance saturation estimates
exist for the LECs of order 2 \cite{bkm93,bkm97}. 
Let us look in more detail in two of these approaches:

\begin{itemize}

\item[I)] {\underline {ROLE OF RESONANCES and RESONANCE SATURATION}}

\end{itemize}
LECs correspond to coefficients of the Taylor expansion, with respect to
the momenta, of some QCD correlation functions, once the singularities (poles
and discontinuities) associated with the contributions of low-momentum
pseudoscalar intermediate states have been subtracted. The Green's function
involved being order parameters of the spontaneous breaking of chiral
symmetry do not receive contributions from perturbative QCD at large momentum
transfers, thus the LECs are expected to be sensitive to the physics 
in the intermediate regions, that is to the spectrum of mesonic resonances
in the mass region around the hadronic scale. Most attempts up to date
to estimate the values of the LECs from resonance data
are thus based on the so-called principle of
{\it {resonance saturation  which states that the LECs are largely
 saturated by resonance
 exchange}}. 
This is also known as chiral duality. 
 
The procedure relies on the construction of an effective Lagrangian 
with resonance 
degrees of freedom. Determining  the LECs  amounts to 
decoupling these resonances from the effective field theory.
The traces of these frozen particles are then encoded in the numerical 
values of certain LECs.  
A first systematic analysis of the role of
resonances in the CHPT Lagrangian was performed in the meson sector, 
\cite{dono88ed,egpr89,eglpr89} 
 the couplings of meson resonances of the type
$V(1^{--}$), $A(1^{--}$), $S(0^{++}$) and $P(0^{-+}$) being studied. In fact
inclusion of vector particles follows many years of phenomenological analysis
in both nuclear and particle physics.  
They have therefore been considered
in chiral Lagrangians from the early days on, usually with the assumption 
that vector and axial vector mesons are gauge bosons of local chiral
symmetry (for reviews see \cite{bky88}). Note that in the baryon sector
the role played by the $\rho$ meson is, in that case, the $\Delta$(1232). 
We will discuss this later. 
In order to determine local operators from the resonance Lagrangian
one lets the resonance masses $M_R$ become very large with fixed ratios of 
coupling constants to masses.   
Typically the resonance propagator is replaced by its
corresponding momentum expansion in $t/M_R^2$ where $t$ is a typical 
momentum transfer squared. Thus the exchange of virtual resonances generates 
pseudo-Goldstone boson couplings proportional to powers of $1/M_R^2$.
Clearly the lowest lying resonance will thus be the most important one, a fact 
which is supported by phenomenology. 
In \cite{egpr89,eglpr89} the ${\cal O}(p^4)$ couplings of 
${\cal L}_{\pi \pi}$ 
have been determined this way. It is found that whenever the vector
meson contributes it almost saturates the LECs which is in agreement
with the vector meson dominance principle.  
 



Recently this analysis has been generalized and Lagrangians of this resonance 
chiral theory ($R \chi T$) have been 
constructed following the ideas of Weinberg \cite{ceekpp06}, however
having no power counting. They  enlarge the 
range of validity of the 
effective field theory in the region $M_\rho <E <2$ GeV and aim
at determining the LECs up to ${\cal O}(p^6)$
in the meson sector. 
They are guided by 
the following principles:
 
(a) they use large 
$N_c$ arguments which say that the correlators of colour-singlet
quark-antiquark currents are given by tree-level exchanges of infinite
towers of narrow resonances \cite{th74}.

(b) the appropriate
QCD short-distance constraints whose  importance has been stressed 
in \cite{kn01}
and recognized since then, are implemented
 in the calculation. 
This additional input leads to the so called
Minimal Hadron Ansatz which states that only those resonances which are
needed to fulfil the constraints are taken into account.

Integration 
of the resonance fields then leads to LECs parametrized in terms of resonance
masses and couplings. Information on those couplings and on the LECs have
been extracted analysing QCD Green functions of currents both for large
and small momenta, for example $\langle VAP \rangle$ and $\langle SPP
\rangle$ Green functions 
\cite{ceekpp06}.
Note that though it has some advantages one could work without a Lagrangian 
formulation and only
use Green's functions \cite{kn01,ceekpp05,ceepp04}. 



Most calculations have at present only considered the large $N_c$ limit. 
Incorporating next-to-leading contributions in the $1/N_c$ counting 
is not straightforward. Indeed quantum loops including virtual resonance
propagators constitute a major technical challenge. Some studies
have been done but mostly within models. In order to 
gain some understanding it seems worth to perform some explicit
one-loop calculations of well chosen amplitudes. A first detailed 
investigation of the pion vector form factor at 
next-to-leading order in the $1/N_c$ expansion has thus been done in
\cite{rsp04}.

One problem has to be raised, namely the scale dependence of the
LECs which is absent in this resonance picture at leading order. 
It has however been 
shown \cite{egpr89} that if one decomposes the LECs into a resonance
part and a remainder  which takes the scale dependence $\mu$ into account
there is a region in $\mu$ where the LECs are almost entirely given 
by the resonance part, $\mu \sim M_R$. Recently this problem has 
been considered within the $R \chi T$ theory at order $1/N_c$ and first
determinations of the scale dependence of certain  LECs  
have been obtained \cite{cape02,rsp07}.

Resonance saturation has proven 
very successful in  determining the values of certain LECs in the meson
sector at one loop.
However, the validity
of the resonance saturation from the light resonance sector 
for the ${\cal O} (p^6)$ LECs \cite{km06} has
been recently examined with the conclusion 
that some of these 
couplings are in fact not dominated by resonance contributions. 
In the nucleon sector only the LECs of order 
two have been studied so far. As an example,
let us look  at $c_3$. 
The dominant contribution comes from
the $\Delta(1232)$, there are additional smaller corrections 
from the $N^*(1440)$ resonance and also contributions from a scalar 
meson exchange.  The $\Delta$ contributes 
$-2.54 \cdots - 3.18$ GeV$^{-1}$, the $N^*$ $-0.06 \cdots -0.22$ GeV$^{-1}$
and the scalar $-1.33$ GeV$^{-1}$ leading to $c_3$ varying between -3.6
and -5.0 GeV$^{-1}$ in the range of  the empirical value
quoted in Table \ref{tab:lecsn}. In view of its success,   
extension of the idea of resonance saturation
has also been done in the two-nucleon sector \cite{emge02}.
Clearly more studies are needed on the role of resonances in the 
determination of the LECs, especially in the nucleon sector. 

\begin{itemize}
\item[II)] {\underline {LATTICE}}
                                                              
\end{itemize}
As will be stressed in  Section \ref{lat} there exists today some lattice 
evaluation of the ${\cal O}(p^4)$
LECs in the meson sector, mostly the SU(3) ones. Let me give here the results 
for the SU(2) LECs  $l_3$ and
$l_4$ which belongs to the symmetry breakers' class. 
These are
the ones of most concern in the baryon sector since they enter the pion mass 
and decay constant at next-to-leading order, respectively. 
They are also of particular importance since they contribute to  
fundamental quantities as the $\pi \pi$
scattering lengths.
As can be seen from Table \ref{tab:lecsm} where the scale-independent bar 
quantities related to the $b_i^r$ by 
\be
b_i^r = \frac{\kappa_i}{32 \pi^2}\bigl(\bar b_i + \ln \frac{M^2}{\lambda^2}
\bigr)
\label{eq:renct}
\ee   
are given, there is a  rather good agreement 
between the SU(2) lattice results \cite{bal07,ddal06} and the one obtained in 
standard CHPT.
Note that the latter are in good agreement with the expected size of the
LECs. 
One can also infer values for these LECs from  SU(3) calculations 
using the relations derived in \cite{gl85}.  
The MILC collaboration \cite{beal06} found smaller value for $\bar l_3$
from such a calculation.   
Two remarks are in order here. First 
the lattice values are obtained for 
large values of the strange mass and what is really needed are the values
for $m_s=0$ or $m_s$ physical. Second OZI-rule violating vacuum fluctuations 
would strongly affect the standard CHPT values \cite{dsg02}, especially
$\bar l_3$ whose value could be pushed towards larger negative value, as
much as -17.8 \cite{dsg02}.
In the following the CHPT values quoted in the table will be 
used. 
\begin{table}
\begin{minipage}[hbt]{18 cm}
\caption{Two LECs of ${\cal L}_{\pi \pi}^{(4)}$. $r_\pi^S$ is the scalar pion 
radius. Values determined as 
explained in the text.} 
\label{tab:lecsm}
\end{minipage}
\begin{center}
\vskip 0.4truecm
\begin{tabular}{|c c|c|c|}
\hline
&&&\\[-0.8mm]
&& $\bar l_3$ & $\bar l_4$ \\
\hline
&&$\kappa_3 =-1/2$ &$\kappa_4 =2$ \\
\hline
sources&&mass ratios, $\pi \pi$ scat. & $r_\pi^S$, $F_\pi$, $\pi \pi$ scat.  \\
\hline
 \phantom{     }standard CHPT&&$2.9 \pm 2.4$ \cite{gl84}& $4.4 \pm 0.2$ \cite{cgl01}\\
\hline
lattice&  \cite{bal07}&$3.65 \pm 0.12$   & $4.52 \pm 0.06$ \\
 &\cite{ddal06}        & $3.5 \pm 0.5\pm0.1$ & \\
&\cite{beal06}        & $0.6 \pm 1.2$ & $3.9 \pm 0.5$ \\
\hline
\end{tabular}
\end{center}
\end{table}

\vspace{0.2cm}
\noindent ii) {\bf {leading order LECs and their determination 
($F_\pi, \,F_K, \, g_A$)}}
\vspace{0.2cm}

The most accurate experimental information on $F_\pi$ and $F_K$ come
from the semi-leptonic transition $P \to \mu \nu$ while the axial-vector
coupling constant $g_A$  
is measured in (polarized) neutron beta-decay
\cite{pdg06}. 
Unfortunately the result of these measurements do
depend on (a priori unknown) axial electroweak (EW) couplings
of the $u$ and $d/s$ quarks to the $W$. In turn these EW
couplings require a knowledge on these QCD quantities as well as  others like 
transition form factors which are
also measured  in semi leptonic transitions of the type $P' \to P l \nu$
where $P=\pi, K, D, B$.
At present the only well known EW quantity is the vector coupling
${\cal V}_{ud}$ of the $u$ and $d$ quarks to W. It is very accurately 
determined 
from $0^+ \to 0^+$ transitions in nuclei  assuming conservation of the vector 
current,
 ${\cal V}_{ud}=0.97377(26)$ \cite{mas05}. What can presently be given 
very precisely are
the values of the three LECs $F_\pi$, $F_K$ and $g_A$ within the 
Standard Model. Indeed in this framework the axial and vector couplings
are equal and the CKM matrix $V_{ij}^{CKM}$ is unitary 
(${\cal V}_{ud}= V_{ud}^{CKM}$
in the SM). 
All over this review I will use the Standard Model result for the 
$SU(2)$ quantities $F_\pi$ and $g_A$  namely:

 \vspace{0.2cm}
\begin{center}
\framebox{
$
F_\pi \equiv F_\pi|_{SM}= 92.4(3) \,{\rm MeV} \, ,  \quad \quad \quad 
g_A \equiv g_A|_{SM} =1.2695 \pm 0.0029
$}
\end{center}
but the reader should keep in mind that physics beyond the Standard Model
such as non standard EW couplings to quarks (for example RHCs \cite{s06})
would modify these values.  
It turns out that  
the meson sector would be more  affected than the 
nucleon one since in that case the loops are proportional to $g_A/F_\pi$,
quantity which is independent of the axial effective couplings to W. 
Note that in the standard model  the ratio
\be 
F_K/F_\pi|_{SM}=1.182(7)
\ee
(obtained from  the experimental ratio of the radiative inclusive decay rates 
for $K \to \mu \nu (\gamma)$ and
$\pi \to  \mu \nu (\gamma)$ \cite{mpdg06}) is considerably smaller 
than the value 
which has 
been used up to now in CHPT, namely $1.22$ (obtained by taking the
ratio of the central values of the experimental results for $F_K$ and $F_\pi$ 
\cite{spdg06}. There, as stressed above values for the axial EW couplings had
to be inferred). Apart from CP-PACS/JLQCD \cite{goeck98} recent SU(3) lattice 
determinations of this ratio \cite{beal06,ukqcd05,bbos06} lead to central 
values larger than the SM one, between 1.20 and 1.24 with rather small
error bars. For
a comparison of different lattice results see
\cite{kkk07}.
In Section \ref{3flav}
the exact value of this ratio will be irrelevant for the applications 
discussed.
Indeed, an average value between $F_\pi$ and $F_K$ has usually been taken, the
differences between these two quantities being in those particular 
cases of higher order.

\vspace{0.2cm}
\noindent  iii) {\bf {higher order LECs of $\cal{L}_{\pi N}$}}
\vspace{0.2cm}

Essentially two processes have been used to pin them down: 
$\pi N \to \pi N$ and
$\pi N \to \pi \pi N$. These two processes will be discussed
in more detail in Section \ref{pind}. Here I will restrict the discussion to 
what is directly related to the determination of the counterterms.

Several works have concentrated their efforts in determining the 
counterterms of order two and three in $\pi N$ scattering. Different 
methods have been used:

a) fit to the data in the physical region \cite{fm00}.

b) use of subthreshold coefficients \cite{bkm97,bl01,bkm95}. Note that 
in \cite{bl01,bkm95} they use the tree level formula which give the
LECs in terms of the subthreshold coefficients whereas in \cite{bkm97}
the next order terms are taken into account. It turns out that the
corrections are not that small leading to somewhat different values
for the counterterms. 
  
c) It has been advocated by B\"uttiker and Mei{\ss}ner 
\cite{bum97} that a much better 
determination
can be obtained working in an unphysical region of the Mandelstam plane,
namely in the inside of the Mandelstam triangle defined by  $s<(m+M_\pi)^2$, 
$u<(m+M_\pi)^2$ and $t<4 M_\pi^2$. Indeed in this region the scattering
amplitude is purely real and furthermore the kinematical variables $t$ and
$(s-u)/4m$ take their smallest values. However in this region which is by
definition
unphysical there is no direct access by experimental data. Use of 
dispersion relations allows to circumvent this problem. In this framework
each LEC appears in one particular invariant amplitude only but for
$c_4$ and $\bar d_{18}$ present in two different ones. Thus most of 
the dimension three LECs cannot be pinned down accurately since they appear
in the amplitude with small prefactors. The problem of this framework is
that it is unfortunately hard to determine
the theoretical uncertainties.

\begin{table}
\begin{center}
\begin{minipage}[htb]{18 cm}
\caption{4 LECs of ${\cal L}_{\pi N}^{(2)}$ and 2 LECs of 
${\cal L}_{\pi N}^{(3)}$.
 Values determined as explained in the text.} 
\label{tab:lecsn}
\end{minipage}
\vskip 0.4truecm
\begin{tabular}{|c|c|c|c||c|c|}
\hline
\multicolumn{4}{|c||}{}& \multicolumn{2}{|c|}{} \\
\multicolumn{4}{|c||}{order 2 $[{\rm{GeV}}^{-1}]$
 }& \multicolumn{2}{|c|}{order 3 $[{\rm{GeV}}^{-2}]$} \\
\multicolumn{4}{|c||}{}& \multicolumn{2}{|c|}{} \\
\hline
$c_1$& $c_2$
&$c_3$& $c_4$&$d_{16}$&$d_{18}$\\[1mm]
\hline
&&&&$\kappa_{16}= g_A(4-g_A^2)/8$&$\kappa_{18}=0$ \\
\hline
&&&&&\\[-1mm]
$\sigma_{\pi N}$ & $\pi N \to \pi N$ &$\pi N \to \pi N$ 
&$\pi N \to \pi N$& 
$\pi N \to \pi\pi N$&
GT rel. \\
\hline
&&&&&\\[-1mm]
$-0.9^{+0.2}_{-0.5}$ & $3.3 \pm0.2$ & $-4.7^{+1.2}_{-1.0}$ & 
$3.5^{+0.5}_{-0.2}$&$-3.4 \cdots -0.92$&$-0.72 \pm 0.27$
\\

&&&&&\\\hline
\end{tabular}
\end{center}
\end{table} 

%
%

A summary of the by now admitted values of the LECs of order 
two \cite{m05}
 are given in 
Table~\ref{tab:lecsn}. They have been determined combining results from 
\cite{fms98,bkm97,bum97,ral03} and using
a value of the sigma term of 45~MeV consistent with \cite{gls91} (\cite{ral03}
is a determination of $c_3$ and $c_4$  from $p p$ and
$n p$ scattering). 
We have discarded here the values obtained at
tree level.  
Note that attention has to be paid to the definition used for the finite
parts of the counterterms $b_i^r$ (see Eq.~(\ref{eq:renct})) when doing a 
calculation,
 their values 
depending on the regularization scheme used \cite{af07}. This
of course does not hold for the $c$'s.

Two order three counterterms enter many of the studied quantities. 
These
are $d_{16}$ and $d_{18}$. The first one has been studied in $\pi N \to 
\pi \pi N$ \cite{fbm00} but not very precisely determined.
$d_{18}$ is calculated  via the so-called
Goldberger-Treiman discrepancy i.e the deviation from the 
Goldberger-Treiman relation which states that the $\pi N N$ coupling constant
deviates from the ratio of the axial vector coupling and the pion
decay constant by terms of order $M_\pi^2$. It has been found \cite{gss88,fm00,
bl01}
that the non-analytic terms in the ratio $g_A m/F_\pi$ and in $g_{\pi NN}$
are the same up to $M_\pi^3$ so that one has: 

\begin{equation}
g_{\pi N}= \frac{g_A m}{F_\pi}\Bigl(1-\frac{2 M_\pi^2 \bar d_{18}}{g_A}\Bigr)
+{\cal O}(M_\pi^4)
\label{eq:gtrel}
\end{equation}
(the Golderger Treiman 
discrepancy has also been studied within SU(3), see \cite{glsz99}).
The values of these two counterterms are given in Table~\ref{tab:lecsn},
the one for $\bar d_{18}$ uses $g_{\pi N N}
= 13.18$. Note that this number is subject to some uncertainty, 
having decreased from the 1983 H\"ohler's value of $f^2=g_{\pi N N}^2
M_\pi^2/16 \pi m^2 =0.079$ to nowadays somewhat smaller values in 
the range from 0.075 to 0.076, see for example \cite{srt98}. A very
recent determination of $f$ from the Goldberger-Miyazawa-Oehme sum rule can
be found in \cite{ams07}.  
  
The discussion of the LECs related to the inclusion of photons is given in 
Section \ref{isoviol}.

\subsubsection{HBCHPT\label{hbchpt}} 

Let us come back briefly to the power counting, for more details see
for example \cite{BKM95r}. We have seen that in the 
meson sector loops and chiral order are intimately connected. In the
nucleon case however, an extra scale appears, the nucleon mass which is of the
order of the symmetry breaking scale and which does not vanish in the 
chiral limit. It was first pointed out by \cite{gss88} that this new scale
destroys the power counting, namely an arbitrary number of loops contributes
to a certain chiral order (apart for the first order).

The first to bring a solution to this problem were Jenkins and Manohar 
\cite{jmc91} following
methods from heavy quark physics. The idea  in this framework called heavy
baryon chiral perturbation theory (HBCHPT)
is to consider the nucleon as
extremely heavy, thus only the baryon momenta relative to the 
rest mass will be relevant and can be small. 
One has
\be
p_\mu =m v_\mu + k_\mu  \, \quad  \quad
\ee
with $k_\mu \ll v \cdot p$.
The baryon field is split into velocity-dependent ``heavy'' and
``light'' components
\be
N_v(x)= e^{imv \cdot x} P_v^+ \psi (x) \, \, , \, \, \, \, \, \,   
H_v(x)= e^{imv \cdot x} P_v^- \psi (x) \, \, , \, \, \, \, \, \,  
P_v^\pm =\frac{1}{2}(1 \pm \not v)
\label{eq:hbchpt} 
\ee
and the heavy component $H_v(x)$ is integrated out from the 
Lagrangian. One finally gets \cite{bkkm92}
\bea
{\cal L}_{\pi N}=
 &&\bar N_v \bigl\{ {\cal A}^{(1)} + {\cal A}^{(2)} + {\cal A}^{(3)} + 
 (\gamma_0 
 {\cal B}^{(1)\dagger} \gamma_0) \frac {1}{2m}  {\cal B}^{(1)} \bigr.
 +\frac {(\gamma_0
  {\cal B}^{(1)\dagger} \gamma_0 )  {\cal B}^{(2)} + (\gamma_0
    {\cal B}^{(2)\dagger} \gamma_0 )  {\cal B}^{(1)}}{2 m}
     \nonumber \\
 &&  \bigl. -(\gamma_0
   {\cal B}^{(1)\dagger} \gamma_0 ) \frac {i(v \cdot D) + g_A(u \cdot S)}
   {(2m)^2}   {\cal B}^{(1)} \bigr\} N_v + {\cal O}(p^4) \,\, .
\label{laghbchpt}
 \eea
In this equation ${\cal A}^{(1)}$ is given by 
\be
{\cal A}^{(1)} =i(v \cdot D + g_A u \cdot S) 
\label{eq:a1}
\ee
and the spin operator $2S_\mu =i \gamma_5 \sigma_{\mu \nu} v^\nu$
obeys the following relations (in $d$ space-time dimensions)
\be
S \cdot v =0, \, S^2=\frac{1-d}{4},\, {S_\mu,S_\nu}=\frac{1}{2}(v_\mu v_\nu
-g_{\mu \nu}),\, [S_\mu,S_\nu] = i \epsilon_{\mu \nu \alpha \beta } v^\alpha
S^\beta \,\, .
\ee
Expressions for the other operators can be found in \cite{bkkm92}.
As can be seen from Eqs.~(\ref{laghbchpt}), (\ref{eq:a1})
the heavy nucleon mass is, in this way, shuffled from the propagators
to the vertices. HBCHPT is thus a double expansion in $q/\Lambda$
and $q/m$. At a given order any observable is given as a sum of a
finite amount of terms. We will see that this is not the case for
the other regularizations discussed. Note that an advantage of this
method is its extreme computational simplicity.   

\begin{figure}[htb]
\epsfysize=9.0cm
\begin{center}
\includegraphics*[width=3.cm]{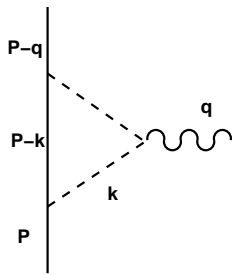}
\end{center}
\begin{minipage}[htb]{19cm}
\caption{Triangle graph. The solid, dashed and wiggly lines represent the 
nucleons, pions and an external source, respectively. \label{triangle}}
\end{minipage}
\vspace{-0.2cm}
\end{figure}

As discussed in detail  by Becher and Leutwyler\cite{BL00}, 
let us look at the convergence of this $1/m$ expansion in the case of one
specific scalar loop function corresponding to the triangle graph shown in 
Fig.~\ref{triangle}. 
This graph enters into a whole series of processes as for 
example, the nucleon electroweak and scalar form factors. The external sources
depicted by the wiggly line
are in these cases the photon and the weak bosons  or a scalar source, 
respectively.
It is given in the relativistic formulation by:
\be
\gamma (t)=\frac{1}{i} \int \frac {d^4k}{(2\pi)^4}
\frac{1}{(M^2-k^2-i \epsilon)(M^2-(k-q)^2-i \epsilon)(m^2-(P-k)^2-i \epsilon)}
\, .
\ee
This function is analytic in $t=q^2$ except for a cut along the positive 
real axis starting at $t=4 M^2$.
Its imaginary part is 
expressed in terms of elementary functions:
\be
{ \rm Im} \gamma (t)=\frac{ \theta (t-4m^2)}{16 \pi m\sqrt t} \biggl\{ {\rm arctan}
\frac{\sqrt{(t-4M^2)(4m^2-t)}}{t-2M^2}\biggr\} \, . \label{eq:igre}
\ee
The analytic continuation  of $\gamma(t)$ to the second sheet contains a
branch point $t_c=4M^2-M^4/
m^2$ just below threshold coming from the arctan. This turns out to be
very important in the calculation of the isovector charge and
magnetic radii for example, for a quantification of the effect see \cite{hp75}.  
In fact Frazer and Fulco\cite{frf} long time ago pointed out 
that
extended unitarity leads to a drastic enhancement of the isovector 
electromagnetic spectral function on the left wing of the $\rho$ resonance,
exactly due to this singularity at the anomalous threshold $t_c$.
Note, however, that even though a similar branch point appears in the isoscalar
spectral function\cite{bkm96}, the three-body phase factors suppress its
influence in the physical region. In that case, the spectral function
rises smoothly up to the $\omega$ pole.
  
What happens within HBCHPT? As was shown in \cite{bkkm92} the heavy baryon 
perturbation series  corresponds to the $1/m$ expansion of the
relativistic result. Due to the power counting the coefficient of the arctan 
is a quantity of order
${\cal O}(1/p)$ so that to ${\cal O}(p^3)$
\be
{ \rm Im} \gamma (t)=\frac{ \theta (t-4m^2)}{32 \pi m {\sqrt t}}  \biggl\{
\pi  - \frac{ (t-2M^2)}{ m\sqrt{(t-4M^2)}} +{\cal O}(p^2) \biggr\}\, . \label{eq:ighb}
\ee
Clearly at this order the normal and the anomalous threshold have coalesced. 
There is a breakdown of the expansion
Eq.~(\ref{eq:ighb}) close to threshold. 
The series indeed converges if the quantity $(t-2M^2)/ m \sqrt{(t-4M^2)}$ is 
small which is obviously not the case in that region. 
Thus the loop function Eq.~(\ref{eq:ighb})
is a decent representation only  away from threshold. In HBCHPT an 
infinite series of internal insertions must be summed up to properly describe
the loop function near threshold. As a 
consequence the isovector electromagnetic spectral functions for
example show an abnormal threshold behaviour in this framework. The origin
of the problem is  that for some of the graphs the loop 
integration cannot be interchanged with the nonrelativistic expansion.

Similar problems appear with Born terms\cite{be00} where in HBCHPT the 
positions of the
poles are moved due to the expansion of the nucleon propagator leading
again to a breakdown of the series close to the singularity. 
An example of such a problem will be given in Section \ref{elecpro} when the 
Furlan-Fubini-Rosetti sum rule will be discussed.  

Kaiser proposed  in his calculation of the electromagnetic 
form factors  \cite{k03} to expand all but 
the arctan function in order to
incorporate in a proper way the anomalous singularity, this being taken as
a compromise between the correct analytical structure and the strict chiral 
power counting. This, however, is a minimal prescription. We will now see 
how to deal with the problem in a consistent way.  

\subsubsection{Infrared regularization} \label{ir} 
The idea is then to formulate a theory which has the proper power counting
and at the same time the proper analytic structures. The first attempt
in that direction is due to Tang and Ellis\cite{t96,te96}. While in HBCHPT the 
anti-nucleon field is integrated out in order to recover the proper power
counting, Tang\cite{t96} noted that their contributions are hard-momentum 
effects, 
and that EFT's permit useful low-energy expansion only if all hard momentum
effects are absorbed into the parameters of the Lagrangian. He thus proposed
to deviate from \cite{gss88} at the loop integral level in the following way:\\
i) take the loop momenta to be of order $p$, \\
ii) make a covariant $p/m$ expansion of the integrand, \\
iii) exchange the order of the integration and summation of the resulting
power series.   \\
He showed that this prescription indeed extracts the soft part of a 
Feynman diagram and that it fulfils the power counting. The hard momentum
part which is a local polynomial in the small chiral parameters 
is then absorbed into the parameters of the most general effective Lagrangian.

This method however relies on the chiral expansion of the loop
integrals which is not always convergent. Thus Becher and Leutwyler \cite{BL00}
took up this idea of extracting the
soft momentum parts or so called infrared singular parts
and proposed a more formal 
scheme which is known as Infrared 
Regularization of baryon CHPT. It relies on the
fact that these infrared singular
parts of the loop graphs can unambiguously be separated from the remainder 
for non integer-values of the space-time dimension, leading to a unique, i.e process
independent result in accordance with the chiral Ward identities of QCD.
\begin{figure}[htb]
\begin{center}
\begin{minipage}[t]{6 cm}
\includegraphics*[width=4cm]{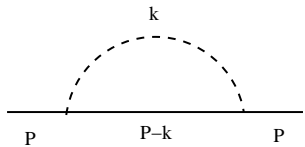}
\end{minipage}
\begin{minipage}[t]{18 cm}
\caption{Self-energy graph \label{selfene}}
\end{minipage}
\end{center}
\vspace{-0.2cm}
\end{figure}
In order to illustrate the method let us look at the simplest case, namely the self-energy, Fig.~\ref{selfene}. The corresponding scalar loop integral
is given by
\be
H (P^2)=\frac{1}{i} \int \frac {d^4k}{(2\pi)^4}
\frac{1}{(M^2-k^2-i \epsilon)(m^2-(P-k)^2-i \epsilon)} \label{eq:scalo} \,\, .
\ee
This integral develops an infrared singularity as $M$ goes to zero coming
from the low momentum region of integration. The high momentum one is
free of infrared singularities and thus leads to a contribution
which can be expanded in an ordinary power series. These corresponds to the
soft and hard momentum parts of Tang, respectively. In order to isolate the
infrared parts one introduces the standard Feynman parametrization
\be
\frac{1}{ab}=\int_0^1 dz \frac{1}{\{(1-z)a+zb\}^2}
\ee
where in our case $a$ corresponds to the pion propagator and performs a change
of variable $z=\alpha u$ with $\alpha =M/m$. The upper limit of integration
becomes 
large as $M$ goes to zero
so that one can extend the integration to infinity. One thus extracts
the infrared singular part:
\bea
I=\kappa \alpha^{d-3} \int_0^\infty du \, D^{\frac{d}{2}-2} \,\, , & &
 \nonumber \\
D= 1 - 2  \Omega u+ u^2 +2\alpha u(\Omega u -1)+\alpha^2 u^2-i\epsilon,
\,\,\, \,\,\,\, \,& &\kappa = (4\pi)^{-\frac{d}{2}} m^{d-4} \Gamma\biggl(2-\frac{d}{2}\biggr) \, \, ,\label{eq:lamm} 
\eea
where $\Omega=(P^2-m^2-M^2 )/2m M$ is a dimensionless quantity of order one. 
Clearly the chiral expansion of $I$ exclusively 
contains fractional powers of $M$ for non integer-value of the dimension. 
One thus gets the following decomposition of the scalar loop integral, 
Eq.~(\ref{eq:scalo}):
\be
H= \kappa  \int_0^1 dz \, C^{\frac{d}{2}-2}  =  \kappa \, \, 
\biggl(\int_0^\infty dz-\int_1^\infty dz \biggr) \, C^{\frac{d}{2}-2} 
\, \, \, 
= \,\,\, I \,\,\,\,\,\,\,\,\,\,\,\,\,+\,\,\,\,\,\,\,\,R \,\,\,\,\,\, ,
\label{eq:irbl}
\ee
where the explicit expression for $C$ can be found in \cite{BL00}.
The infrared singularity coming from the small values of $z$ is clearly 
excluded in the expression for $R$. This quantity contains the fractional 
powers of $m$ 
and its expansion is an ordinary Taylor series. Both parts $I$ and $R$ are
chirally symmetric by themselves so that {\it $R$ can be absorbed in
the low-energy constants} of the effective Lagrangian. 
In the Feynman parametrization,
the only difference between $H$, $I$, and $R$ is that the integrations extend
over different intervals. The one relevant for $R$ is $1<z<\infty$ which 
is mapped onto $-\infty <s<0$. Accordingly {\it $R$ has a cut along the 
negative real axis}.   
The infrared singular part contains 
a whole string of $1/m$ suppressed terms which corresponds to {\it the 
resummation of the kinetic energy corrections} to the nucleon propagator.
To leading order it reproduces the HBCHPT result. 
Different possibilities were explored in the treatment of these $1/m$ terms. 
Becher and Leutwyler who were mostly concerned with the correct treatment
of the relativistic properties kept in their applications the full denominators
of loop integrals while expanding the numerator to the desired chiral order
only. However they tried to choose the kinematic variables to be kept fixed 
when performing the chiral expansion so as to lead to the best possible 
convergence. 
Others more concerned with expanding the range of validity of the chiral 
series to 
somewhat higher energies  kept all
terms. Differences between the two approaches are of higher order in the 
expansion. 
   
What has happened to the regularization scale $\lambda$ ?
\begin{itemize}
\vspace{-0.2cm}
\item Due to the presence of the extra scale $m$ the quantity $\kappa$
in Eq.~(\ref{eq:lamm})
which in dimensional regularization is defined in terms of the regularization
scale $\lambda$, is   proportional to $m$ so that one 
naturally gets here $\lambda=m$. In fact it was already stressed
 in \cite{bkkm92}
that such a condition led to a  proper matching
of the relativistic 
approach to the heavy baryon one which, as was argued in that paper, is not 
quite unexpected 
since the heavy field of mass $2m$ were integrated out of the latter theory. 
It has one  advantage. As we have just seen the infrared singular parts 
contain 
a whole string of $1/m$ suppressed terms which 
may contain infinite pieces.
These divergences are accompanied by
the appearance of a logarithmic dependence on the scale $\lambda$. Both 
these infinite parts and the logarithms cannot be taken care of as long as
one introduces running coupling constants to a finite
order. Thus 
using the natural scale $m$ as the regularization scale removes the
otherwise unphysical scale dependence in the physical results. For the
higher order divergences they have to be removed by hand. 
\item  Another alternative
is  to keep $\lambda$ free and study the $\lambda$ dependence of the results
as was done for example in \cite{ku02}. 
\item Another philosophy has been proposed
in \cite{phw03}. The idea is  to promote to the order one is working 
counterterms of effectively higher order when
getting  $1/m$ suppressed divergences compared to this order, 
thus getting rid of the $\lambda$ dependence. In the
calculation of the nucleon mass to ${\cal O}(p^3)$ for example, 
only one 
divergence proportional to $M_\pi^4$ is found and absorbed  by adding a 
counterterm of ${\cal O}(p^4)$.
\end{itemize}

The analysis of the self-energy  can be generalized to arbitrary one-loop 
graphs which carry factors of the loop momentum in the numerator. They can
always be reduced to combinations of the scalar ones:
\be 
H_{mn}= \frac{1}{i} \int \frac{ d^dk}{(2\pi)^d} \frac{1}{a_1\cdots a_m 
b_1 \cdots b_n} \,\, .
\ee
The representation of the corresponding infrared parts in terms of 
Feynman parameters
coincides with the one obtained for $H_{mn}$ except that the integration 
over  the parameter which combines the meson propagators with the nucleon
ones runs from 0 to $\infty$ instead of 0 to 1. One has the special cases:
\be
I_{m0}=H_{m0} \,\, , \,\,\,\,\,\, R_{m0}=0  \,\, , \,\,\,\,\,\,
I_{0m}=0      \,\, , \,\,\,\,\,\, R_{0m}=H_{0m}  \,\, .
\ee
For more details on the calculations see \cite{BL00}. 

It was shown in 
\cite{BL00} that the procedure just discussed can be viewed as a 
manifestly Lorentz-invariant alternative
regularization thus the name infrared regularization given to it.

This IR method has been extended to the case of two heavy particles 
\cite{glps01} and referred to as EFT dimensional regularization. 
Recently the case where spin 1 fields are accounted 
for explicitly in the theory\cite{bm04} has been considered in the same
framework. Indeed the appearance of the large 
mass scale, the mass of the vector fields, lead to the same problem as in the
nucleon case when calculating loop diagrams. This new scale destroys the
one to one correspondence between the chiral expansion and the loop expansion
as discussed before. There is an additional complication here, namely that
the vector particles can decay into Goldstone bosons and thus appear in loops 
without appearing in external lines.

\subsubsection{Other regularizations} \label{or} 

The most important step made by Ellis and Tang was to realize that the 
power violating terms were
just polynomials. This makes it possible to come back to the relativistic
theory and within this framework to get rid of these unwanted terms. We have 
just seen one possible systematic 
way to do so namely IR which directly calculates the infrared 
terms obeying the power counting.
Soon after some other approaches have
been derived which, contrary to IR,
evaluate the power counting violating regular part $R$ which is then 
subtracted to 
the relativistic contribution.  
One of these is the extended 
on-mass-shell (EOMS) scheme~\cite{ge99,fu03}. The central idea 
consists in performing additional substractions beyond the ${\overline {\rm MS}}$
scheme. It was shown in ~\cite{geja94} that $R$
can be
obtained by first expanding the integrand in small quantities and then 
performing the integration for each term in much the same way as the expansion 
of the infrared part in Ellis and Tang. One has in the massless pion case

\bea
R=  && \int \frac{d^n k}{(2\pi)^n} 
\sum_{l=0}^\infty \frac{(p^2-m^2)^l}{l!}\left[ 
\left(\frac{1}{2p^2}p_\mu\frac{\partial}{\partial p_\mu}\right)^l 
\frac{1}{(k^2+i0^+)[k^2-2k\cdot p+(p^2-m^2)+i0^+]}\right]_{p^2=m^2}\nonumber\\ 
&=&\int \frac{d^n k}{(2\pi)^n} \biggl[ 
\left.\frac{1}{(k^2+i0^+)(k^2-2k\cdot p+i0^+)}\right|_{p^2=m^2}\nonumber\\ 
&&+(p^2-m^2)\left[\frac{1}{2m^2}\frac{1}{(k^2-2k\cdot p+i0^+)^2} 
-\frac{1}{2m^2}\frac{1}{(k^2+i0^+)(k^2-2k\cdot p+i0^+)}\right.\nonumber\\ 
&&\left. -\frac{1}{(k^2+i0^+)(k^2-2k\cdot p+i0^+)^2} 
\right]_{p^2=m^2} 
+\cdots \biggr] \, \, . \label{eq:reoms}
\eea

The use of the subtraction point $p^2=m^2$ in
Eq.~(\ref{eq:reoms}) gave rise to the 
name EOMS for the renormalization condition in analogy with the on-mass-shell 
renormalization scheme in renormalizable theories. In this example only
the first term of the series violates the power counting and is
thus a priori the only term which one needs to subtract to $H$ in order to 
get a renormalized expression $H^R$ which fulfils the proper power counting. 
Thus in the EOMS 
one has $H^R= H -H^{subtr}$ with :

\be
H^{subtr}=\left. 
-i\int \frac{d^n k}{(2\pi)^n} 
\frac{1}{(k^2+i0^+)(k^2-2k\cdot p+i0^+)}\right|_{p^2=m^2} 
\ee 
the first term in the expansion of $R$. In IR, the full series is subtracted,
I will comment on that in the summary. Note that the infrared 
regularization has been formulated in \cite{sgs04} in a form analogous to the 
EOMS
renormalization. Also it has been explicitly demonstrated within a toy 
model \cite{sgsc04} that theses two schemes can be applied to the calculation 
of multiloop diagrams. Extending the  EOMS a consistent power counting 
in manifestly Lorentz invariant baryon CHPT including 
vector mesons as internal fields\cite{fgjs03} or the $\Delta$ resonance\cite{hwgs05} has been obtained.       

Based on these same ideas, namely to extract from a loop integral $H_G$
\be 
H_G=\mu^{L(4-d)}\int \prod_{l=1}^L \frac{d^dk_l}{(2\pi)^d}\prod_{i=1}^I\frac{i}
{q^2_i-m_i^2+i\epsilon} \, \, ,\label{eq:hg}     
\ee 
the regular part $R_G$ proportional to fractional power of the heavy mass
$m$ and thus responsible for the violation of power counting,  
a Lorentz covariant regularization scheme for
effective field theories with an arbitrary number of propagating
heavy and light particles is derived in \cite{lp01}. In Eq.~(\ref{eq:hg})
$I$ is the number of internal lines, $L$ the number of loops, $\mu$ the
renormalization scale of
dimensional regularization and $q_i^\mu$ the momentum of each internal line 
$i$.   Based on the Landau
equations \cite{iz80} which gives a necessary but not sufficient condition
for the occurrence of singularities in $H_G$, the separation $H_G=I_G+R_G$
is achieved by systematically separating out subgraphs of the original 
graphs $G$ representing singularities that lie outside of the low energy 
region. These subgraphs are called regular subgraphs. Let us consider here
the case of one loop graphs and let us, following \cite{lp01}, define a 
minimally contracted regular subgraph (MCR) as a regular graph $g$ which is 
not a subgraph of
any other regular graph with more lines than $g$. The regular part is
given as a sum on these MCR as follows:
\bea    
R_G&=&- \kappa_I \sum_g \int_1^\infty d\lambda \lambda^{|g|-1}
(1-\lambda)^{I-|g|-1} 
\int_0^1\biggl(\prod_{i \in g} dz_i \biggr) \delta(1-\sum_{k \in g} z_k)
\int_0^1\biggl(\prod_{j \not\in g} dz_j \biggr) \delta(1-\sum_{k \not\in g} 
z_k) \nonumber \\
& & \times 
\biggl[ (1-\lambda)^2 \sum_{i,j \not\in g} z_i \Omega_{ij} z_j +\lambda^2
\sum_{i,j \in g} z_i \Omega_{ij} z_j  +2 \lambda(1-\lambda) \sum_{i \in g,j
\not\in g}
 z_i \Omega_{ij} z_j -i\epsilon
\biggr ]^{\frac{d}{2}-I}
\eea  
where 
\be
\kappa_I =(-)^I\frac{i^{I+1}}{16 \pi^2} \frac{ \Gamma(I-d/2)}
{(4\pi\mu^2)^{d/2-2}}, \,\,\,\,\,\,\,\,\,\,\,\,\,\,\,  \Omega_{ij}=
\frac{1}{2}\biggl(m_i^2+m_j^2-(q_i-q_j)^2\biggr) \,\, .
\ee

\begin{figure}[htb]
\begin{center}
\begin{minipage}[htb]{10 cm}
\epsfig{file=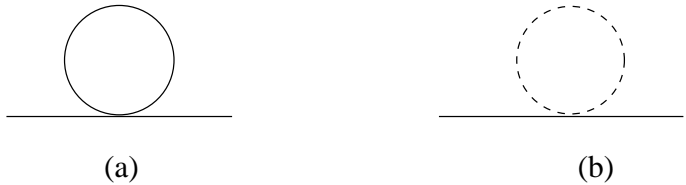,scale=0.5}
\end{minipage}
\begin{minipage}[htb]{18 cm}
\caption{Subgraphs of the triangle graph \label{subleadtri}}
\end{minipage}
\end{center}
\end{figure}

As an example let us look one more time at the triangle graph 
Fig.~\ref{selfene}.
It has two subgraphs Fig.~\ref{subleadtri} obtained by contracting $n=1$ 
internal lines of the original graphs.  
They correspond to
a 
(sub)$^n$-leading singularity with $n=1$.
Studying the Landau 
equations one finds that the tadpole Fig.~\ref{subleadtri}a yields a 
singularity
$m^2=0$ which lies outside the low energy region defined by
$p^2-m^2 \ll \Lambda$, $M_\pi \ll \Lambda$ and $m \sim \Lambda$ 
while the  other one
has a singularity $M_\pi^2=0$ well inside the low energy region. Thus graph
\ref{subleadtri}a is the regular graph and the expression for $R_G$ reads:
\be
R_{G}=-\kappa \int_1^\infty d\lambda\biggl[(1-\lambda)^2 M_\pi^2+
\lambda^2 m^2 +2 \lambda (1-\lambda)\frac{(m^2+M_\pi^2-P^2)}{2} \biggr]^{
\frac{d}{2}-2}
\ee 
which is nothing but the result obtained by Becher and Leutwyler 
Eq.~(\ref{eq:lamm}).

In Ref.~\cite{lp01} the Taylor series expansion of $R_G$ in powers of $p$ is 
truncated at the order of the EFT Lagrangian. For a more detailed discussion 
we refer the reader to that paper.

Up to now we have discussed schemes which used dimensional regularization.
In fact any other scheme 
can be used as long as it respects the Ward identities. For 
example cut-off methods have a long history in CHPT \cite{ga81}.  
However it has
been demonstrated in \cite{esma94} that the Ward identities severely
constrain the choice of regulators. Also, 
doubts have been issued in the literature \cite{dh98,dhb99,ylt03,lty04} 
whether  
an effective field theory when  utilized in connection with
dimensional regularization is ``effective'' enough to be applied 
to extended objects with complicated internal structure like baryons.
These are of course not acceptable since from the viewpoint of field 
theory  one scheme cannot provide superior 
results over the other. I will come back to this issue when
discussing recent analysis of the nucleon mass as well as calculations 
within SU(3). We will see that in some of these analysis Ward identities
are not respected. A way of implementing a cut-off regularization 
which do not break chiral symmetry or gauge invariance has been discussed 
in \cite{bhm04}. Also
Djukanovic et al. \cite{dsgs05} have introduced a method of improving the
ultraviolet behavior of the propagators which is a smooth cutoff 
regularization scheme and which preserves all symmetries. 
    
\subsubsection{Summary}  \label{sum}

A summary of the differences between the relativistic, heavy $m$ and infrared
approaches is given in Table~\ref{tab:addi}. As seen there, while in HBCHPT
the series is finite -- it sums terms up to the order one is working --  it is 
infinite in the other approaches, higher order 
terms compared to the order 
one is calculating being included. However if one expands 
a physical quantity obtained in  these different 
approaches up to a certain order
the result will of course be the same. What can be different in these 
different regularizations are the numerical values of the 
counterterms since the treatment of the analytic terms differs. 
Let us come back to the higher order terms.  In the infrared regularization 
one knows exactly what they are. They come from a 
resummation of the $1/m$ kinetic energy terms.  
However the prize to
pay is the appearance of additional unphysical cuts (see discussion after 
Eq.~(\ref{eq:irbl})). 
These can give rise to rapidly growing quantities when one goes toward larger 
momentum or larger pion masses that is when one goes beyond the validity of 
the theory. This is thus not a problem in itself. It only becomes one if
{\it one wants to push the theory too far}.  
The problem comes merely from the fact that whereas functions like  
arccos($M/m$) 
appear in the relativistic calculation they are transformed into
arccos(-$M/m$) in IR. We will show this in one application in section
\ref{magm}.
   
\begin{table}
\begin{center}
\begin{minipage}[htb]{14.5 cm}
\caption{Properties of diverse approaches to CHPT with nucleons.} 
\label{tab:addi}
\end{minipage}
\vskip0.4truecm
\begin{tabular}{|l|l|l|l|}
\hline
& & & \\
& Relativistic
&Heavy $m$& IR\\[2mm]
\hline
&&&\\[-2mm]
Power counting & no & yes & yes\\[2mm]
$\to$ series & infinite: &  finite & infinite: $1/m$ kinetic energy included.
 \\
contributions at $q^N$   &all orders & the order $N$ &   
 orders $\ge N$  \\[2mm]
analyticity & yes & not always & yes (modulo unphys. cuts at large \\
& & &energies)\\[2mm]
regularization & dim. reg.: & dim. reg. & infrared  reg.:\\
&fractional powers of $m$& &eliminates
fractional powers of $m$\\[2mm] 
order by order & exact &exact &
higher order divergences   \\
renormalization & & &
removed by hand, 
 $\lambda = m$\\[2mm]       
convergence & property not applicable  &not good in  
& improved with resp. to HBCHPT, \\
& &{\bf certain} cases& larger energy range in {\bf certain} cases \\
[2mm]
leading non- & yes& yes & yes\\
anal. terms &&& \\[2mm]
\hline
\end{tabular}
\end{center}
\end{table}     

\subsection{\it Effective theory with $\Delta$'s\label{eftdel}}

Up to now we have considered the nucleon as the only degree of freedom
in the baryon sector. Any QCD state of higher
mass appeared only implicitly in the values of certain LECs. 
A reason for that is the existence of the decoupling theorem~\cite{gz80}
which states that in the chiral limit all S-matrix elements and transition
currents are given in terms of the Goldstone bosons and the ground-state
baryon octet. However it is well-known that the $\Delta(1232)$
resonance plays an important role in many phenomenological descriptions
of low- and medium-energy processes as we will see examples in the 
next sections. One can understand this very well at least for three
reasons: first, there is a very small mass gap between this particle and
the nucleon, second it is rather strongly coupled to the $\pi N$ channel 
and finally it has relatively large photon decay amplitudes.  
It  thus seems quite legitimate
to ask oneself whether one should not include this particle as an
explicit dynamical degree of freedom of an effective field theory, 
 as was first stressed by Jenkins and Manohar\cite{JM91},  for
describing processes which are not restricted to the threshold domain
but which cover energies up to the resonance region.   
The idea is that the convergence of the chiral expansion could be 
improved by shuffling to lower order some terms which would appear
later in the series in conventional BCHPT. The problem is again to define 
a scheme with a proper power counting. 

\vspace{0.3cm}

$\bullet$ {\it {Small scale expansion}}

\vspace{0.2cm}

It turns out that it is possible to do so if one counts the nucleon-delta
mass splitting $\Delta$ as an additional small parameter, as was first 
shown by Hemmert, Holstein and Kambor\cite{hhk96}. Indeed 
\begin{equation}
\Delta \equiv m_\Delta-m =294 \, {\rm MeV} \sim 3 F_\pi
\label{eq:ssdel}
\end{equation}
where $F_\pi$ was taken as scale since $\Delta$ is not vanishing in the 
chiral limit. Thus, 
one can formulate an  extended
EFT in which any matrix
element or transition current has  a low energy expansion in power
of $\epsilon$ where 
\begin{equation}
\epsilon  \in \left\{ \frac{p}{\Lambda_\chi},  \frac{M_\pi}{\Lambda_\chi}, 
\frac{\Delta}{\Lambda_\chi}  \right\}~,
\end{equation}
with $\Lambda_\chi \simeq 1\,$GeV the scale of chiral symmetry
breaking. This power
counting scheme  
is often called $\epsilon$-- or small scale expansion (SSE).
Such an  EFT however
does not have the same chiral limit as QCD as it is well known\cite{gz80} 
since in the chiral limit of 
vanishing quark masses, neither $\Delta$ nor $F_\pi$ vanish. Strictly
speaking it is a phenomenological extension of HBCHPT.

The $\Delta$ is described within the Rarita-Schwinger (R-S) formalism. 
A few points in that respect have to be remembered. The R-S spinor
$\psi_\mu$ has more degrees of freedom than required which has 
a certain number of consequences. Indeed the need of subsidiary conditions
that the spinor must fulfil leads to the non-uniqueness of the classical
$\Delta$
Lagrangian. The family of one parameter Lagrangians ${\cal L}(A)$ which can be
constructed is invariant under the following point transformation: 
$\psi_\mu \to \psi'_\mu= R_{\mu \alpha}(a) \psi_\alpha$
where $A \to A'= (A-2a)/(1+4a)$ and $R_{\mu \nu}(a)=g_{\mu \nu}
+a \gamma_\mu \gamma_\nu ,\,\,\, a \neq -\frac{1}{4}$, the operator $R_{\mu
\nu}$
acting only on the spin 1/2 content of $\psi_\mu$. The physical quantities
should of course be $A$-independent which is assured by the Kamefuchi-
O'Raifeartaigh-Salam theorem. This last property as well as the fact that
only the correct number of degrees of freedom have to be left constrain the 
construction of any interaction terms. This leads to the  appearance of the
so-called ``off-shell parameters'' in these interaction terms.
 
As in heavy baryon CHPT one has to identify 
the ``light'' and ``heavy''
degrees of freedom, the problem 
being somewhat more challenging due to these off-shell spin-1/2
degrees of freedom associated with the R-S field. 
In order to separate the spin-3/2 from the spin-1/2 components 
it is convenient to introduce a complete set of orthonormal spin projection 
operators for fields with {\it fixed velocity} $v_\mu$. For example the
projector on the 3/2 component reads: 
\begin{equation}
P^{3/2}_{(33)\mu \nu}  =  g_{\mu \nu} - \frac{1}{3} \gamma_{\mu} \gamma_{
                               \nu} - \frac{1}{3} \left( \not\!{v} \gamma_{\mu} 
                               v_{\nu} + v_{\mu} \gamma_{\nu}\not\!{v} \right).
\end{equation}
The ``light'' spin-$3/2$ degrees of freedom are then defined by
\begin{equation}
T_{\mu}^i (x) \equiv  P_{v}^{+} \; P^{3/2}_{(33)\mu\nu} \; \psi^{\nu}_i (x) 
                    \; \mbox{exp}(i m_B v \cdot x) \label{eq:T}
\end{equation}
where $m_B$ is a baryon mass scale and  
$P_{v}^{+}$ is the projection operator of the heavy mass formalism,
see Eq.~(\ref{eq:hbchpt}).
$T_{\mu}^i$  satisfy the constraints
\begin{equation}
v_\mu T^{\mu}_i=\gamma_\mu T^{\mu}_i=0 \label{eq:subsidiary}
\end{equation}
and correspond to the SU(2) version of the decuplet field introduced in 
Ref.\cite{JM91}.
The remaining degrees of freedom -the heavy 3/2 components
and the four off-shell spin-1/2 contributions - form the heavy baryon fields. 
They are combined in a  five  
component vector denoted by $ G_{\mu}^i (x)$
and are integrated 
out. 

Let us now
consider the most general Lagrangian 
involving relativistic spin-1/2 ($\psi_N$) and spin-3/2 ($\psi_\mu$) fields: 
\begin{equation}
{\cal L}={\cal L}_N + {\cal L}_\Delta + \left( {\cal L}_{\Delta N} + h.c. 
\right) \, \, .\label{eq:4L}
\end{equation}
The part describing the nucleon will not be discussed here since it is 
the same as in heavy baryon CHPT.
Rewriting ${\cal L}$ in terms of the ``light'' and ``heavy'' components one 
finds \cite{hhk96}
\begin{eqnarray}
 {\cal L}_{\Delta N}&=& \bar{T} {\cal A}_{\Delta N} N + \bar{G} {\cal B}_{\Delta N} N +
                 \bar{H} {\cal D}_{N \Delta} T + \bar H {\cal C}_{N \Delta} G +
                  h.c. \, \, ,
\nonumber\\
{\cal L}_{\Delta}&=&\bar{T} {\cal A}_{\Delta} T + \left( \bar{G} {\cal B}_{\Delta} T +
             h.c. \right) - \bar{G} {\cal C}_{\Delta} G \, \, ,
\label{eq:ldeltan}
\end{eqnarray}
where the ${\cal B}$ are five components vectors and the ${\cal C}$
$5 \times 5$ matrices. $\cal A$,  ${\cal B}$,  ${\cal D}$ and  ${\cal C}$
have a low energy expansion which starts at 
${\cal O}(p)$ for the first three quantities and at ${\cal O}(p^0)$ for the
last one. Explicit expressions for these quantities 
can be found in \cite{hphd97}.
Note that the definition of the
heavy nucleon field must involve the same mass $m_B$
as in the heavy $\Delta$ one in order that all exponential
factors drop out in Eq.~(\ref{eq:ldeltan}).
Choosing the heavy baryon mass parameter $m_B=m$ the leading order matrices
are given by 
\begin{eqnarray}
{\cal A}_{\Delta N}^{(1)}  &=& g_{\pi N\Delta} w_\mu^i \, \, ,   
 \nonumber\\
{\cal A}_{\Delta}^{(1)} &=& - \left[ i \; v \cdot D^{ij} - \Delta \; 
\delta^{ij} + g_{1} \; S \cdot u^{ij} \right] \; g_{\mu \nu}  \, \, ,
\label{eq:llead}
\end{eqnarray}  
where the chiral invariant couplings are defined as:
\be
D_{\mu}^{ij} \; \psi^{\nu}_j  =  \left( \partial_{\mu} \; \delta^{ij} + 
                                   \Gamma_{\mu}^{ij} \right) \psi^{\nu}_j \,\,\, , 
u_\mu^{ij}         =  u_\mu\delta^{ij}-i\epsilon^{ijk}w_\mu^k  \,\,\,\,\,\,\,\,  ,
w_\mu^i            = \frac{1}{2} Tr 
                         \left[ \tau^i u_\mu \right] \,\,\, .
\ee
Shifting variables and completing the square  in analogy to the 
heavy mass formalism one gets:
\begin{equation}
S_{\rm eff}= \int d^4x \left\{ \bar T \tilde {\cal A}_{\Delta} T
+\bar N \tilde {\cal A}_{N} N
+\left[ \bar T \tilde {\cal A}_{\Delta N} N + h.c.\right] \right\}
\label{Seff}
\end{equation}
with
\begin{eqnarray}
\tilde {\cal A}_{\Delta } &=& {\cal A}_{\Delta} 
+ \gamma_0 \tilde {\cal D}_{N \Delta}^\dagger \gamma_0 \tilde {\cal C}_N^{-1} 
\tilde {\cal D}_{N \Delta} 
+ \gamma_0 {\cal B}_\Delta^\dagger \gamma_0 {\cal C}_\Delta^{-1} 
{\cal B}_{\Delta } \,\,\,\, ,
\nonumber \\ 
\tilde {\cal A}_{N} &=& {\cal A}_{N} 
+ \gamma_0 \tilde {\cal B}_{N}^\dagger \gamma_0 \tilde {\cal C}_N^{-1} 
\tilde {\cal B}_{N } 
+ \gamma_0 {\cal B}_{\Delta N}^\dagger \gamma_0 {\cal C}_\Delta^{-1} 
{\cal B}_{\Delta N} \,\,\,\,\, ,
\nonumber \\ 
\tilde {\cal A}_{\Delta N} &=& {\cal A}_{\Delta N} 
+ \gamma_0 \tilde {\cal D}_{N \Delta}^\dagger \gamma_0 \tilde {\cal C}_N^{-1} 
\tilde {\cal B}_{N} 
+ \gamma_0 {\cal B}_\Delta^\dagger \gamma_0 {\cal C}_\Delta^{-1} 
{\cal B}_{\Delta N} \,\,\, ,
\label{eq:Atilde}
\end{eqnarray}
which represents the master formula of the treatment of a
coupled spin-1/2 - spin-3/2 system in HBCHPT. 
It enables one to construct the $1/m_B$ corrections to the leading order
contributions  Eq.~(\ref{eq:llead}). One of course has
to add at each order the most general counterterm Lagrangian
consistent with the symmetries of the system. It has been given to 
${\cal O}(\epsilon^2)$ in \cite{hphd97,hhk98}. 

\vspace{0.3cm}

$\bullet$ {\it{Infrared and other regularization}}

\vspace{0.25cm}

As in the case without  $\Delta$  other regularizations than the one
discussed here can be used. 
I will not report here on the attempts to develop new regularizations in the
line of what has been done in the pure nucleon sector 
but just refer the reader to the papers 
by \cite{te02} and \cite{bhm03,hwgs05,shgs04}. 

\subsection{\it Lattice QCD and CHPT} \label{lat}

We have just seen that CHPT is the low energy effective theory of the 
Standard Model and that it allows for analytical calculations of low-energy
QCD processes in terms of the light pseudoscalar mesons masses. 
Another model-independent way to try and solve QCD is lattice QCD.
Since Wilson first introduced this method
\cite{wil74}
enormous progress
in that field has been made due to advances in computer power as well as
improvements in algorithms and actions. First of all the question of how
to deal with chiral symmetry on the lattice has been solved with 
the Ginsparg-Wilson relation  \cite{gw82}.  
There exists thus today essentially
two categories of fermions 
depending on whether they obey (type II) or not (type I)
this relation. Contrary to the latter the former preserves chiral 
symmetry for zero lattice spacing. Also an important feature is that 
chiral fermions are automatically ${\cal O}(a)$ improved. For 
a review on chiral fermions  see  for example \cite{cw04}. 
Second 
from the quenched era where the 
fermion 
determinant was taken as a constant one has now  come to the dynamical era
where full unquenched calculations are slowly becoming available. 
Different sea and valence
quark masses are also taken leading to partially quenched theories. 
There are, however, still problems to be solved essentially linked to the
cost of the simulations. Let me summarize the 
different realizations  of fermions which are actually used  each of them 
having their own 
advantages
and disadvantages. The Wilson fermions, historically the first, 
${\cal O}(a)$ improved ones \cite{sy80}
and the staggered fermions \cite{ko75}, see also \cite{sh94,gs85})
are of type I. 
They have the advantage of being 
simple to deal with. However they have some drawbacks. The 
staggered fermions for example
naturally appear with four copies (tastes). For the sea 
quarks a
fourth root 
trick has thus to be used to reduce the determinant to one flavour 
which may lead to errors arising from non-localities.
Lattice QCD with Wilson quarks and a chirally twisted term
represents a promising alternative regularization which does not
suffer from unphysical fermion zero modes, it is referred as twisted mass 
QCD (tmQCD)
\cite{fre01}. However some technical complications arise due to the fact that
some of the physical symmetries such as flavour symmetry and parity are 
only restored in the continuum limit. 
Domain wall fermions (II) \cite{ka92,sha93} have the great advantage that 
chiral symmetry can be 
preserved to a very high accuracy depending on the size of the 
fifth dimension $L_5$, the problem being that the computational 
cost grows with $L_5$. Calculations with overlap fermions (II) which have 
exact ${\cal O}(a)$ 
improvement and exact chiral symmetry \cite{neu98}  turn out to be 
expensive and thus involve dynamical simulations on  small 
volumes. 
In order to try to optimize all these problems the use of  mixed-action
calculational schemes have been recently developed \cite{re04}. An example 
of the latter is the use of
Domain-Wall or Overlap fermions as valence quarks and staggered or
Wilson-like as cheaper sea quarks \cite{be06}. However the fact that valence 
quarks
do not match with the sea quarks leads to scaling violations. A summary of 
some collaborations cited in the following  
and the type of fermions they use is given in Table \ref{tab:coll}.

\begin{table}
\begin{center}
\begin{minipage}[htb]{18 cm}
\caption{Collaborations cited in the text and corresponding action. 
References are given in the text.} 
\label{tab:coll}
\end{minipage}
\vskip 0.4truecm
\begin{tabular}{|l|l|l|l|l|}
\hline
\multicolumn{1}{|c|}{}& \multicolumn{4}{|c|}{} \\
\multicolumn{1}{|c|}{action}& \multicolumn{4}{|c|}{Collaboration} \\
\multicolumn{1}{|c|}{}& \multicolumn{4}{|c|}{} \\
\hline
Wilson& JLQCD 
&CP-PACS  &QCDSF \& UKQCD  &LPHC/SESAM \\
\hline
Kogut-Susskind&MILC&  &&\\
\hline
Domain wall& RBC &CP-PACS & & \\
\hline
Hybrid&LPHC/MILC  & & & \\
\hline
\end{tabular}
\end{center}
\end{table} 

Every lattice calculation is done in finite volume, finite lattice
spacing and finite quark mass. The state of the art calculation involves
$L > 2.5$ fm, $a < 0.1$ fm and $M_\pi> 250 \, {\rm {MeV}}$ which 
is still far from the  real situation.
Actually the Kogut-Susskind (staggered)
formulation enables to go to the smallest 
quark masses.  
To improve on these numbers demands much more computational efforts which 
turns out to be  quite 
expensive and are not foreseen for the near future. 
It thus seems very appreciable to have an analytical method which
would help understanding the dependence on these three extrapolation 
parameters as well as on the diverse approximations which have to be done 
when one works on the lattice. 

In fact CHPT is such a method. For a recent review on the applications of 
CHPT to lattice QCD see \cite{sh06}. Here I will just briefly sketch the
diverse extensions of this method  which have been performed 
to estimate the size of systematic effects in lattice calculations:

$\bullet$ It was for example proposed \cite{sh98,lee99} to include  
non-zero lattice spacing in CHPT. In that case the two-step matching 
procedure to effective field theories (Lattice $\to$ Symanzik $\to$ CHPT)
has proven to be an appropriate tool \cite{baer04}. Presently the
chiral Lagrangian including the $O(a^2)$ lattice artifacts has been derived 
\cite{barush04}. It is indeed important to go up to that order since
lattice spacings in current unquenched simulations are not very small. 
Also it will allow to learn how the continuum limit is approached in the 
case of  the improved Wilson fermions since for these fermions 
the leading corrections are precisely of that order. 

$\bullet$ 
Finite volume effects have also been examined within CHPT
\cite{col04,da03}. 
Gasser and Leutwyler \cite{galeu87,galeu88} were the first to extend 
the CHPT framework to a finite volume: it becomes a systematic
expansion in both the quark masses and the inverse box size. In order
for CHPT to be valid $L$ has to fulfil  the following condition: 
$L \gg 1/2F_\pi \sim 1 \, {\rm {fm}}$, where $4 \pi F_\pi$ has been
taken as the chiral symmetry breaking scale. There is another important
relation which has to be considered namely the one between the pion mass 
$M_\pi$ and $L$. Indeed it 
determines the importance of the Goldstone boson zero modes in the 
 evaluation of the path integral in CHPT.   
$M_\pi L \ll 1$ defines the so called $\epsilon$-regime, there
the contribution from the zero-modes are non-perturbative and one
has to modify the usual power counting. Approaching the  
chiral limit on the
lattice implies working in this regime. In the other extreme, $M_\pi L \gg 1$, 
one is in the  $p$-regime, standard CHPT can be applied. In that case, 
finite-volume effects are typically small, deviations are
exponentially suppressed by factors of $M_\pi L$ and relevant for very precise
lattice calculations. Calculations done 
within the $\epsilon$ and $p$ regime can be found for example in 
\cite{ghl04,bcjns04,cod04} in the meson case and 
in \cite{begrru05,bea04} in the nucleon one. Note that finite volume
effects start only at the one loop level requiring two-loop calculations to
check the convergence of the series.  
As has been pointed out in \cite{desa04} another regime could be of interest
for heavy objects as the nucleon, the so-called  $\epsilon'$ regime. There, 
the  behavior of observables are explored in highly asymmetric volumes, long 
in the time-dimension allowing these heavy objects to be near their 
mass-shell, and short in the spatial dimensions.

$\bullet$ Staggered CHPT has been developed in order to control the extrapolation
and the taste breaking. 
The chiral Lagrangian 
for one single staggered fermion was first derived by Lee and Sharpe \cite{lesh99}
considering the quark mass and $a^2$ to 
be of the same order. Aubin and Bernard have generalized it to $n$ staggered
flavors and shown how to accommodate the  root square trick in loop 
calculations \cite{ab03}. A discussion of the validity of the procedure
can be found in \cite{cbe06}.
Fitting their lattice data to NNLO calculations \cite{amilc04} the MILC 
collaboration has provided a sensitive test of the lattice simulations and 
especially of the chiral behavior, including the effects of chiral logarithms.
Most of the studies so far involve the mesons, 
however staggered lattice artefacts have also recently been incorporated into
heavy baryon CHPT \cite{babe05}. Extension of CHPT to mixed actions have 
also been performed \cite{bbrs05}. 

$\bullet$ All these extensions are done using dimensional regularization
or some other continuum regularization method. An interesting development
is the definition of lattice CHPT as an effective theory that exists
directly in the same discrete space-time where lattice QCD resides
\cite{mr94,ss99,blo03}.

We have demonstrated that CHPT  is a useful tool to understand some of the 
limitations linked with calculations done on the lattice. However the 
inverse is also true: lattice
calculations are a useful tool for determining hadron properties within CHPT. 
Indeed we have seen in Section \ref{eft}  that in this last framework
low-energy constants appear which are not fixed by symmetry considerations
but are in principle calculable in QCD, the underlying fundamental theory. 
This is where lattice QCD enters. Already a lot of
effort has been made  to pin down some of the ${\cal {O}}(p^4)$ low-energy 
constants in the meson sector \cite{amilc04}
especially the combination $2 L_8 - L_5$ whose knowledge is required to test 
whether the up quark mass is massless \cite{immsw01,fnk02,fgmss03}, see also 
section \ref{lec}. It was
suggested in \cite{begrru05} that the low-energy constant $c_3$ of the nucleon
chiral Lagrangian  could be determined by fitting to the nucleon mass
at a magic ratio $\beta=1.22262 L$ where $\beta$ is the temporal extend of the
box and $L$ the size in the spatial direction,
however, there exists at present no lattice 
determination of any LECs in the baryon
sector.

\section {Chiral extrapolations} \label{chextrapol}

Numerous works on chiral extrapolation have been done  
in the last years  in the meson sector as well as in the  
baryon one. In this section  I will essentially report on the quark mass
extrapolation of some physical nucleon properties in continuum CHPT. One will
have of course to keep in mind that the lattice artifacts might not be small.
I will also touch upon the question of how to treat  unstable particles 
in a finite volume. 
In the next subsections I will concentrate on SU(2) and leave the discussion
of the SU(3) case  for the last section.
 
\subsection {{\it Nucleon mass}} \label{nucm}

The nucleon mass is particularly interesting from the lattice point of view
since, once chiral extrapolation will be under control, it will be one
of the best ways to estimate the lattice spacings. 
The general expression for the nucleon mass to fifth order \cite{mgb06} is 
given in
terms of the leading term in the chiral expansion of the pion mass $M$ by 
(the fourth order 
calculation can be found in \cite{smf98,km99}): 
\begin{equation}
m=m_0-4 c_1 M^2 -\frac{3 g_0^2  M^3}{32 \pi F^2}
+k_1 M^4 \ln \frac{ M}{m} +k_2  M^4 + k_3   M^5 \ln \frac{ M}{m  }
+ k_4 M^5 + {\cal {O}}(M^6)
\label{eq:mextra}
\end{equation}
where $m_0$ and $g_0$ are respectively the nucleon mass and the axial coupling 
in the chiral SU(2) limit. For a recent full two-loop 
expression involving terms up to $M^6$ see \cite{sdgs06}.    
Replacing in Eq.~(\ref{eq:mextra}) $M^2$ by its one loop result brings an extra $M_\pi^4$
contribution proportional to the LEC $l_3$ from the
 ${\cal {L}}_{\pi\pi}^{(4)}$ 
Lagrangian. 
The coefficients
$k_1$ and $k_2$ are combinations of second ($c_1,c_2,c_3$) and fourth order
LECs while $k_3$ and $k_4$ involves the third order ($d_{16}, d_{18}, d_{28}$)
as well as the fourth order LECs from the meson sector $l_3$ and $l_4$:
\begin{eqnarray}
k_1&=&-(3/32 \pi^2 F^2)(-8 c_1+c_2+4c_3 +g_A^2/m_0) \,\,\,\,\,\,\,\, , 
k_2= -4 e_1+(3/128 \pi^2 F^2)(c_2-2 g_A^2/m_0)  \, \, ,\cr
k_3&=&(3/1024 \pi^3 F^4) (16 g_0^2-3) \, \, ,\cr 
k_4&=&(3/32 \pi F^2)((2l_4^r 
-3 l_3^r)/F^2 -4(2d_{16}^r-d_{18})/g +16 d_{28}^r +g_0^2/(8 \pi^2 F^2)
+1/8 M^2)\, \, .
\label{eq:coeffmextra}
\end{eqnarray}
 
\begin{figure}[htb]
\epsfysize=9.0cm
\begin{minipage}[b]{6 cm}
\vskip -2.cm
\includegraphics*[width=7cm]{msu2new.eps}
\vskip -5.9cm
\hskip 10cm
\includegraphics*[width=5.3cm,angle=270]{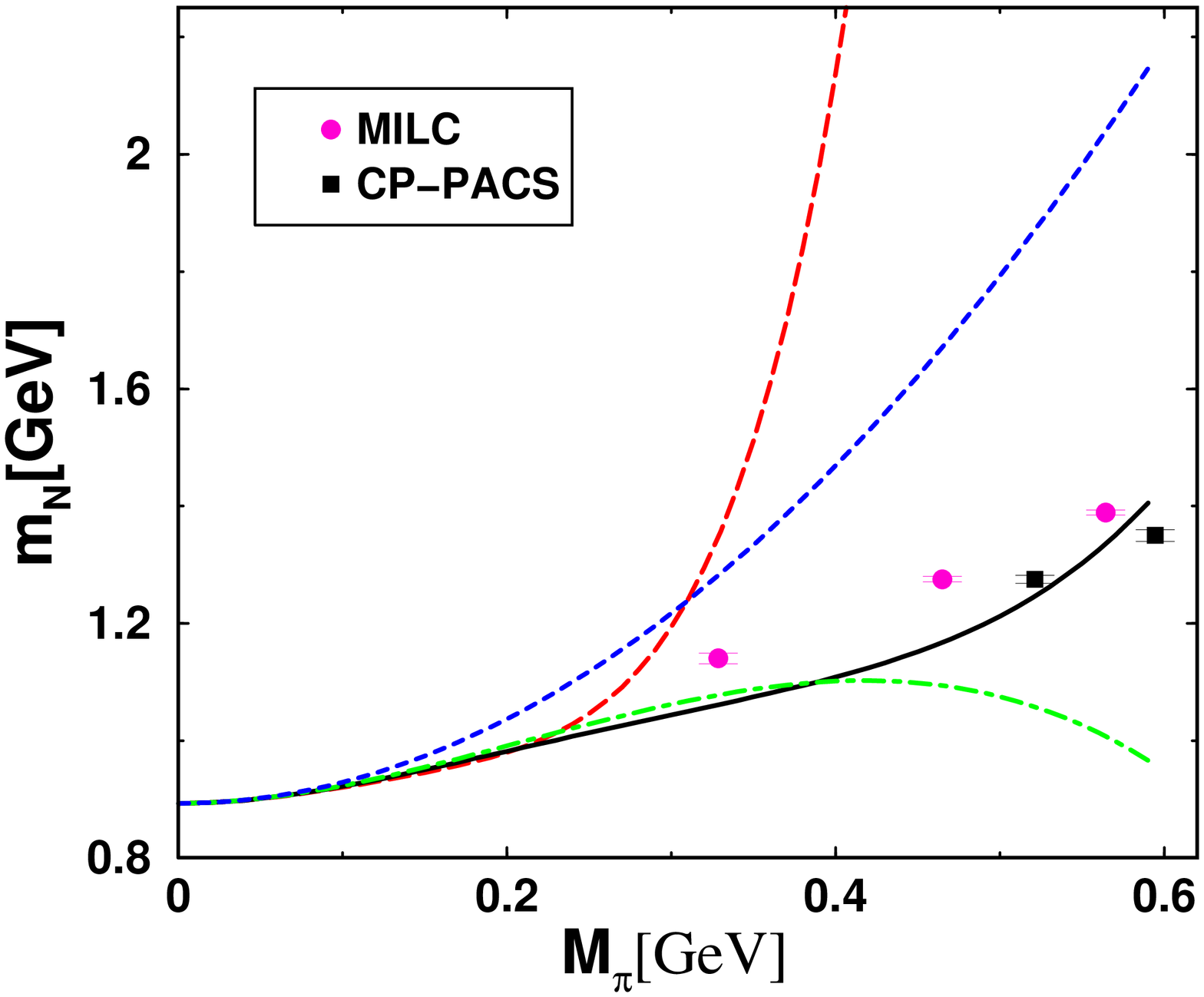}
\end{minipage}
\caption{Nucleon mass. Left panel: the  solid line refers to the best fit
and the dashed  line to the theoretical uncertainty. Results are from 
Ref.\cite{bhm04}. Right panel: convergence of the chiral series. The 
long-dashed, solid, dot-dashed and dashed lines correspond to 
$n=5,4,3,2$, respectively, where $n$ is the order of the series (without
theoretical errors). 
 \label{fig:MN2}}
\end{figure}

Apart from a combination of three different fourth order LECs denoted by 
$e_1$, the 
values of all these low-energy constants are known within some error bars as 
has been discussed in Section \ref{lec}. Thus the coefficient of the $M^5$ term
is rather well determined.  
The ${\cal {O}}(p^4)$ result for the nucleon mass \cite{bhm04}
is shown on the l.h.s. of Fig.~\ref{fig:MN2} in comparison to 
the 
lattice data from MILC \cite{ber01} and CP-PACS \cite{alik02}. There are also 
published
results concerning the hadron masses from the JLQCD \cite{aok03} and  
UKQCD and QCDSF\cite{all02,alik04} collaborations not shown here, 
which   
remarkably
fall (with a rather good accuracy) onto the same curve as the CP-PACS one 
even though they correspond to 
different lattice actions and algorithms (see for example \cite{goec04}).
An extremely
good fit of the lattice data up to amazingly large values of the pion
mass is obtained with $c_1=-0.9$, $c_2=3.2$, $c_3=-3.5$ (all in GeV$^{-1}$)
and $e_1(m  )=-1 \,{\rm{GeV}}^{-3}$ where $g_0$ and $F$ have been taken at 
their
physical values. 
With these parameters the chiral limit value of the nucleon mass
is  $0.89$ GeV. In this fit the constraint that $m  $ gets its
physical value at the physical value of $M_\pi$ was imposed. 
The fact that the $c_i$'s are consistent
with their expected values and that $e_1$ is of natural size  
has led in some literature (see for example \cite{phw03}) to the 
claim that the chiral expansion is valid up to rather high pion masses. 
This is of course not correct. First of all as shown on Fig.~\ref{fig:MN2}
there is a rather large theoretical uncertainty as $M_\pi$ becomes large
due to our rather bad knowledge of $e_1$. Secondly as naively 
expected the
convergence of the series worsens  as $M_\pi$ increases as is clearly seen 
on the r.h.s. of Fig~\ref{fig:MN2}. There $c_3$ is taken at its somewhat bigger central value and $F$ and 
$g$ as obtained from chiral extrapolation (see Eq.~(\ref{fpi}) and 
Section \ref{ga}), $F=86.5$ MeV, $g=1.2$,
$\bar d_{16}=-1.76\,{\rm{GeV}}^{-2}$ and $e_1=5\,{\rm{GeV}}^{-3} $.  One gets the following
 result (with coefficients in appropriate units of powers of~ GeV):
\begin{equation}
m=0.893+3.6M^2 -5.74M^3
-20.01 M^4 \ln \frac{ M}{m} +8.62  M^4 + 55.99   M^5 \ln \frac{ M}{m}
+ 213.38 M^5 + {\cal {O}}(M^6) \, \, ,
\label{eq:mexp}
\end{equation}
where the large contribution to the $M^5$ terms comes essentially from the 
$d_i$. In this expression all but the $M^4$ and the $M^5$ terms (the one
independent of the $\ln$ terms)  are known to a 
very good approximation.
At the physical value of the pion mass this leads to 
$m=0.893(1+0.078-0.017-0.015 +0.006)$~ MeV
which is a rather well converging series.
Note that one should
compare the $O(p^2)$ with the full one loop $O(p^4)$ and full two-loop  
$O(p^6$) (not yet analysed) results. An early attempt to quantify the errors 
associated with terms of order higher than $M_\pi^4$  can be found in 
~\cite{bea04}. In view of   
Fig.~\ref{fig:MN2} the chiral extrapolation 
can be trusted for pion masses below $\sim350$ MeV. Calculations of the 
nucleon mass
including the $\Delta$ as an explicit degree of freedom have also been 
performed \cite{bhm05,proc06} leading to similar~ fits. 

\begin{figure}[b] 
\epsfysize=9.0cm
\begin{center}
\begin{minipage}[tb]{6 cm}
\epsfig{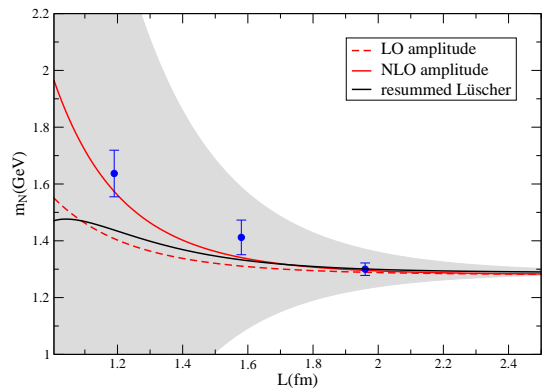}
\end{minipage} 
\begin{minipage}[tb]{18 cm}
\caption{ Finite volume effect for the nucleon mass as a
  function of $L$ for $M_\pi=0.545$ GeV. The lattice data are from
  Ref.~\cite{Aoki:2002uc}. The shaded band represents the uncertainty of 
the  calculation. Figure taken from \cite{cofuha06}. \label{fig:MNL}}
\end{minipage}
\end{center} 
\end{figure}

Here we have studied the nucleon mass as  an expanded series
in $M_\pi$ discussing the range of validity of the calculation
keeping in mind that CHPT is by construction a ``{\bf {low energy
effective field theory}}''.    
In the literature
another philosophy has also  been pursued based on the observation that 
the lattice data show a slow variation as the pion mass becomes moderately
large presenting a clear signal that the higher order terms must cancel
as $M_\pi$ increases toward big values.  
The idea is thus to try and fit the data
in the mass range covered by the simulation by not expanding the series
but keeping all possible higher order terms within one regularization 
(this automatically excludes HBCHPT and the IR regularization). 
The first to have done so in a series
of paper is the Adelaide group \cite{lein05}, introducing 
some model-dependence by the use of some regularized functions. More recently 
Pascalutsa and Vanderhaeghen \cite{pasc05}
have obtained a good fit of the nucleon mass within the EOMS framework
where the regularized one-loop functions indeed
tend to zero for large masses. Let me point out  a few problems in these 
works. First, 
there is no systematic studies of baryon properties within these 
frameworks so that the values of the LECs are unknown in these cases.
As already discussed 
they can be in principle somewhat different from the one discussed in 
Section \ref{lec} if the higher order terms turned out to be important.
As we have seen in that section  LECs
are {\it not} independent but relate many observables so that
it is mandatory to check whether the values obtained via a fit to the lattice
data are relevant for other processes. 
Furthermore these calculations are done at 
one loop-order only. Even though the function obtained has the requested
proper smoothness property it is not clear how the result would be modified
would a two-loop calculation be done.  For example let us expand the nucleon
mass as given in \cite{young02}. 
One finds with a monopole regulator: 
$ m=0.898+3.09M^2 -5.74M^3
+5.77 M^4 \ln (M/m) +23.6  M^4 -45.0   M^5$. The non-analytic 
$M^5 \ln(M)$ term coming from a two-loop calculation is totally absent in this 
framework. This has, however, 
a fixed coefficient independent of the regularization as long as this one does not violate chiral symmetry as can be seen from
Eqs.(\ref{eq:mextra}) and (\ref{eq:coeffmextra}). 
Its contribution at moderate $M_\pi$ is, however, non-negligible. Note
also that the $M^4 \ln( M)$ term is far from its expected value, cf. 
Eq.~(\ref{eq:mexp}).

An important quantity related to the nucleon mass is the so-called sigma term
defined as:
$\sigma_{\pi N} = \langle N(p)|\hat m(\bar u u +\bar d d)|N(p) \rangle$.
Best fits of the nucleon mass to the lattice data give  $\sigma_{\pi N} \sim 
40-50$ MeV whatever the regularization used or whether one works with $\Delta$
degrees of freedom or not \cite{phw03,bhm05}. This value is consistent with 
\cite{gls91,bum97}, see the discussion in Section \ref{iso}.

The question is: what is the validity of the comparison between the CHPT
calculation in the infinite volume limit and the lattice calculation? 
In the case  of a stable particle 
of mass $m$
the finite-size corrections to the lowest energy level given by the poles
of the propagator vanish 
exponentially, $E_1(L)-m =\exp(-{\rm const} \cdot L)$ and in the large 
$L$ limit this
level yields the value of the stable particle (see e.g. \cite{galeu88,lu86b}).
In Fig.~\ref{fig:MNL} where the nucleon mass
is drawn as a function of  $L$ for $M_\pi=0.545$ ~GeV
a comparison of a lattice versus a CHPT calculation is made. As can be seen, 
for 
rather large values of $L, \, L>2 \,$fm, good agreement is obtained between the
two calculations. It has been shown \cite{fuh04} using a resummed  L{\"u}scher 
formula \cite{col04,cdh05} that the finite volume
effects for nucleon mass inside a $L=2$ fm box are below $4\%$ within the 
error bars for a pion mass of the order of $M_\pi=0.5$ GeV. What happens for
lower values of $L$?
NLO results have been obtained in \cite{alik04}
which show a  very good agreement with the lattice data. The significance
of this agreement has been questioned in \cite{cofuha06} since no estimate
of the uncertainties had been made. Using a resummed L{\"u}scher formula
it was found that the inclusion of higher orders spoiled the good agreement
as is illustrated on  Fig.~\ref{fig:MNL}. 
Even worse, 
the uncertainties turned out to be extremely large which can be
seen from the shaded area in the figure, so that at
present it appears difficult to make reliable predictions for the size of
the finite volume effects in the nucleon sector in contrast with the meson
sector. This is due to the fact that chiral symmetry restricts $\pi N$ 
interactions less severely than pionic ones.
However, it is necessary to pursue the effort since such studies  
give further constraints on the LECs, different combinations entering the
finite volume corrections~\cite{proc06}.

\vskip -0.1cm
\subsection{{\it axial-vector coupling}} \label{ga}

\begin{figure}[htb]
\vspace{-0.4cm} 
\epsfysize=9.0cm
\begin{center}
\begin{minipage}[b]{6 cm}
\includegraphics*[width=6cm,angle=270]{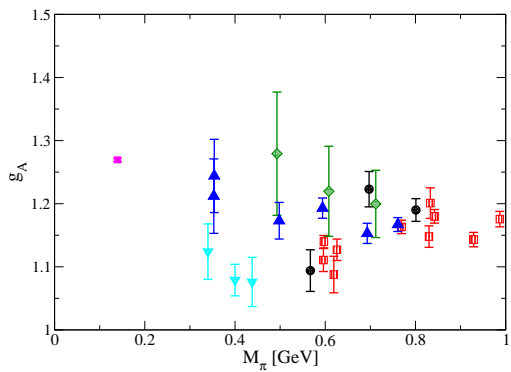}
\end{minipage} 
\begin{minipage}[tb]{18 cm}
\caption{Lattice data: up triangle are from the LPHC/MILC collaboration, 
Ref.\cite{lphcmilc06}, the square from QCDSF/UKQCD \cite{qcdsfukqcd05}, 
the lozenge from 
RBC \cite{rbc02}, the circle from LPHC/SESAM \cite{lphcsesam02} 
and the down triangle from QCDSF \cite{qcdsfprel}.
} \label{fig:gadat}
\end{minipage}
\vspace{-0.3cm}
\end{center}
\end{figure}

\begin{figure}[htb]
\epsfysize=9.0cm
\begin{minipage}[htb]{10 cm}
\epsfig{file=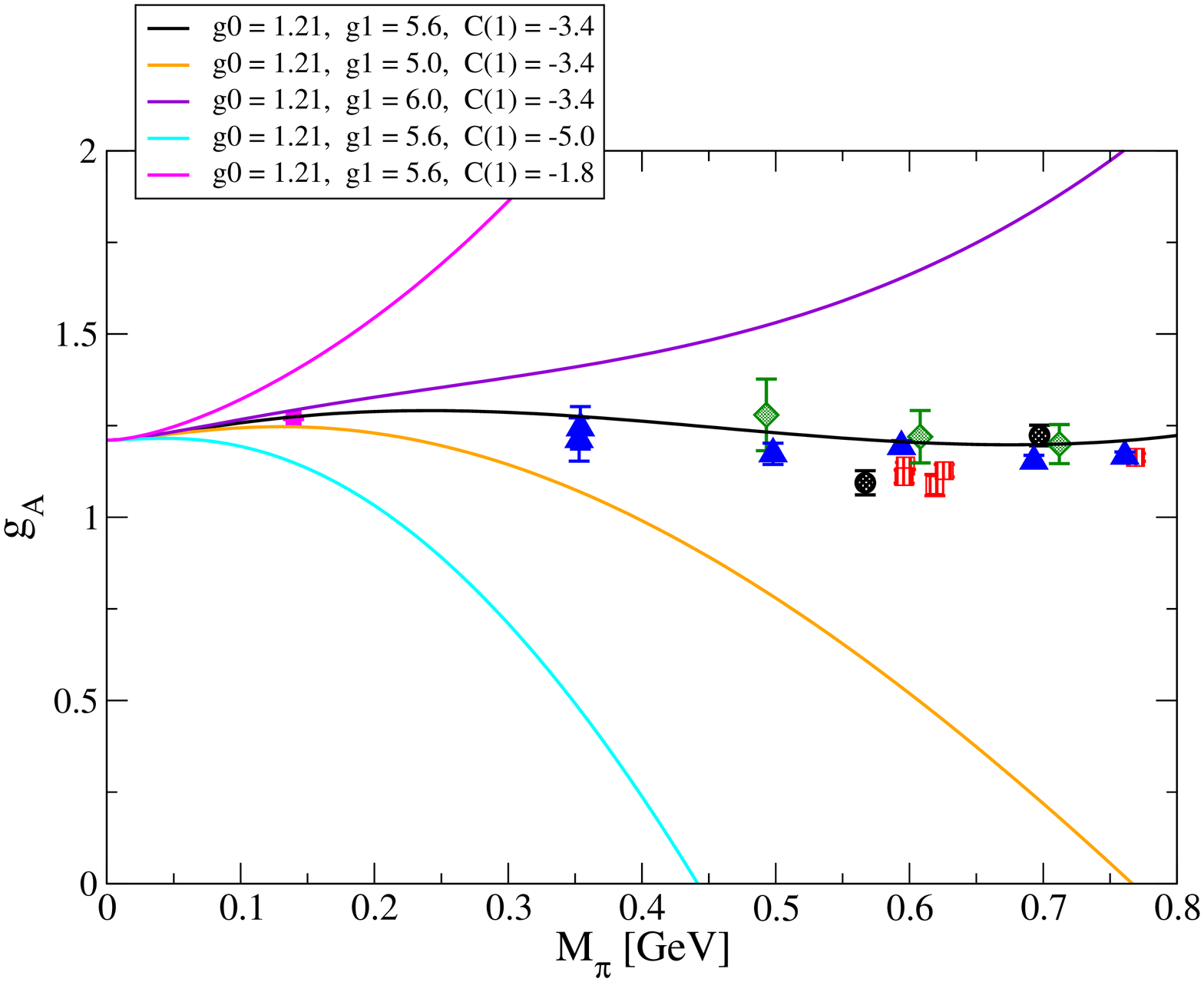,scale=0.3,angle=270}
\vskip -6.5cm
\hskip 10cm
\epsfig{file=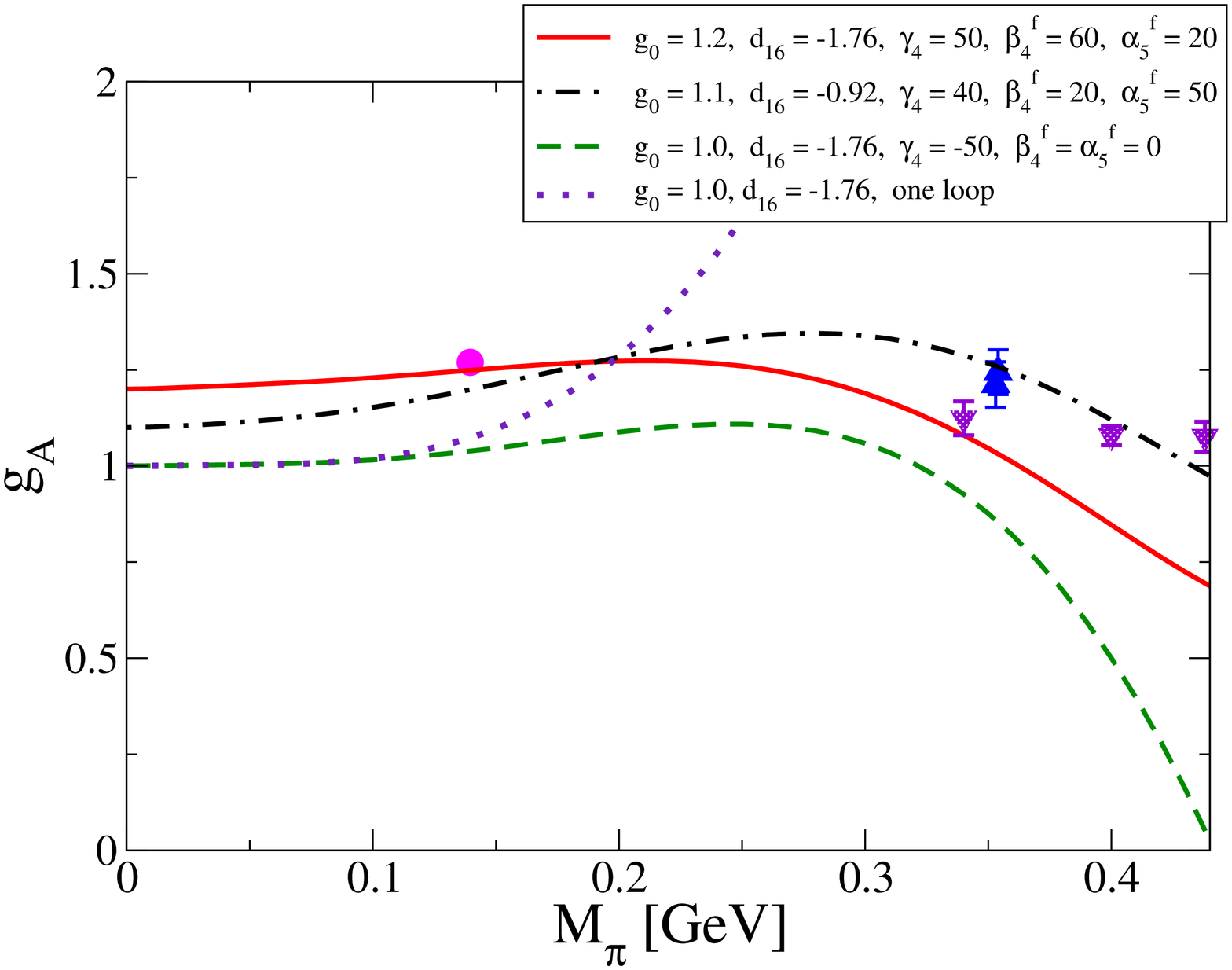,scale=0.3,angle=270}
\end{minipage}
\vskip 0.2cm
\begin{minipage}[htb]{18cm}
\caption{The axial vector coupling as a function of the pion mass. 
Left panel: SSE result to ${\cal O}(\epsilon ^3)$. Right Panel: 
two-loop results as described in the
text. The triangle are the lowest mass data from \cite{lphcmilc06} and the 
inverted triangles are recent results from QCDSF \cite{qcdsfprel}.}
\label{fig:gares}
\end{minipage}
\end{figure}

The axial-vector coupling constant $g_A$ is a fundamental property of
the nucleon that can, e.g. be determined in neutron $\beta$-decay.
It is directly related
to the fundamental pion-nucleon coupling constant by the Goldberger-Treiman
relation, Eq.~(\ref{eq:gtrel}) and thus of great importance for the problem of nuclear binding.
Its formal expression up to two-loop is given by: 
\begin{eqnarray}
g_A &=& g_0 \,\, \biggl\{ 1 + \left( \frac{\alpha_2}{(4\pi F)^2} \ln
\frac{M_\pi}{\lambda} + \beta_2 \right) \, M_\pi^2 + \alpha_3 \, M_\pi^3
\nonumber\\
&& \quad + \left(\frac{\alpha_4}{(4\pi F)^4} \ln^2\frac{M_\pi}{\lambda}
+  \frac{\gamma_4}{(4\pi F)^2} \ln\frac{M_\pi}{\lambda} + \beta_4
\right) \, M_\pi^4 + \alpha_5 \, M_\pi^5 \biggr\} + {\cal O}(M_\pi^6)~,
\nonumber \\
&=&  g_0 \,\, \biggl\{ 1 + \Delta^{(2)} + \Delta^{(3)} + \Delta^{(4)} +
    \Delta^{(5)} \biggr\}   + {\cal O}(M_\pi^6)~.
\label{eq:ga}
\end{eqnarray}
The coefficients up to one loop have been first worked out by \cite{km99}:
\begin{eqnarray}
\alpha_2 &=& -2 -4g_0^2~, ~~ 
\beta_2 = \frac{4}{g_0} \biggl({d}_{16}^r(\lambda) - 2d^r_{28}(\lambda)
\biggr) - \frac{g_0^2}{(4\pi F)^2}~, 
\nonumber\\ 
\alpha_3 &=& \frac{1}{24\pi F^2 m_0}
\left(3+3g_0^2-4m_0c_3+8m_0c_4\right) \,\, .
\label{eq:gacoff1}
\end{eqnarray}
At the physical pion mass and for the central values of the LECs one has
$g_A=g_0 (1-0.15+0.26+\cdots)$. The corrections   $\sim M_\pi^2$
are of natural size, however the corrections $\sim M_\pi^3$
are unnaturally large. This comes from the rather large values of each of the 
$c_i$ combined with the fact that they are of opposite sign so that their
effects get reinforced from their relative minus sign. 
Clearly if the pion mass increases the effect
will be strengthened. One indeed observes a sharp rise of the function beyond
$ M_\pi \sim 300$~MeV (see Fig.4 in \cite{m05}). However lattice data
show a rather flat dependence in the pion mass as is illustrated in 
Fig.~\ref{fig:gadat}
where a compilation of these data are presented. Note again the consistency
of the lattice
results for which different lattice
actions have been used. Only the preliminary QCDSF values are somewhat 
below the others
but this could come from a different treatment of the renormalization.  
A first try to understand these lattice data was made in \cite{hpw03}. The idea
was that the $\Delta$ degree of freedom should be important in understanding
the axial-vector coupling. This is corroborated for example by the 
Adler-Weisberger relation
which relates the deviation of $g_A$ from 1 to the excess of $\pi^+ p$ cross
section over the $\pi^- p$ one where the $\Delta$ dominates. Also as we have
seen previously the convergence of the chiral expansion could be improved
in that way. In this framework some additional terms appear which to 
$O(\epsilon^3)$ are of the form $M_\pi^3/\Delta_0$, $M_\pi^2 \ln R$ and 
$\Delta_0^2 \ln R $ with $R=\Delta_0/M_\pi +\sqrt{\Delta_0^2/
M_\pi^2-1}$ and $\Delta_0$ is the small scale parameter defined in 
Eq.~(\ref{eq:ssdel})
in the chiral limit. In this framework one combination of LECs, $C(\lambda)$, 
appear which 
is unknown. In order to determine it, a matching with HBCHPT has been made 
at a scale  
$\lambda=2 \Delta_0$, the results being insensitive to the exact choice of this
scale between 0.4 and 0.8 MeV. This leads to $ C(1 \, {\rm {GeV}})=(-3.4 \pm 1.2)$
GeV. A flat dependence of $g_A$ was found as is shown on Fig.~\ref{fig:gares}
for the set of parameter $g_0=1.21, \, g_1=5.6, \, {\rm {and}} \, C(1\,
{\rm {GeV}})=-3.4$ GeV, 
where 
$g_1$ is the axial $\Delta \Delta$ coupling constant. However this nice
fitting up to rather large pion mass results from a fine tuning of the LECs.
First $g_1$ turns out to be much larger than its $SU(6)$ value in
contradiction to what is suggested in the study of the $\Delta$ mass
(see next section). 
Furthermore if one lets the LECs vary within their error bars one finds 
rather large uncertainties in the result as can be seen on 
Fig.~\ref{fig:gares}. Thus the result of \cite{hpw03} cannot support the
claim of a controlled and precise determination of $g_A$. 

\begin{figure}[htb]
\epsfysize=9.0cm
\begin{center}
\epsfig{file=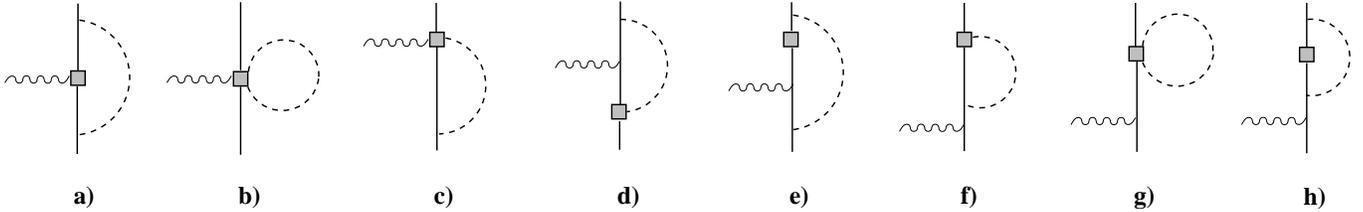,scale=0.3}
\caption{Topologies of the one-loop graphs that generate the coefficient
of the double log at two-loop order. The hatched square denotes a dimension
three insertion proportional to some of the LECs $d_i$.} \label{fig:gatop}
\end{center}
\end{figure}

In view of all
this it became clear that a two-loop
calculation of $g_A$ was needed. The coefficients $\alpha_{4,5}, \,
\beta_4, \, \gamma_4$ were determined in \cite{beme06a}. The double logarithm
can easily be obtained using renormalization group techniques as was shown 
in the meson sector by \cite{we79}. Indeed the non-local pieces which 
appear in a two-loop calculation
\begin{equation}
k(d ) \, \frac{\lambda^{2\epsilon}}{(4\pi)^4} \, \left[ \frac{1}{\epsilon^2} +
 \frac{2}{\epsilon} \, \ln \frac{M_\pi}{\lambda} + \ln^2 \frac{M_\pi}{\lambda}
+ \ldots \right]~,
\end{equation}
have  necessarily to be cancelled by terms coming from one loop 
graphs with vertices involving LECs, in the case at hand coming from the 
$\pi N$ Lagrangian of dimension 3
\begin{equation}
-\frac{h_i(d)}{2} \, \frac{\lambda^{2\epsilon}}{(4\pi)^4} \, \left[ 
 \frac{\kappa_i}{\epsilon^2} +  \frac{\kappa_i}{\epsilon} \ln
 \frac{M_\pi}{\lambda} +  \frac{(4\pi)^2 d_i^r (\lambda)}{\epsilon}
+ (4\pi)^2 \, d_i^r  (\lambda)  \ln  \frac{M_\pi}{\lambda} 
+ \ldots \right]   \,\,\,.
\end{equation}
In fact the renormalization group condition $k_0=h_i^0 k_i$ ensures
that this indeed happens. 
The topologies of the one loop graphs that 
generate the coefficient of the double log are shown in Fig.~\ref{fig:gatop}.

Altogether 8 LECs from the non-equation-of-motion terms contribute, from
which $d_{16}$.  
From the knowledge of the $\beta$ functions $\kappa_i$\cite{fms98,em96} 
one deduces:
\begin{equation}
\alpha_4 = \alpha_4^{\rm irr} + \alpha_4^{\rm red} + \tilde\alpha_4 =
-\frac{16}{3} - \frac{11}{ 3} g_0^2 + 16 g_0^4 \, \, .
\nonumber\\ 
\end{equation}
Taking into account the contribution from the LEC $d_{16}$ to $\gamma_4$,
the $1/m_0$ and $1/m_0^2$ corrections to the large contribution $\alpha_3$
and induced terms from the quark mass expansion of the pion decay constant,
the numerical values of the two-loop coefficients for the central 
values of the LECs are:

\begin{equation}
{\alpha}_4 =7, \quad {\gamma}_4 = 115.3 \,{\rm GeV^{-2}}+{\gamma}_4^f, \quad 
{\beta}_4 = 14.2 \, {\rm GeV^{-4}} + {\beta}_4^f, \quad
{\alpha}_5 = -20.3 \, {\rm GeV^{-5}}+{\alpha}_5^f \, \,,
\label{eq:ga2loop}
\end{equation}
where ${\gamma}_4^f, \, {\beta}_4^f$ and ${\alpha}_5^f$ denote the
additional contributions from unknown LECs which have to be determined, e.g.
from an analysis of lattice data and assuming naturalness.
Some typical examples of such a fit  are shown on the l.h.s. of
 Fig.~\ref{fig:gares}. 
One finds that the pion mass dependence of $g_A$ stays flat for 
$M_\pi \le 350$ MeV.
One obtains at the physical pion mass a good convergent representation,
typically $g_A=1.21(1-0.15+0.26-0.06-0.001)$. The finite volume dependence
of $g_A$ which I will not discuss here can be found in \cite{bs04}.

\subsection{{\it magnetic moment}}  \label{magm}

The magnetic moments provide a nice illustration of the difference between 
different regularizations. They have been calculated in
\cite{hpv05}   using a sum rule, for more details
see that paper. The proton magnetic moment is given by:
\be \kappa_p 
= \kappa_0 + \frac{g_A^2 m^2}{(4\pi F_\pi)^2 } \left\{1 -
  \frac{\mu \,\left( 4 - 11{\mu }^2 + 3{\mu }^4 \right) }{\sqrt{1 - 
{\mu }^2/4}}
  \arccos \frac{\mu }{2} - 6{\mu }^2+
  2{\mu }^2\left( -5 + 3\,{\mu }^2 \right) \ln \mu \right\} \,\, .
\label{mags} 
\ee
where $\mu=M_\pi/m$ and 
 $\kappa_0$, the chiral limit value of $\kappa_p$, is given by a 
combination of two LECs of order two: $\kappa_0 =c_6 + c_7$ \cite{bkkm92}.
Now let us look at the IR result \cite{kubm01}. One has
\be
 \kappa_p 
= \kappa_0 + \frac{g_A^2 m^2}{(4\pi F_\pi)^2 } \left\{
 - \frac{\mu \,\left( 4 - 11{\mu }^2 + 3{\mu }^4 \right) }{\sqrt{1 - 
{\mu }^2/4}}
  \arccos \bigl(-\frac{\mu }{2}\bigr) - \frac{3}{2} {\mu }^4+
  2{\mu }^2\left( -5 + 3\,{\mu }^2\right) \ln \mu \right\},
\label{magir} 
\ee

\begin{figure}[htb] 
\epsfysize=9.0cm
\vspace{0.1cm}
\begin{center}
\begin{minipage}[htb]{6 cm}
\hskip -4cm
\includegraphics*[width=6cm]{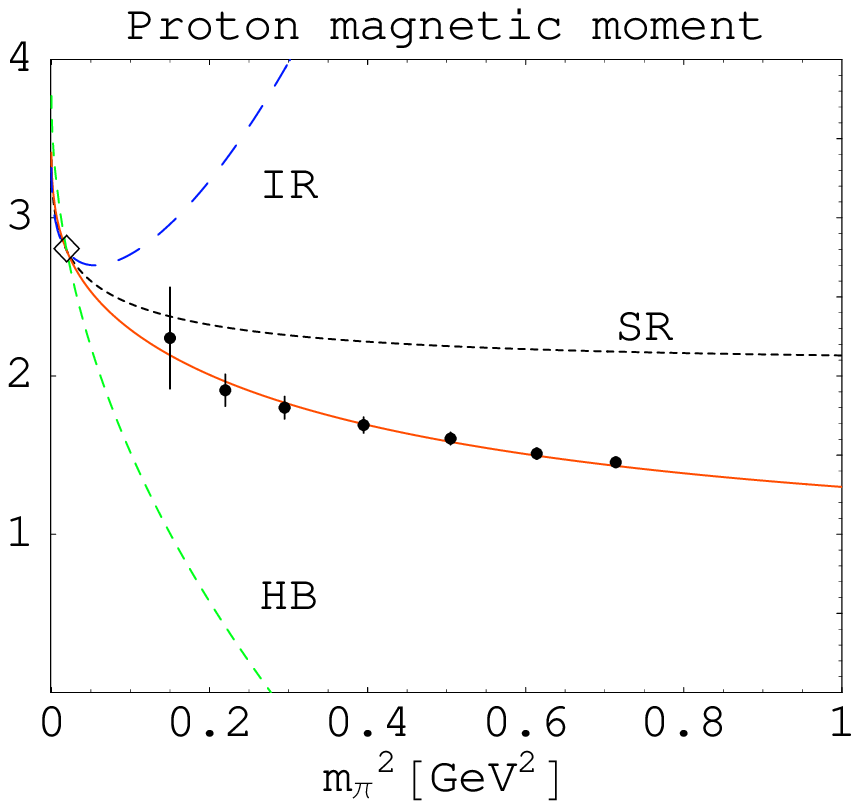}
\vskip -4.8cm
\hskip 4cm
\includegraphics*[width=6cm]{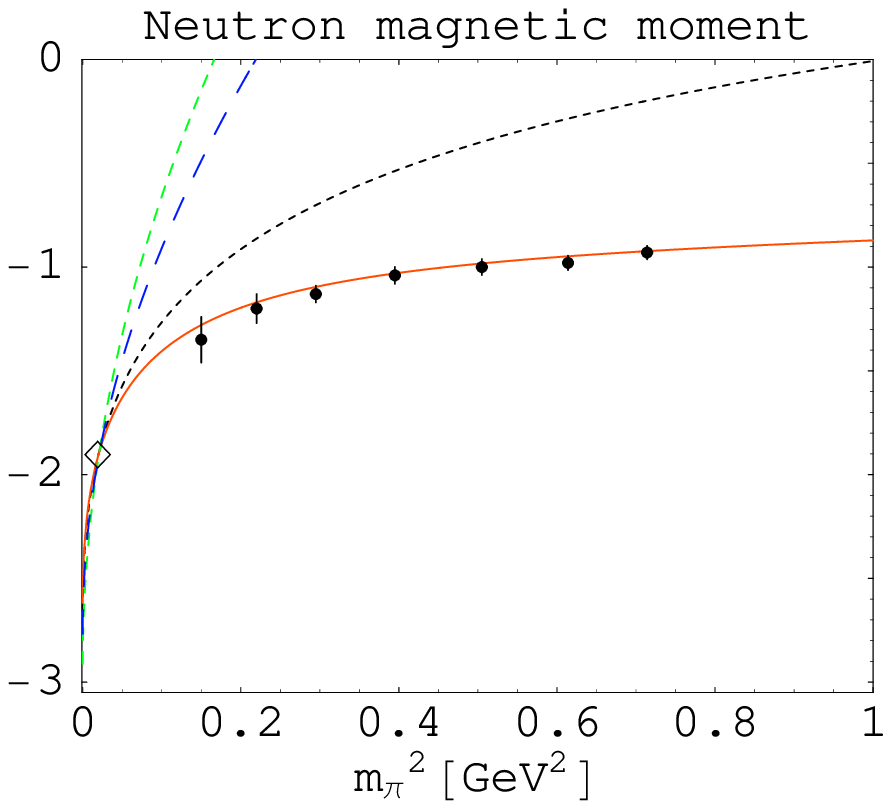}
\end{minipage} 
\begin{minipage}[htb]{18 cm}
\caption{Proton and neutron magnetic moments to one loop 
compared with lattice data. The open diamonds represent the 
experimental values at the physical pion mass. For a discussion of the 
solid line see \cite{hpv05} from which the figures are taken.
\label{fig:magmo}}
\end{minipage}
\end{center}
\end{figure}
\noindent The difference in the two expressions only appear -- as it should -- in the 
terms that are analytic in the quark mass. In the sum rule approach 
which is equivalent to the relativistic framework terms violating power 
counting
appear. The  term $\sim 1$ in Eq.(\ref{mags}) is such a term. As explained
before it can be absorbed in the LECs, illustrating the fact that the 
counterterms can have different values in different regularizations.
Once this is done,   
expanding Eqs.(\ref{mags}) and (\ref{magir}) around $\mu=0$, one recovers the 
HBCHPT result namely, 
\be
\kappa_p = \kappa_0+ \frac{g^2 m^2}{(4\pi F_\pi)^2 }\left\{ -2\pi
\mu-2\,(1+5\ln\mu)\,\mu^2+\frac{21\pi}{4}\,\mu^3
+ O(\mu^4)\right\} \, \, .
\label{maghb}
\ee
As pointed out before the 
argument of the $\arccos$ term is  positive within the sum rule
approach while it is 
negative in IR. This difference of course affects the high
energy behaviour of the magnetic moments. While 
$ \delta \kappa_p=\kappa_p-\kappa_0 $ 
vanishes in the large $M_\pi$ limit in the first calculation 
behaving like $1/M_\pi$, it 
diverges for $M_\pi = 2m$ in the IR case due to these unphysical cuts
discussed previously. Let me stress again that this is not a problem
since the region where it becomes large is outside the range of validity of 
the theory.  
The $M_\pi$ dependence of the three calculations, Eqs.~(\ref{mags}), 
(\ref{magir})
and \ref{maghb}) is shown on Fig.\ref{fig:magmo}. It coincides up to roughly 
350 MeV
and starts to depart then. In \cite{m05} an investigation of the convergence
of the series was made with the same conclusion, namely the theoretical
uncertainty is modest up to 350 MeV and starts to increase then. Thus, as for
the two other observables discussed above, the chiral extrapolation can be 
believed up to 
roughly 350 MeV.    

\subsection{{\it $\Delta$ and Roper masses}} \label{delm}

\subsubsection{infinite volume limit} \label{infvol}
Other baryon properties have been determined on the lattice. Let me review 
briefly results on the $\Delta$ and the Roper mass.

\vskip 0.35cm
\begin{figure}[htb]
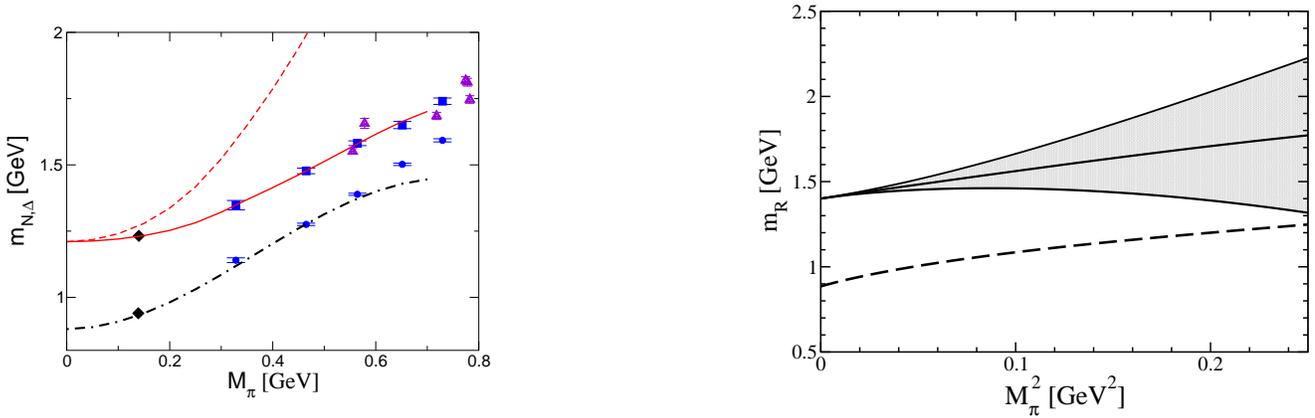

\epsfysize=9.0cm
\begin{minipage}[htb]{10 cm}
\epsfig{file=mdelta4.eps,scale=0.28}
\vskip -5.3cm
\hskip 10cm
\epsfig{file=mrobw.eps,scale=0.3}
\end{minipage}
\vskip 0.3cm
\begin{minipage}[htb]{18 cm}
\caption{ 
Left panel: The nucleon (dot-dashed line) and the (real part) of the $\Delta$ 
mass (solid line) as a function of the pion mass. The filled diamonds denote
their physical values at the physical pion mass. The dashed line is the chiral
extrapolation with $a_1$ fixed at its SU(6) based value. The filled squares
and circles are the MILC data \cite{ber01} and the filled triangles the recent
data from QCDSF. Figure from \cite{bhm05}.
Right panel:
The Roper mass as a function of the square of the pion mass. The grey band indicates the theoretical uncertainties. The nucleon mass, dotted line, is shown for
comparison. For more 
details see \cite{bbml06}. Figure courtesy of Ulf-G. Mei{\ss}ner.} 
\label{fig:mdelta}
\end{minipage}
\end{figure}

The dependence of the $\Delta$ mass on the pion mass can easily be obtained
within the SSE framework \cite{bhm05}. 
To fourth order in the $\epsilon$ expansion one obtains:

\begin{equation}
m_\Delta =m_0^\Delta -4 a_1 M_\pi^2  -4 e_1^\Delta M_\pi^4 +m_\Delta^{N-{\rm 
loop}}
+m_\Delta^{\Delta-{\rm loop}}
\end{equation}
where $m_\Delta^{N-{\rm loop}}$ and $m_\Delta^{\Delta-{\rm loop}}$ are the
nucleon/$\Delta$ loop contributions respectively. $a_1$ is the symmetry 
breaker term analog to $c_1$ for the nucleon. The vertices involve
the axial N$\Delta$ coupling $c_A$ and the $\Delta \Delta$ coupling $g_1$
as well as 4 unknown LECs, 2 from the $N \Delta$ Lagrangian and two 
combinations from the $ \Delta \Delta$ one. In \cite{bhm05} a combined
study of the nucleon and $\Delta$ mass was performed. Thus
6 combinations of LECs have to be fitted to the lattice data imposing 
again the constraint from their physical values at the physical 
$\pi$ mass . The 
dependence of these masses on the pion mass is shown on the l.h.s. of 
Fig.~\ref{fig:mdelta}. A good fit is obtained for $g_1=2$ not far from its 
SU(6) value
$9 g_A/5=2.28$ and $a_1=-0.3$ GeV$^{-1}$ markedly smaller than $c_1$
although both couplings should be equal in the SU(6) limit. As can be seen
from the dashed line
in Fig.~\ref{fig:mdelta} the assumption of strict SU(6) symmetry is clearly
at odds with the lattice data. 
The smallness of $a_1$ compared
to $c_1$ leads to a $\pi \Delta$ sigma term much smaller than the nucleon
one, $\sigma_{\pi \Delta}=20.6$ MeV. Before drawing any conclusion, 
constraints from 
other physical processes would be needed
to confirm the values of the LECs obtained. Also more detailed
precise
fits to the lattice data including error and correlation analysis should be 
done.   

\begin{figure}[t] 
\epsfysize=9.0cm
\vspace{0.1cm}
\begin{center}
\begin{minipage}[htb]{6 cm}
\includegraphics*[width=7.2cm]{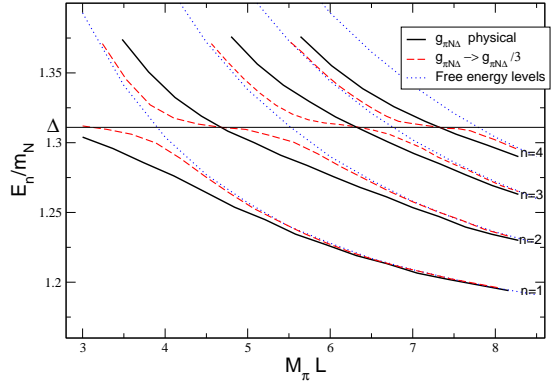}
\end{minipage} 
\begin{minipage}[htb]{18 cm}
\caption{The dependence of the energy levels in a finite box size $L$
for a small value of the coupling constant $g_{\pi N \Delta}$ (dashed line) 
and its physical value (solid line).
For comparison, the free energy levels are also displayed (dotted lines). Figure taken from \cite{rutmeibe06}. \label{fig:delfv}}
\end{minipage}
\vskip -1cm
\end{center}
\end{figure}
 
A formalism has been developed in \cite{bbml06} which is suited to study 
systems with two heavy mass scales in addition to a light mass scale. 
It has been applied to the case of the Roper resonance $N^*(1440)$,
a very intriguing excited state of the nucleon. This first even-parity state 
is lighter than the first odd-parity one, the $S_{11}(1535)$ and has a 
significant branching ratio into two pions. The relevant effective Lagrangian 
needed to study this particle has
been given in \cite{bbml06} as well as the modification of the regularization
scheme due to the appearance of an extra large mass scale $m_R$ in addition
to the already present one, $m$. The case $m^2/m_R^2
\ll 1$ (in nature this ratio is  $\sim 1/2.4$) has been described. 
Results of the one loop
calculation of the quark
mass dependence of the Roper mass are shown on the r.h.s. of
 Fig.~\ref{fig:mdelta},
using naturalness arguments as well as some bounds from the nucleon
to determine the new LECs which appear in the Roper Lagrangian. 
As can be seen  the dependence is similar to the
one of the nucleon. It should be emphasized again that the one loop 
formula cannot be trusted for pion masses beyond 350 MeV.  
On the lattice side there is not yet a clear picture on the nucleon
resonance spectrum. In \cite{mcddhllz03} a rapid cross over of the first 
positive
and negative excited nucleon states is found close to the chiral limit,
whether this does not seem to be the case in \cite{gps04} for example. 
Simulations closer to the chiral regime are certainly needed.

\subsubsection{finite volume} \label{finvol}

Again arises the question of the comparison between the CHPT calculation
discussed here and the lattice data. As we have seen lattice
calculations are not only done in a finite volume but also at rather 
high pion masses so that 
the ``lattice $\Delta$ particle'' contrary to the physical one does not decay. 
The calculation thus proceeds in 
exactly the same way as for the nucleon discussed above. 
At some point, however,  lattice calculations will reach values of 
pion masses for which the $\Delta$ will be unstable. Clearly
unstable particles are more complex to treat than the stable ones.
The question of identification of the resonances on the lattice has 
been addressed in a series of papers \cite{lu88,wi89,rg95,cky05,kss05}. It has 
been demonstrated
that, in the presence of a narrow resonance, the dependence of the energy
spectrum of the system on the box size $L$ exhibits a very peculiar
behavior near the resonance energy, where the so-called ``avoided
level crossing'' takes place, for more details see the references above.   
In this case the dependence of the energy levels on $L$ is 
governed by a power 
rather than by an exponential law. In \cite{rutmeibe06} the specific 
question of 
how to determine the mass of the $\Delta$ in an effective field theory 
approach has
been discussed. Several questions arise:

$\bullet$ Lattice data being alway real, does one get only the real part
of the resonance pole mass as a result of a chiral extrapolation?
 
$\bullet$ Is there a way to determine the decay width of a resonance?

Typical energy levels obtained in the ${\cal O}(\epsilon^3)$
SSE calculation \cite{rutmeibe06} are shown on Fig.~{\ref{fig:delfv}}.
It is seen that the energy levels in the presence of an interaction
interpolate between different free energy levels. An abrupt change
emerges in the vicinity of the resonance when 
$g_{\pi N \Delta}$ is taken at 1/3 of its
physical value, nicely illustrating the phenomena of avoided level crossing. 
In the case of the physical value, the width of the
$\Delta$ is too large and the phenomenon is washed out. In that case
the mass cannot be determined so easily. However the lowest
energy level $E_1$ and the difference between the first two lowest energy
levels $E_2-E_1$ show some significant dependence in the $\Delta$ mass
and in the $\pi N \Delta$ coupling respectively. 
It was thus suggested in \cite{rutmeibe06}
to use these properties to determine
the mass and $g_{\pi N \Delta}$:
having fixed a value for
the coupling constant the $\Delta$ mass is obtained from 
the $M_\pi L$ dependence of $E_1$. Then
a fit of $E_2-E_1$ to the data gives a new value for the coupling constant,
the finite volume results for  $m_\Delta$ and $g_{\pi N \Delta}$
being obtained once convergence is achieved.

\vspace{0.35cm}

{\underline{CONCLUSION:}}
From the analyses done above it is fair to say that {\it {chiral extrapolations
of nucleon properties can be trusted for pion masses below $\sim 350$ MeV.}}

\section{Pion-nucleon and Pion-deuteron scattering} \label{pind}
Let us come back to the real world and look at some applications of the 
CHPT machinery.
A systematic investigation of processes involving pions will allow to 
understand in a precise manner how the chiral symmetry violation 
takes place. In this section  I will concentrate on elastic $\pi N$
and $\pi d$ scattering.

\subsection {\it $\pi N$ scattering} \label{pin}
Elastic pion-nucleon scattering
has been one of the most intensively studied process in hadron
physics with a long history in theory  
and experiment (e.g. \cite{ho83}). In
chiral perturbation theory 
it has been studied within different regularization approaches~\cite{m98,dp97,
te03,fm00,fms98a,bl01} 
and isospin
symmetry breaking effects have been evaluated. 
As we will see the experimental situation is unfortunately not yet 
totally satisfactory. There has been  two generations of $\pi N$ measurements. 
The first two decades (1957 through 1979) of
experiments focused on the $\pi N$ system and nonstrange baryon resonances
and  produced a large amount of data below 2.6 GeV. These were analysed 
for example in
\cite{KA85}. 
A second generation 
of $\pi N$ measurements (both unpolarized and polarized) were carried out 
at  high-intensity facilities such as LAMPF, TRIUMF, and PSI. These more 
recent measurements generally have
small statistical and systematic uncertainties. Some of them are included
in the analysis discussed here \cite{EM97,SP98}. Other very recent ones
\cite{dal06,mal04,bal06}
are still being analysed \cite{absw06}. Particular interest in these recent
experiments comes from the fact that they were done at very low energy
where information is still missing and there are regions where existing
experimental data are contradictory. Clearly more precise data at  energies 
below 
100 MeV are still needed to get a precise knowledge of $\pi N$ scattering 
amplitude and to finally
settle the problem of the $\sigma$ term,
the $\pi N N$ coupling as well as the isopin breaking effects which will 
be discussed here. However as we have seen in Section \ref{lec} some LECs
could still be obtained with rather good precision.
Note that in the most recent analysis of these data \cite{absw06} 
where an overall fit up to 2.6 GeV is done no important
changes were found in the low energy region which we are concerned here,
so that results reported in this review based on older data should not be 
altered. 

The scattering amplitude for the process $\pi(p,a) N(q) \to \pi(p',a') N(q')$
is given by (with $a,a'$ isospin indices)
\bea
T_{a'a} &=& \delta_{a'a} T^+ + \frac{1}{2} [ \tau_{a'}, \tau_a] T^- \, \, ,
\label{eq:scattam} \\ \nonumber
T^{\pm} &=&\bar u' \Bigl(A^\pm+
\frac{1}{2}  (\not\!q'+ \not\!q) B^\pm \Bigr) u 
=\bar u' \Bigl(D^\pm-\frac{1}{4m} [ \not\! q', \not\!q] B^\pm
\Bigr) u \, \, , 
\eea
where $D= A + \nu B$ with $\nu=(s-u)/4m$, $s$ and $u$ being the usual 
Mandelstam variables.
Note that the usual decomposition in terms of the $A$ and $B$ amplitudes
is best suited for a dispersive analysis  while it is not useful 
when  performing 
a low energy expansion
since the leading contributions from these amplitudes cancel \cite{gss88}.

\subsubsection{isospin symmetric case} \label{iso}
Calculations of $\pi N$ scattering in the isospin symmetric case have been 
performed in CHPT up to 
fourth order. The amplitudes decompose into three pieces, the tree
and  counterterm parts of polynomial type and  pion loop corrections
which start at order $q^3$. The number of counterterms are 4, 9 and 14 to
orders $q^2$, $q^3$ and $q^4$, respectively. As already noted this is 
much less than the total number of terms in the Lagrangian allowed at the 
various orders. In fact these numbers can be determined by looking at the
most general polynom for  $A^{\pm}$ and $B^{\pm}$. The 
values of these countertems have already been discussed in  Section \ref{lec}.
Explicit expressions for the amplitudes can be found in \cite{fm00}
for HBCHPT.  
An analysis of these amplitudes using dispersion relations has been performed
by \cite{bl01} as an alternative regularization procedure. 
It has indeed been shown in this paper that the result of the 
dispersive representation agrees with the representation 
in terms of infrared regularized loop integrals up to 
terms that are beyond the accuracy of a one loop calculation. It has the
advantage that it exhibits the structure of the amplitude in a more
transparent way and involves only the expressions for the imaginary parts
which are determined from the tree level amplitude due to unitarity. 
For a very recent review on the relation between CHPT and dispersive analysis, 
see \cite{bm06}.

\vspace{0.2cm} 
$\bullet$ 
{\underline {phase shift analysis}}
\vspace{0.2cm}

\begin{figure}[htb] 
\epsfysize=9.0cm
\begin{center}
\hskip -6cm
\begin{minipage}[tb]{1.cm}
\vskip-3.5cm
\epsfig{file=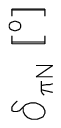,scale=0.7}
\vskip 3.cm
\epsfig{file=fitka85leg.eps,scale=0.7}
\end{minipage} 
\hskip -0.4cm
\begin{minipage}[tb]{8cm}
\vskip -3truecm
\vskip 3truecm
\epsfig{file=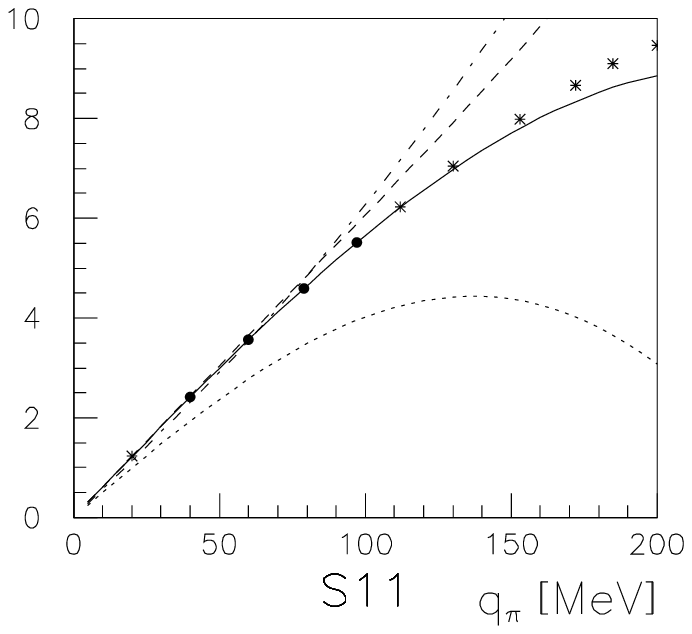,scale=0.7}
\vskip -5.cm
\hskip 7cm
\epsfig{file=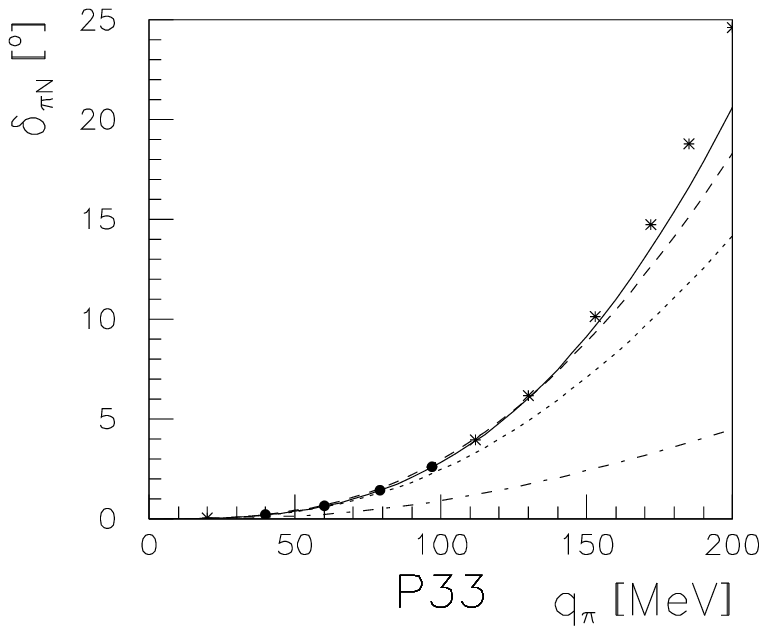,scale=0.7}
\epsfig{file=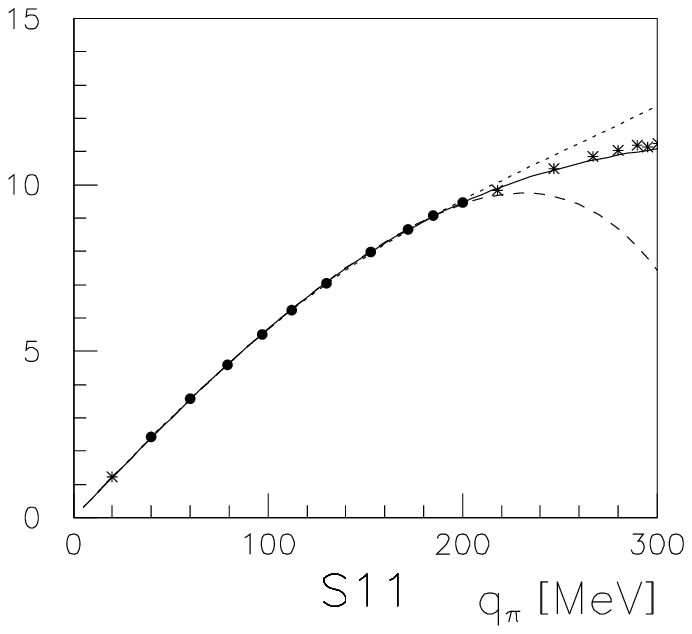,scale=0.7}
\vskip -5.cm
\hskip 7cm
 \epsfig{file=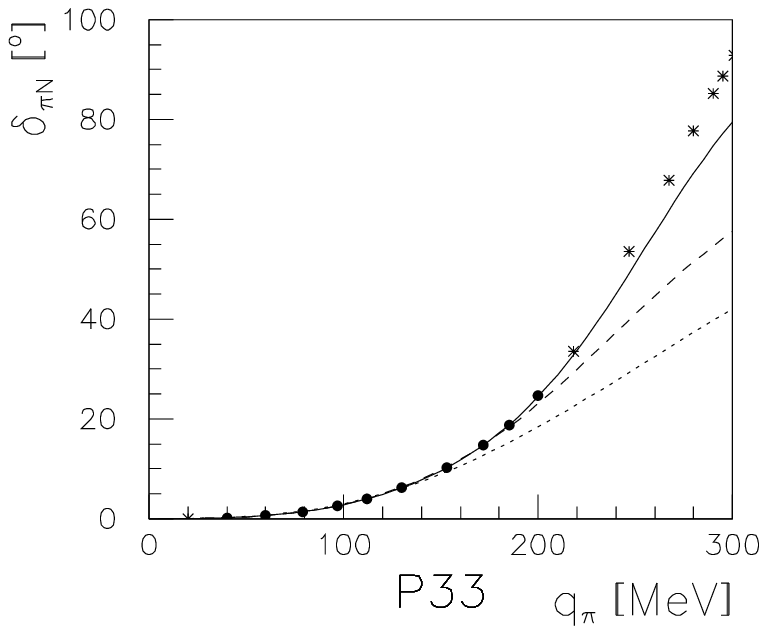,scale=0.7}
\end{minipage}   
\begin{minipage}[tb]{18 cm}
\caption{ Fit  to the KA85 phases (solid 
dots) between 40 and 97 MeV and predictions for higher
and lower energies for two phase shifts as a function of the
pion laboratory momentum $q_\pi$. The various lines refer to the contributions
from the various order as explained in the text. Note the 
different scales on the different panels. Figure from \cite{fm00,fm01xy} 
\label{fig:kafit}}
\end{minipage}
\end{center}
\end{figure}

The strategy in \cite{fm00} is to fit to the phase shifts 
provided by three different partial wave analysis\cite{KA85,EM97,SP98}
for pion momentum in the laboratory frame typically in the range from 40 to
100 MeV. This allows
a determination of the LECs as well as a prediction of the threshold
parameters and of the phase shifts at higher energies. On the upper panel of 
Fig~\ref{fig:kafit} we show the result of such a fit for two
partial waves, namely $S_{11}$ and $P_{33}$ in the case of the KA85 analysis.
The two other analysis give comparable results. As can 
be seen the fits are rather good with a $\chi^2$/dof of 0.5.
In these plots
the dot-dashed, dotted, dashed and solid lines show the contributions from the 
amplitude up to first, second, third and fourth order, respectively. The 
convergence is rather good, the fourth order contribution is mostly
rather small. Other partial waves can be found in \cite{fm00} with the same 
conclusion for the quality of the fit and the convergence of the series.    

The representation of the scattering amplitude to $O(q^4)$ incorporates
the pole and cuts generated by the exchange of one or two
stable particles (nucleons or pions) but account for all other
singularities only through their contributions to the effective coupling
constants. However, resonances generate poles on unphysical sheets of the
scattering amplitude, the first resonance in the $s$-and $u$- channels 
being the $\Delta(1232)$ and the $\rho$ in the t-channel. Whereas in the
region $|t| \le 10 M_\pi^2$ the $\rho$ meson pole which occurs at $t\sim 
30 M_\pi^2$ is well represented by a polynomial the situation is different
for the $\Delta$. A description of the $\pi N$ amplitude has thus been 
considered within an effective theory with explicit $\Delta$ degrees of 
freedom \cite{fm01xy}. On the lower panel of Fig~\ref{fig:kafit} is shown
a  fit comparable to the upper one just discussed 
but now with the $\Delta$ included in
a third order calculation in the small scale expansion as 
described in Section \ref{lec}. As expected a clear improvement on the 
description of the resonant $P_{33}$ phase-shift is obtained, an extremely 
good fit being extended up to pion momentum in the laboratory of 300 MeV.
For the other partial waves the fits are comparable to the fourth order
one although the overall description is still better here.  
Note that to this order the number of counterterms is the same as in 
the effective theory without $\Delta$ up to fourth order. Also in all
these fits with and without $\Delta$ the counterterms are all of natural 
sizes. 

A different strategy has been used in \cite{bl01}. There the four LECs 
are determined from four equations relating them either to  
threshold or subthreshold parameters, but, as noted in
Section \ref{lec}, at tree level. Differences are in principle of higher 
orders. However, this leads to values for the $c's$ about 1.5 times smaller.
For example
\be
c_3=-a_{01}^{+} F_\pi^2 - 
\frac {g_A^2 M_\pi}{16 \pi F_\pi^2}(g_A^2+ \frac{77}{48}) \, ,
\label{eq:c2t}
\ee
which leads to a correction of  $40 \%$ compared to the tree result.
In this equation  $a_{01}^{+}$ is the standard notation for one 
coefficient  in 
a subthreshold expansion around $\nu=t=0$ of the $\pi N$ amplitude.
Hence, using the Karlsruhe
analysis the one-loop result \cite{bl01} for the total cross section  starts
deviating from the experimental values already at rather low energies.      
It has thus been stressed in this reference that the work of \cite{fm01} 
would not describe well  the subthreshold amplitudes. This would of course
be true if one would do a tree level calculation. However taking into account 
higher order corrections as given in \cite{bkm97} one 
obtains an overall good description of the subthreshold quantities.   
Discrepancies for some quantities  clearly survive
especially in the isoscalar case. As discussed in
~\cite{bl01} the one loop result for 
the contribution of the dispersion integrals to $D{^+}$
at $\nu=  M_\pi\, , \,t=0$  is a factor of two smaller than
the experimental result. Even tough these are small effects they do indeed
matter at the level of accuracy needed in the discussion of some low-energy
theorems of chiral symmetry (see below).  
 
\vspace{0.2cm} 
$\bullet$  
{\underline {low energy theorems and scattering lengths}}
\vspace{0.2cm}

Chiral symmetry imposes two constraints on $\pi N$ scattering. 
There are
different ways of implementing them. Either one writes two relations for the
subthreshold coefficients or two low-energy theorems. The first variant of
these theorems
concerns  the scattering
amplitudes $D^{+}$ and  $D^{-}$ both evaluated 
at the Cheng-Dashen point $\nu=0$, $t=2 M_\pi^2$ \cite{cd71,bpp71}.
The first low-energy theorem
\begin{equation}
\Sigma=\sigma(2 M_ \pi^2) + {\cal O}( M_\pi^2)
\label{eq:let}
\end{equation}
relates the quantity   
$\Sigma=F_\pi^2 \bar D^+(0,2 M_\pi^2)$,
where $D$ is defined in Eq.~(\ref{eq:scattam}) and the bar means that the 
pseudoscalar
Born term is subtracted,
to the scalar form factor of the nucleon
\begin{equation}
\langle N(p')| m_u \bar u u +m_u \bar d d| N(p) \rangle =\sigma(t) \bar u'u
 \,\, , \,\,\,\,\,\,\,\, t=(p'-p)^2  \,\, .
\label{eq:sig}
\end{equation}
The difference between $\Sigma$ and  $\sigma(2 M_ \pi^2)$ has been evaluated 
in \cite{bkm96x} and found to be of order ${\cal O} (M_ \pi^4)$ 
and 
free of infrared singularities up-to-and-including that order. This result
was later confirmed \cite{bl01}. 
It is expected
to be a small correction since one is comparing the properties of the 
amplitude at the Cheng-Dashen point with those at $s=u=m^2+M_\pi^2$, $t=0$.
In this region, the amplitude does not contain branch points and the relevant
distance is small, of order $M_\pi^2$. In fact it was estimated 
to be of the  order of 1 MeV so that the deviation to the low-energy 
theorem is rather small.
 
The second of the two low-energy theorem reads
\begin{equation}
C=F_\pi^2 \frac{\bar D^-(\nu,t)}{\nu}\Big|_{\nu=0,t=2 M_\pi^2} =1 + {\cal O}
(M_\pi^2) \,\, .
\end{equation}
The difference to one contains in that case infrared singularities and starts
already at order $M_\pi^2$. It was evaluated in \cite{bl01} and found to 
be equal to $0.02 \pm 0.01$.
 
Another variant of these low-energy theorems are the 
Weinberg's current algebra prediction  \cite{wei66} for the 
scattering lengths:
\begin{equation}
a^+= {\cal O}(M_\pi^2), \,\,\,\,\,\,\, a^- = \frac{M_\pi}{8 \pi (1+M_\pi/m)
F_\pi^2} + {\cal O}(M_\pi^3) \, \, , \label{eq:wein}
\end{equation} 
where $a^\pm =1/4 \pi (1+M_\pi/m)^{-1} T^\pm(M_\pi)$.
In this case the corrections are expected to be large (for a first calculation
see \cite{bkm93}). Indeed the amplitude is analyzed
at a singular point, 
the threshold
sitting on top of the branch point required by unitarity.  
The $\pi N$ scattering analyses of the scattering lengths
~\cite{bl01,fm01} lead to a  
range of values,
\be
a^+= -0.83  \cdots 0.5 \,\, , \,\,\, \,\,\,\,\,\,\, a^- = 7.7 \cdots 9.2 \,\, ,
\ee 
in units of inverse powers of the pion mass times $10^{-2}$
compared to 
the Weinberg's results $ a^+ = 0 \,, \,
a^-=8.76$, Eq.(\ref{eq:wein}) in the same units. In ~\cite{fm01} good fits of these quantities have been 
obtained and it was shown that the series
converges well.

However the most direct way to get a handle on these threshold
quantities experimentally is the measurement of the strong interaction shift 
and the
decay width in pionic atoms. A whole effort is being pursued in PSI in order 
to get to these quantities as well as on the $\pi N N$ coupling constant
to better than a percent. I will come back to the discussion 
of the scattering lengths in more detail when discussing pionic deuterium 
and isospin breaking. 

\vspace{0.2cm} 
$\bullet$  
{\underline {$\sigma-term$}}
\vspace{0.2cm}

One quantity of great interest is the $\sigma$-term, $\sigma(t=0)$, see 
Eq.~(\ref{eq:sig}) 
since it is related to the strange quark content in the nucleon, $y$: 
\be
\sigma(t=0) = \frac{\hat \sigma}{1-y} \, , \,\,\,\,\,\,\,\,\,\,\,\,
 y=\frac { 2 \langle
N|\bar s s | N \rangle}{ \langle
N|\bar u u + \bar d d | N \rangle}
\ee
where the canonical result $\hat \sigma =35 \pm 5$ MeV 
which measures the nucleon mass shift away from the chiral limit ($m_u=m_d=0$)
is due to Gasser
\cite{ga81} based on SU(2) CHPT. A more recent determination ~\cite{bm97}
gives   $\hat \sigma =36 \pm 7$ MeV.
The determination of the $\sigma$-term and
thus of $y$ has spawned a whole generation of $\pi N$ scattering experiments.
Indeed the canonical value obtained by Koch in the 
80's \cite{k82} yielded $y=0.11 \pm 0.07$ a value considered large in light
of results from e.g. neutrino scattering.

Up to $O(q^3)$ the $\sigma$-term can be directly 
calculated. However at next order appears a combination of LECs
denoted $e_1$ in Section \ref{nucm}  which has no pion 
matrix-element and therefore cannot appear in the scattering 
amplitude. One can then evaluate the $\sigma$- term  indirectly 
using the low energy theorem, Eq.~(\ref{eq:let}) and the fact that the 
difference 
$\sigma(2M_\pi^2)-\sigma(0)$
is well understood -- the evaluation within CHPT confirms the result of the 
dispersive analysis \cite{glls85}, $\sigma(2M_\pi^2)-\sigma(0)=15.2 \pm0.4 
{\rm {MeV}}$:
\be
\sigma=\Sigma-\sigma(2M_\pi^2)+\sigma(0)+\Delta 
=\Sigma-(15.2 \pm0.4 \pm 1) {\rm {MeV}} \, \, .
\ee
In this equation use has been made of $\Delta \sim 1 \, {\rm {MeV}}$ as 
discussed previously.
A rather large range of values for $\Sigma$ at $O(q^4)$ has been obtained 
depending on the way of evaluating it. In \cite{gls91} it is divided into 
two subthreshold parameters plus a curvature term $\Delta_d$ which is 
determined using $\pi \pi N \bar N$ phase shifts extrapolated 
from the KH80 $\pi N$ phases. A variant which contains some fourth order 
pieces has been given in \cite{o00}.  
One gets $\Sigma \sim 61 \cdots 104 \, {\rm {MeV}}$
which can be translated into $\sigma \sim   46 \cdots  89 \, {\rm {MeV}}$.
There is still controversy concerning the empirical value of $\Sigma$ . 
The latest analysis of the VPI/GW
group advocates  a rather large value $\Sigma= 79 \pm 7$ MeV while the older
Karslruhe one
gives a smaller value $64 \pm8 $ MeV. Such a large value of $\Sigma$ and
thus of $\sigma$ leads to an astonishingly large value of $y$.
It would certainly be of great interest to have a lattice determination of 
the $\sigma$-term, the question being how long one will have to wait for
having the desired precision to settle the problem. For an attempt,
see e.g. \cite{phw03}. 


\vspace{0.3cm}
Even though a lot of work has been done,   
the analysis of $\pi N$ scattering has not yet reached the high precision
of the $\pi \pi$ scattering case where the amplitude is now known in the
threshold region to an amazing degree of accuracy solving Roy equations
below 800 MeV \cite{cgl01}. Such an analysis remains to be done for $\pi N$
using the Roy type equations which have been derived in \cite{bl01},
allowing to link the physical and the subthreshold region.

\subsubsection{isospin violating case} \label{isoviol}

Having obtained a rather accurate representation of the isospin 
symmetric 
amplitude one can attack the more subtle problem of isospin violation.
Indeed it has been pointed out by Weinberg already in 
the seventies in his
seminal paper ~\cite{w77} that reactions involving nucleons and neutral pions
might lead to gross violations of isospin symmetry. This was reformulated
in more modern terminology in ~\cite{w95}.  
However in order to 
really pin down isospin breaking due to the light quark mass difference
one needs a machinery that allows to {\it simultaneously} treat the
electromagnetic and the strong contributions. CHPT is such
a machinery, it is in fact the only known framework at present which allows  
to do so. 

\vspace{0.2cm}
i) {\underline{formalism}}

\vspace{0.1cm}
Efforts to include virtual photons were fist done in the meson sector. An
extended power counting~\cite{egpr89,u95,nr95,bkw96,ku98} is introduced since
one has  to deal now with an expansion in the small momenta/masses
(the chiral expansion) and a second one in the electromagnetic coupling. 
A priori, these
two expansions can be treated separately. It was proposed in ~\cite{u95}
based on the observation that $e^2/4 \pi \sim M_\pi^2/(4 \pi F_\pi)^2
\sim 1/100$ to count the electric charge as a small momentum
\be
e={\cal O}(p) \, \, ,
\ee
which is the most economic way of organizing the double expansions in terms
of one. 
Along the same line, the construction of the generating functional of two 
flavor HBCHPT
in the presence of virtual photons with all finite terms 
up-to-and-including third order in the chiral dimension was then done in 
~\cite{mms97,ms98} and extended to fourth order in ~\cite{mm99} where the 
relativistic case was also considered. It turns out that contrary to the 
pions where $G$ parity
forbids a term of the type $(\bar u u -\bar d d)$ at leading order  
giving a pion mass difference dominated by electromagnetic effects, such terms
are allowed in the pion-nucleon sector allowing for large isospin breaking 
effects. 

The inclusion of virtual photons proceeds as follows. One works
with the nucleon charge matrix
\be
Q = e 
\left( \begin{array}{cc}
1 & 0 \\ 
0 & 0 
\end{array} \right)
      =\frac{e}{2}(\tau^3 +1)
\ee
and introduces spurions      
\be 
Q_\pm=\frac{1}{2} (u Q u^\dagger \pm u^\dagger Q u)
\ee
which under chiral $SU(2)_L \times SU(2)_R$ symmetry transform
as any matrix-valued matter field, 
\be
Q_\pm \to K Q_\pm K^\dagger
\ee
where $K$ is the compensator field representing an element of the 
conserved subgroup $SU(2)_V$. This leads for example to the generalized 
pion covariant derivative, see Eq.~(\ref{eq:deffield}) 
\be
d_\mu U = \partial_\mu U -i(v_\mu +a_\mu +QA_\mu)U +i U(v_\mu -a_\mu +QA_\mu)
\ee
and to additional local contact terms. At second order for example one has
\be
{\cal L}_{\pi N , {\it {em}}}^{(2)} =\sum_{i=1}^3 F_\pi^2
f_i \bar N {\cal O}^{(2)}_i N
\label{eq:isolag}
\ee
with the operators ${\cal O}^{(2)}_i$,
\be
{\cal O}^{(2)}_1 =\langle \tilde Q^2_+ - \tilde Q^2_- \rangle \, , \,\,\,\,
{\cal O}^{(2)}_2 =\langle Q_+ \rangle \tilde Q_+  \, , \,\,\,\,
{\cal O}^{(2)}_3 =\langle \tilde Q^2_- - \tilde Q^2_- \rangle
\ee
where $\tilde Q_\pm =Q_\pm - \langle Q_\pm \rangle /2$.
Apart from 
$f_2$ which can be deduced from the neutron-proton mass difference, 
$-e^2 F_\pi^2 f_2=(m_p-m_n)^{\rm{em}}$ \cite{ms98} not much is known about the 
new LECs.
$f_1$ has been recently determined from a combined analysis of  pionic hydrogen
and pionic deuterium 
\cite{mrr06}. The values for these two LECs are:

\vspace{0.2cm}
\begin{center}
\framebox{
$
f_2= -(0.97 \pm 0.38) \, , \,\,\,\,\,\ f_1=-2.1^{+3.2}_{-2.2} \,\,\,\,\,\, 
{\rm {in \, units \,\, of}} \, {\rm {GeV}}^{-1}
$.}
\end{center}

\vspace{0.2cm}

\noindent Note that the central value of $f_1$ agrees with a model-based 
estimate
\cite{lgfv02}. Also its errors  do not include the 
uncertainty
coming from the higher orders in ChPT and should thus be considered as
preliminary.
An estimate of the size of the other LECs has been done in ~\cite{mm99} 
by dimensional analysis. It was argued there that, measured
in appropriate powers of the inverse scale of chiral symmetry breaking,
these should be of order $1/4 \pi$ or $1/(4 \pi)^2$.
The detailed expressions of the Lagrangian up to ${\cal O}(p^4,e^2 p^2)$
as well as the divergent parts of the low energy couplings
is given in \cite{gilmr02}. As already stressed care has to be
taken on the definition of the $\pi \pi$ Lagrangian. 
For more details see \cite{gilmr02}. Subtleties on the separation between 
QCD and QED can be found in ~\cite{grs03}.

\begin{figure}[htb] 
\epsfysize=9.0cm
\begin{center}
\begin{minipage}[tb]{12 cm}
 \epsfig{file=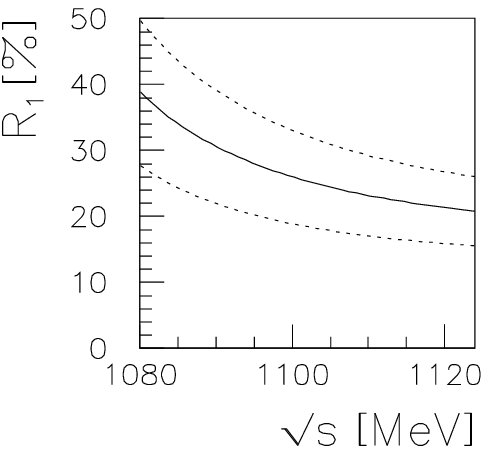,scale=0.9}
\hskip 3cm
\epsfig{file=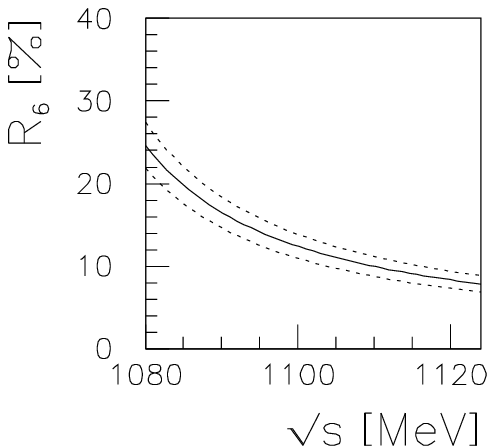,scale=0.9}
\end{minipage}
\begin{minipage}[tb]{18 cm} 
\caption{Isospin violation in the S wave projection of the ratios $R_1$ and 
$R_6$. The dashed line indicate the range for isospin 
violation. Figure taken from \cite{fm01xx}. \label{fig:triangle}}
\end{minipage}
\end{center}
\end{figure}

\vspace{0.1cm}
ii) {\underline{Studies of the isospin violation of $\pi-N$ scattering}} can be 
summarized as follows:

\vspace{0.1cm}
$\bullet$ Isospin-violating contributions to the $\sigma$-term can be as
large as $10\%$ but are negligible for the shift to the Cheng-Dashen point
\cite{ms98}.

$\bullet$ The corrections to Weinberg's time-honored prediction for the
difference of the $S$-wave scattering for the neutral pions off nucleons
are small, their numerical value lies in the range from 4 to 18$\%$
\cite{mm99}.

$\bullet$ A variety of triangle relations between measurable channels that 
vanish if isospin were an exact symmetry can be established, two of them
being of purely isoscalar type:
\bea
R_1 &=& 2 \frac{T_{\pi^+ p \to \pi^+ p}+T_{\pi^- p \to \pi^- p}-2
T_{\pi^0 p \to \pi^0 p}} 
{T_{\pi^+ p \to \pi^+ p}+T_{\pi^- p \to \pi^- p}+2
T_{\pi^0 p \to \pi^0 p}} \\ \nonumber
R_6 &=& 2  \frac{T_{\pi^0 p \to \pi^0 p}-T_{\pi^0 n \to \pi^0 n}}
{T_{\pi^0 p \to \pi^0 p}+T_{\pi^0 n \to \pi^0 n}} 
\,\, .
\eea 
In the analysis \cite{ms98,fms99} the strong violation of the triangle
relation $R_6$ predicted by Weinberg
was confirmed,
\be
R_6 \sim 25 \% \, .
\ee 
However an even more
important isospin violating effects 
was found in the isoscalar triangle relation $R_1$ connecting the charged and 
neutral pion scattering off protons ~\cite{fms99}:
\be
R_1 = 36.7 \% \, .
\ee
Interestingly this ratio could be accessible experimentally extracting the 
$\pi^0 p$ scattering by a precise pion photoproduction experiment as proposed
by Bernstein ~\cite{b98}. Note that these results are sensitive to the
exact value of the isoscalar scattering length and thus efforts should be
made to measure it more precisely.
The analysis of these triangle relations  was extended in 
\cite{fm01xx} to the energy range above threshold. In Fig.~\ref{fig:triangle}
are depicted the two previous relations
as a function of the total center-of-mass energies $\sqrt{s}$. The isospin
breaking effect drops by a factor of two over the first 50 MeV. This was to 
be expected since 
the lower the energy, the more important the quark mass difference is compared
to the kinetic energies of the particles and the larger the isospin breaking
should be. 
    
$\bullet$ A {\it complete} analysis of $\pi N$ scattering including the yet
missing part in the previously
described analysis \cite{mm99,ms98,fms99} -- loops with virtual photons as well as electromagnetic
counterterms -- was performed in \cite{fm01}.
A fit to all data (with the exception of the ones generally
considered inconsistent) below pion lab momenta of 100 MeV was done. 
A very pronounced difference in the hadronic amplitude for elastic 
scattering $\pi^- p \to \pi^- p$ compared to the standard PWA's was 
observed. This difference could be traced back to the inclusion of 
non-linear $\pi \pi \bar N N \gamma^*$ couplings as demanded by chiral
symmetry. These effects should be included in any calculations of em
corrections. In the S-wave triangle relation between elastic scattering,
$\pi^\pm p \to\pi^\pm$, and charge exchange, $\pi^- p \to \pi^0 n$ a 
strong isospin violation  of $\-0.75 \%$ was found in the low-energy
region, an order of magnitude smaller than reported in the literature 
\cite{glk95} but consistent with the expected size of isospin violation.     
  
$\bullet$
Use of pionic atoms to measure scattering lengths have developed
throughout the years. 
Let me stress here again the importance of a precise determination of these
scattering lengths. In the case of the $\pi N$ system for example they are 
correlated with the pion-nucleon sigma-term,
the pion-nucleon coupling constant and the Goldberger-Treiman 
relation. Moreover constraining the 
$\pi N$ interactions
at low energies also affects our understanding of more
complicated systems where $\pi N$ interaction serves as input, e.g. $ N N$
interaction, $\pi$-nucleus scattering, three-nucleon forces, etc.  
There are actually three types
of on-going experiments on hadronic atoms
(see for example \cite{hadatom03}): the DIRAC experiment 
which aims at the measurement of the $\pi^+ \pi^-$ atom lifetime
within a $10\%$ accuracy, allowing to determine the difference of 
the $\pi \pi$ scattering lengths $a_0-a_2$ with a 5$\%$ accuracy,
the DEAR collaboration  at the DA$\Phi$NE facility
which measures the energy levels shift and lifetime of the $1 s$ state
in $K^- p$ and $K^- d$ atoms aiming at a precise determination of the
$I=0,1$ $S$-wave scattering lengths and finally the PSI experiments \cite{psi}
which we are interested in here, which aims at a determination of the    
$\pi N$ scattering lengths  by measuring  at the percent level
the strong energy level shift $\epsilon_{1s}$, and the decay width into 
$\pi^0 n$ state, $\Gamma_{1s}$, 
in $\pi^- p$ atom.
Therefore the knowledge
of the relation between the scattering lengths and the threshold amplitude
should match the accuracy of the experiment.

The energy shift in pionic hydrogen is given at leading
order in isospin breaking by the well known formula of Deser et 
al.~\cite{dgbt54}. At next to 
leading order, that is at ${\cal O}(\alpha^4,
\alpha^3 (m_d-m_u))$ it reads \cite{lr00} 
\be 
\epsilon_{1s} =-2 \alpha^3 \mu^2_c (a^+ + a^-) (1+ \delta_\epsilon)
\ee
with $\mu_c=m_p M_\pi(m_p+M_\pi)^{-1}$ the reduced mass of the $\pi^- p$
system, $\alpha \sim 1/137.036$
and $\delta_\epsilon$ describes the isospin breaking effects. This
is the quantity to be determined very precisely in order to be able to 
determine the scattering lengths. Its expression is given by:
\be
\delta_\epsilon = \frac{ \alpha T^\gamma +(m_d-m_u)T^m} { 4 \pi 
(1+M_\pi/m_p) (a^+ + a^-)} +K +\delta_\epsilon^{vac}
\, ,  \, \, \, \, \, \, \, \, \, \, \,
K= -2 \alpha (\ln \alpha -1) \mu_c (a^+ + a^-) \, \, ,
\ee
where $T^\gamma, \, T^m$ are the electromagnetic and strong isospin breaking 
parts of the $\pi N$ 
amplitude and $\delta_\epsilon^{vac}$ is a correction due to the interference
of vacuum polarization and strong interactions which is in principle
of order $\alpha^5$ but is amplified by the large factor $(M_\pi^+/m_e)^2$. 
A study of this quantity  has been first done in \cite{lr00} where the leading
order result was obtained. It was then refined in  \cite{gilmr02} with 
next-to-leading order calculation. In the expression of $\delta_\epsilon$ 
some LECs appear. At next-to-leading order one has two LECs from the isospin 
symmetric part and 
five from the isospin breaking part of the Lagrangian. These are $f_1$ 
(see Eq.~(\ref{eq:isolag}))
and two combinations of two LECS from ${\cal {L}}_{\pi N}(e^2 p^2)$ and 
${\cal {L}}_{\pi N}(e^2 p)$. It turns out that in the calculation of 
$ \delta_\epsilon$~\cite{gilmr02} 
the most important contribution comes from the LEC $f_1$. In this reference
these LECs have been put to zero and their contribution have been 
taken into account in the error bars. They find:  
\be
\delta_\epsilon=(-7.2 \pm 2.9) \cdot 10 ^{-2}
\label{eq:deleps}
\ee
where an important contribution comes from the triangle graph discussed
in Section \ref{hbchpt}. Such a triangle graph has also shown to be important 
in the 
photoproduction process which will be discussed in Section \ref{elecpro}.
The leading order result gives $\delta_\epsilon=(-2.1 \pm 0.5) \cdot 10 ^{-2}$. As we have seen previously a recent determination of $f_1$ has been obtained. 
With it the central value, Eq.~(\ref{eq:deleps}) 
increases since the contribution of $f_1$ is positive and the error bars are 
bigger due to the large error on its value:
\be 
\delta_\epsilon=(-3.0^{+6.4}_{-4.5}) \cdot 10 ^{-2} \, \, .
\ee
This has to be compared with the potential model result~\cite{sbglo96}
which predicts $\delta_\epsilon=(-2.1 \pm 0.5) \cdot 10 ^{-2}$
Note that there no unknown LECs appear and thus the result is much more
precise. However it is well known that potential models do not, in 
general, include all effects of QCD+QED. In view of the error bars
a precise determination of the scattering lengths has clearly at present
to await a more precise determination of $f_1$.

\subsection{\it {$\pi d$ scattering}} \label{pid}
As we have just seen the pionic hydrogen energy shift allows to determine
the combination $(a^+ + a^-)$. From the width which  has not yet been 
discussed $\Gamma=8 \, \alpha^3 \, \mu_c^2 \, p_0 \,(1+1/P)(a^-(1+
\delta_\Gamma))^2$ 
where $P=1.546 \pm 0.009$ is the Panofsky ratio, $p_0$ the
center-of-mass momentum of the $\pi^0n$ pair and $\delta_\Gamma$,
the isospin breaking correction, one can  access  $a^-$. 
However, $a^+$ being
much smaller than $a^-$, a very high accuracy is needed in order to 
determine this quantity from the combination  $(a^+ + a^-)$.
A complementary piece of information comes from a 
measurement of  the ground-state energy shift of  pionic deuterium.
The determination of  $a^+$ and $a^-$ from such a measurement
is a two-step process. First one has to relate the energy shift to 
$\pi d$ scattering and then this scattering to $a^+$, $a^-$. 
As was stressed in \cite{mrr05} this requires different effective theories,
scales being different in the two steps. In the first one the hard momentum
scale is given by the average value of the three-momentum of the nucleons
bound within the deuteron, $\gamma={\sqrt{\epsilon m}} \simeq 45$ MeV 
(with $\epsilon=2.22$ MeV the deuteron binding energy)
and one 
can perform an expansion in the ratio of the scales $\alpha \mu_d /\gamma
 \simeq  {\cal O}(\alpha)$
where  $\alpha \mu_d$ is the momentum scale at which the 
charged pion and the deuteron form an atom whose observables
are measured by the experiment. 

 
Let us look at the second step. 
There are different ways of dealing with the study of very low-energy 
pion-deuteron scattering within the framework of effective field theories.

$\bullet$ the oldest approach is the so called hybrid 
approach. It
follows the seminal paper by Weinberg \cite{w92} where chiral Lagrangians 
were systematically  applied for the description of interactions of pions 
with nuclei. $\pi d$ scattering is calculated by sandwiching the irreducible 
transition kernel for pion scattering on two nucleons obtained by use 
of the chiral Lagrangian between ``realistic'' deuteron wave function  
\cite{k02,dov04}. However, this hybrid approach is only justified for
processes dominated by the long-range mechanism.

$\bullet$ Weinberg's proposal could be improved  
with the development of chiral effective theories for the $N N$
system, for a review see \cite{e06}. Both the kernel
and the deuteron wave function could then be calculated within ChPT. 
This was applied to $\pi d$ in \cite{bbemp03}. It was demonstrated 
there that the usual chiral counting was not suitable for the description
of low-energy $\pi d$ scattering due to infrared enhancements resulting 
from the 
anomalously small deuteron binding energy. Indeed it was shown that 
diagrams describing
processes with a virtual pion emission/absorption were two orders
of magnitude smaller than other diagrams appearing at the same order. 
A modified power-counting was thus introduced, for more details see 
this reference.

$\bullet$ Other approaches were developed, like the framework 
with perturbative pions which has been used for example in \cite{bg03}.
A so called Heavy Pion EFT with the dibaryon field has also been
introduced\cite{bs03}. In these two approaches the magnitude of the
LECs can be estimated to be large leading to  rather large theoretical
uncertainties in the relation between the $\pi d$ and $\pi N$ scattering
lengths.

$\bullet$ Recently another approach \cite{mrr05} 
was developed based on the finding of Ref.\cite{bbemp03}. There the 
absorption and emission of hadrons do not appear explicitly at the level
of Feynman diagrams but is included in the couplings of the effective
Lagrangian much in the spirit of the Heavy Pion approach. A detailed
comparison of these different approaches is discussed in \cite{mrr05}.  

In this second step 
the hard momentum scale is $M_\pi$ and  the
expansion is done in terms of $\gamma/M_\pi \sim 1/3$. This is indeed  
the most efficient expansion of the effective theory 
following the remarks above.
The matching between the theory in the first and in the
second step is then performed for the 
$\pi d$ scattering amplitude at threshold: this quantity must be the same
in both theories.   


Here I will follow Ref.\cite{bbemp03}. In the isospin symmetric case
the relation between the 
pion-deuteron scattering length $a_{\pi d}$ and the $\pi N$ ones is given 
by:
\bea
{\rm Re}\,a_{\pi d}&=&2\,\frac{1+\mu}{1+\mu/2}\,a^+
+2\,\frac{(1+\mu)^2}{1+\mu/2}\,\left((a^+)^2-2(a^-)^2\right)\,
\frac{1}{2\pi^2}\,
\left\langle\frac{1}{{\bf q}^2}\right\rangle_{\rm wf}
\nonumber\\
&+&2\,\frac{(1+\mu)^3}{1+\mu/2}\,\left((a^+)^3-2(a^-)^2(a^+-a^-)\right)
\,\frac{1}{4\pi}\,
\left\langle\frac{1}{|{\bf q}|}\right\rangle_{\rm wf}
+a_{\rm boost}+\cdots\,
\label{eq:apid} .
\eea
\vspace{0.30cm}
\begin{figure}[htb] 
\epsfysize=9.0cm
\begin{center}
\begin{minipage}[tb]{6 cm}
\epsfig{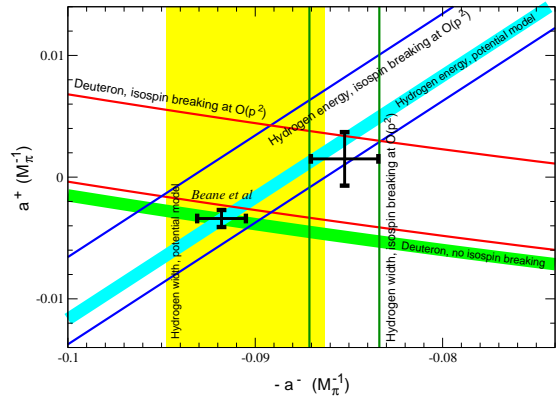}
\end{minipage} 
\begin{minipage}[tb]{18 cm} 
\caption{Determination of the $\pi N$ $S$-wave scattering lengths $a^+$ and
$a^-$ from the combined analysis of the experimental data
on the pionic hydrogen energy shift and width, as well as the pionic deuterium
energy shift. The cross denoted {\it {Beane et al}} is taken from  
Ref. \cite{bbemp03}. The second cross corresponds to the scattering lengths 
obtained in \cite{mrr06} from which the figure is 
taken. \label{fig:pidiv}}
\end{minipage}
\end{center}
\end{figure}

In the above equation, $a_{\rm boost}=(0.00369\cdots 0.00511)M_\pi^{-1}$,
$\left\langle 1 / {\bf q}^2 \right\rangle_{\rm wf} = (12.3\pm 0.3)M_\pi\,$
and $\left\langle 1 /|{\bf q}|\right\rangle_{\rm wf}
= (7.2\pm 1.0) M_\pi^2\,$,   
where  NLO wave functions with  cutoff mass in the interval
$\Lambda=(500\cdots 600)~{\rm MeV}$  \cite{egm00} have been used in order 
to evaluate the above wave-function averages.
Result for $a^+$ and $a^-$, cross denoted Beane et al, using  
Eq.~(\ref{eq:apid}) and the experimental value:
\be
a_{\pi d}^{\rm {exp}} = (-26.1 \pm 0.5 + i(6.3 \pm0.7) \cdots 10^{-3} 
M_\pi^{-1}
\label{apidexp}
\ee
is displayed on Fig.\ref{fig:pidiv} which gives
the state of the art determination of the $\pi N$ scattering lengths. Also
shown are the experimental bands \cite{sal01,hal98}
which uses a potential model
approach \cite{sbglo96} to calculate a set of correction like $\delta_\epsilon$
as discussed before and the new experimental values 
(denoted isospin breaking on the Fig.)
on the width and shift in hydrogen atoms  
with an improved 
accuracy \cite{g05}. In these new shifted bands isospin symmetric corrections at 
leading order
have been applied \cite{lr00,gilmr02}.
Treating the deuteron in the isospin symmetric case 
there is a discrepancy 
between the new pionic hydrogen and the deuterium experimental bands.  
It was a 
serious problem for some time and it
was first argued that the discrepancy was due to the incomplete 
treatment of the deuteron structure. However investigations in this
direction \cite{bbemp03,bs03,bg03,mrr05,nh06}  
showed that the uncertainty in the three-body
calculations cannot be solely responsible for the large discrepancy 
observed. Isospin breaking effects for the deuterium have then been calculated 
at leading
order\cite{mrr06}. As can be seen they are in fact important. 
They shift  the isospin symmetric
result \cite{bbemp03} in 
a non negligible
way. For example corrections to $a_{\pi d}$ are found to be \cite{mrr06}
\be
\Delta a_{\pi d}^{{\rm {LO}}} = -(0.0110 ^ {+ 0.0081}_{-0.0058}) M_\pi^{-1}
\, \, .
\ee
These corrections in the deuterium can thus reconcile, as shown 
in \cite{mrr06} 
the experimental bands coming from 
pionic hydrogen  and deuterium. All these bands 
once the 
isospin breaking is included in the deuterium case 
have now a common intersection 
area in the $a^+$, $a^-$-plane. This leads
to \cite{mrr06}
\bea
a^+ &=&(0.0015 \pm 0.0022)    M_\pi^{-1} \, \, , \\ \nonumber
a^- &=&(0.0852 \pm 0.0018)    M_\pi^{-1} \, \, .
\eea
Estimation of higher orders have been made in \cite{mrr06} but 
clearly the isospin-breaking corrections in pionic deuterium should be 
evaluated at least at ${\cal O}(p^3)$ in CHPT.

Interestingly, the $\pi d$ scattering length is
a complex-valued quantity already at threshold,
see Eq.~(\ref{apidexp}). Diagrams contributing to the 
imaginary parts are of higher order and thus have not been included in 
the calculations I have discussed up to now. These are called dispersive
contributions. However a first estimate by Br\"uckner \cite{b55} already 
speculated
that the real part and the imaginary part of these contributions should be
of the same order, an expectation which was confirmed later \cite{at74,elt02}.
Given the high accuracy of the measurement and the size of the imaginary 
part these dispersive contributions were very recently looked up within
CHPT \cite{lbhhkm06}. What is needed is a controlled power counting
for $NN \to N N \pi$ consistent with the one used for $\pi d$. This 
was developed in recent years \cite{cfmk96,hkm00} 
-- for a review see \cite{h04}.   
Once all diagrams contributing to leading order
to the hadronic part of the dispersive and absorptive corrections
as well as to the transition $\pi d \to \gamma N N \to \pi d$ are included, 
their net effect provides a small correction
to the real part of $a_{\pi d}$ of the order of 6.5$\%$ of the experimental 
value.  
The absorptive part is found to be
\be 
{\rm {Im}}(a_{\pi d})=((4.25 \pm 1.2)+(1.4\pm 0.4)) \cdot 10^{-3}
M_\pi^{-1} \,\, ,
\ee
in very good agreement with the experimental value:
\be
{\rm {Im}}(a_{\pi d}^{\rm {exp}})=((4.7 \pm 0.5)+(1.6\pm 0.2)) \cdot 10^{-3}
M_\pi^{-1} \,\, .
\ee
In both these equations the hadronic and electromagnetic contributions 
are given separately. 

In view of all these developments an improved measurement
of the energy shift in pionic deuterium would certainly be welcome.
In fact in the near future a new measurement with a projected total 
uncertainty of $0.5 \%$ for the real part and $4 \%$ for the imaginary
part of the scattering length will be performed at PSI \cite{gal}. 
For the reader interested on older works there is a review in \cite{tl80}.

\section{Electromagnetic properties} \label{emprop}

Processes involving nucleons interacting with one or several photons are also 
of particular interest. They contain fundamental observables describing
the internal structure of the nucleon. 
 
\subsection{\it {Form Factors}} \label{ff}
\begin{figure}[t]
\begin{center}
\epsfysize=1.8in
\epsffile{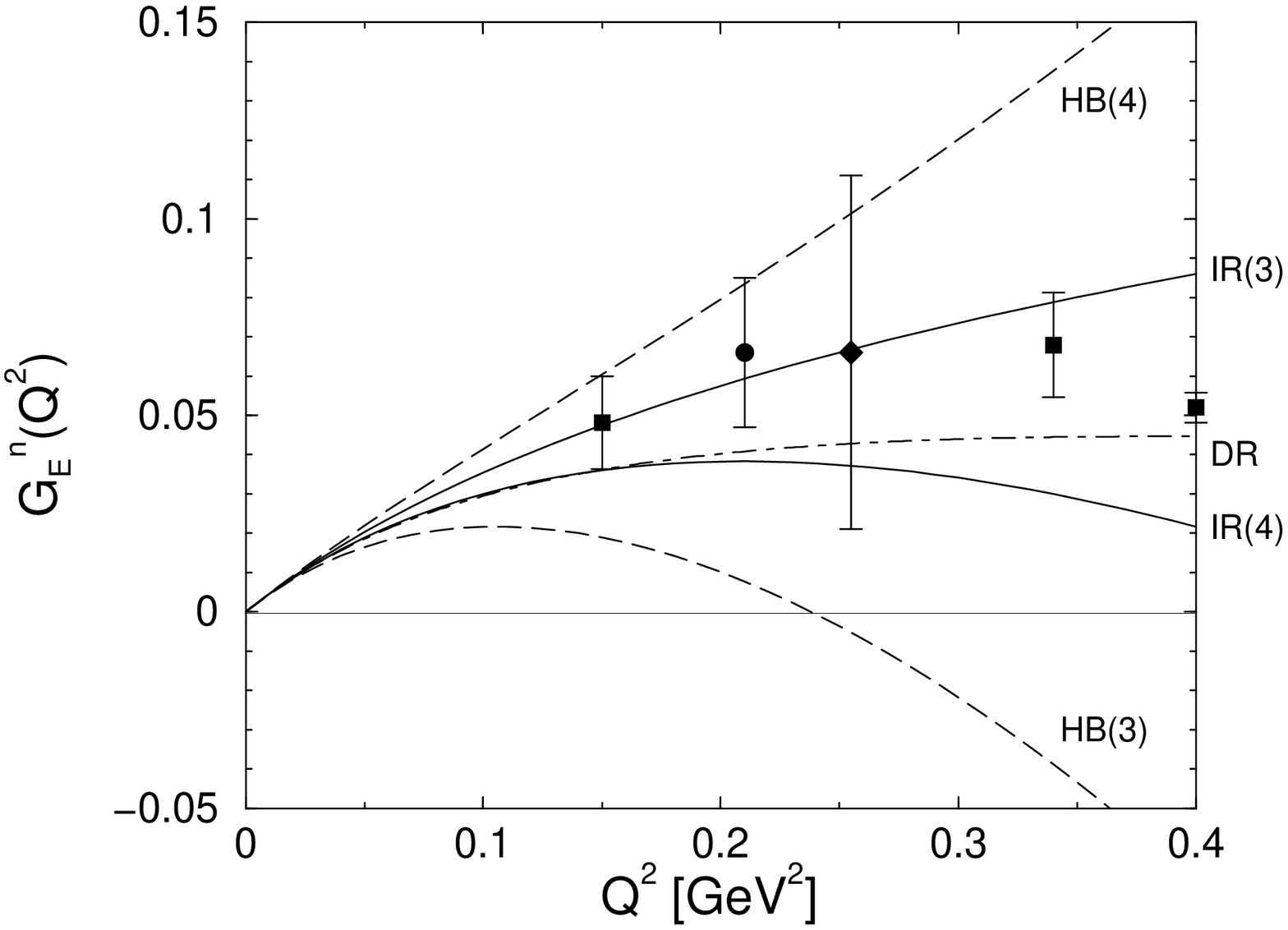}
\hspace{0.5cm}
\epsfysize=1.8in
\epsffile{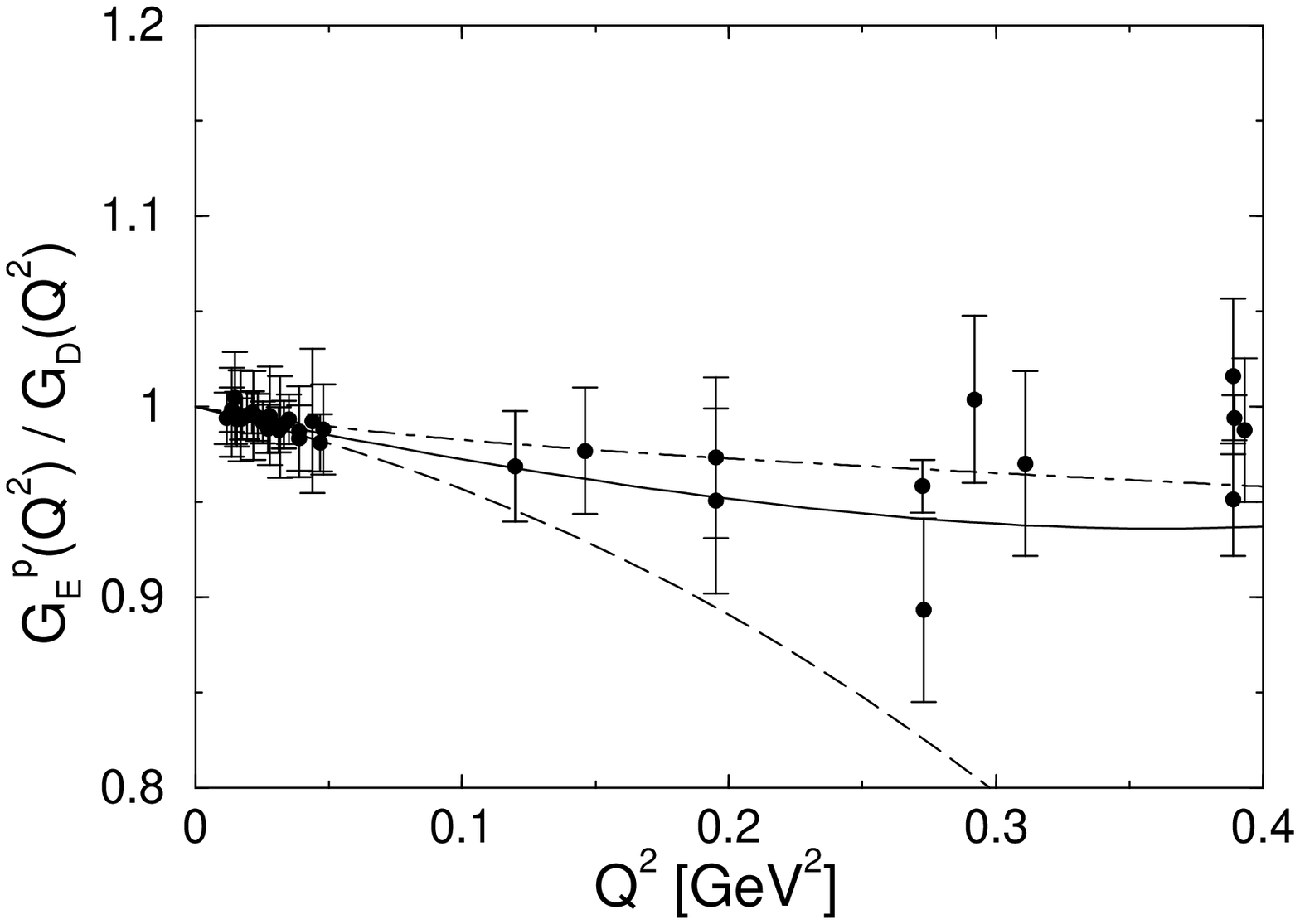}
\end{center}
\vskip 0.1cm
\caption{Left panel: The neutron electric form factor in IR regularization.
All LECs are determined by a fit to the neutron charge radius measured in 
neutron-atom scattering. The data are from \cite{nffdata}. Right 
panel:
The proton electric form factor including vector mesons to third (dashed curve)
and fourth (solid curve) order, divided by the dipole form factor. The data
are world average. 
In both panel the dot-dashed
curve is the dispersion theoretical result. Figures taken from \cite{kubm01}.
\label{fig:emff}}
\end{figure}

The structure of the nucleon  as probed by virtual photons  
is parametrized in terms of four form factors,  
\be  
\langle N(p')\, | \, \bar{q} {\cal Q} \gamma_\mu q \,  | \, N(p)\rangle   
= e \,  \bar{u}(p') \, \biggl\{  \gamma_\mu F_1^{N} (t)  
+ \frac{i \sigma_{\mu \nu} q^\nu}{2 m} F_2^{N} (t) \biggr\}   
\,  u(p) \,, \quad N=p,n \,,  
\ee  
with ${\cal Q}$ the quark charge matrix and $t = q^2=(p'-p)^2$ the invariant 
momentum   transfer squared.
$F_1$ and $F_2$ are called the Dirac and the Pauli  
form factor, respectively, with the normalizations $F_1^p (0) =1$,  
$F_1^n (0) =0$, $F_ 2^p (0) =\kappa_p$ and $F_2^n (0) =\kappa_n$. Here,  
$\kappa$ denotes the anomalous magnetic moment.
One also uses the electric and magnetic Sachs form factors,  
\be  
G_E (t) = F_1 (t) + \frac{t}{4m^2} F_2 (t) \, , \quad G_M (t) = F_1 (t) + F_2 (t) \, .  
\label{sachsdef}
\ee  
In the Breit--frame, $G_E$ and $G_M$ are nothing but the Fourier--transforms  
of the charge and the magnetization distribution, respectively.
The understanding of these form factors is of utmost importance in any theory
or model of strong interactions. They have thus been extensively studied
experimentally as well as theoretically, for a recent review see \cite{ppv06}
and references therein. Experiments have been performed
or are underway at NIKHEF, MAMI, ELSA, MIT-Bates, JLAB $\cdots$. 
On the theory side it was already established a long time ago that 
the isovector charge radii are diverging in the chiral limit of 
vanishing pion mass \cite{bz72}. Since then there has
been several calculations within CHPT with different 
regularization scheme \cite{gss88,bkkm92,flms97,kubm01,fgs04,sgs05} as
well as in the SSE approach \cite{bfhm98}. As has been pointed out in 
Section \ref{hbchpt}, HBCHPT suffers in this case from a distorsion of the 
analytical structure of the isovector spectral function \cite{bkm96,k03}. 
I will thus report here on the calculation done within IR \cite{kubm01}.
As can be seen on the l.h.s. of  Fig.\ref{fig:emff}  a good description 
of the neutron charge form factor is obtained for momentum transfer up to 
about $Q^2=0.4$ GeV$^2$. Also shown  is, as expected, the much better 
convergence of the series compared to HBCHPT.  
For the other three form-factors not shown here the agreement with the data is 
not very good. This however can be easily understood. 
It has indeed long been established that vector mesons
play an important role in these form factors. In \cite{kubm01} the low-lying 
vector mesons $\rho$, $\omega$ and $\phi$ have been included in a chirally
symmetric manner based on an antisymmetric tensor field representation. 
Refitting the LECs by subtracting the vector meson contribution a good 
agreement for all four form factors is obtained. As an example  
the right panel of Fig.\ref{fig:emff} shows the
electric proton form factor with the vector mesons included. 
Similar findings have been obtained in \cite{sgs05}.
 
\subsection{\it {Compton Scattering}}  \label{cs}

Let me now come to processes involving two photons.
We will consider here real, virtual and double virtual Compton scattering 
(V$^2$CS) off nucleons (neutrons or protons) in forward direction,
that is the reaction 
\be
\gamma^{(\star)} (q,\epsilon) + N(p,s) \to \gamma^{(\star)} (q,\epsilon ')+ 
N(p,s^\prime)~,
\ee
with $q~(p)$ the real/virtual photon (nucleon) four--momentum, $s~(s^\prime)$
the nucleon spin (polarization) and $\epsilon\, (\epsilon ')$ the
polarization four--vector of the incoming (outgoing) photon. 

\subsubsection{Real and Virtual Compton Scattering} \label{vcs}
Real Compton scattering has been considered 
since the early 1950s with the determination of the electric $\alpha$ and 
magnetic $\beta$ polarizabilities.  
Note that their determination within CHPT was one of the successes of 
this framework which showed the importance of 
pion loop effects. The ${\cal O}(p^3)$ result \cite{bkm91} 
which is completely given by the pion loops is in remarkable agreement with 
the experimental results. Also interesting is that  these quantities diverge
in the chiral limit which is a challenge for lattice QCD calculations.
Since these first results extensions 
to  ${\cal O}(p^4)$ have been performed \cite{bkms93,bmmpk02} 
and the role of the resonances has been examined \cite{hhkk98}. 
Note that in this SSE calculation a 
discrepancy between theory and experiment for the 
proton magnetic polarizability was obtained due to 
a large contribution from the $\Delta$ driven to lowest order and the absence
of the large negative non-analytic pion loop contribution due to the 
order the calculation was done, see \cite{bkms93}. For the precision aimed at,
a calculation at the next order ${\cal O}(\epsilon ^4)$
within this framework has certainly to be performed.  
For real photons two spin-dependent polarizabilities have been measured. 
Let me mention here the backward spin polarizability, I will come back to 
the forward spin one in the next section. A first experiment by the LEGS group
\cite{leg} has challenged the theoretical predictions with a value
much smaller in absolute value. However there is now contradicting evidence
from recent MAMI data \cite{mami1,mami2}, their new results being well in 
the range from both dispersion theory \cite{dpv02} and CHPT \cite{ghm00,kmb00}.
Extension of these polarizabilities to non-zero energies have been 
defined in \cite{gh02}. These are the so-called dynamical polarizabilities
which test the global low energy excitation spectrum of the nucleon
at non-zero energy. A first analysis of the sensitivity of proton Compton
cross section data to these quantities has been done in \cite{hghp03}.

\begin{figure}[hbt]
\epsfysize=18.0cm
\begin{center}
\vskip -0.3cm
\includegraphics*[width=9cm]{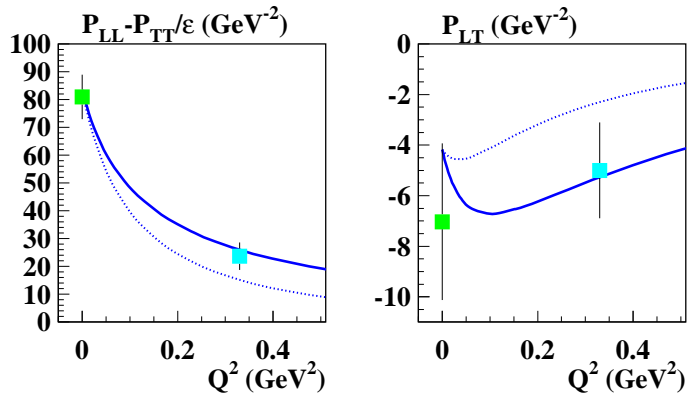}
\end{center}
\caption{Comparison of a HBCHPT (solid line) calculation of two generalized 
polarizabilities with the MAMI measurement at $Q^2=0.33$ GeV$^2$. The dashed
line shows the structure functions obtained by switching off the 
spin-dependent polarizabilities. Figure courtesy of N. D'Hose.   
 \label{fig:gps}}
\end{figure}
It is only by the seventies that it was recognized that VCS would enable one 
to measure generalized polarizabilities (GPs) \cite{ad74}. 
The theoretical framework for this reaction can be found in 
\cite{gui95,dkms97}.
There are altogether 6 GPs that are  independent due to crossing 
symmetry and charge conjugation invariance. From these 6 two are spin 
independent and are extensions of 
$\alpha$ and $\beta$ at finite $Q^2$.  
They provide information about the spatial distribution of charge
and magnetism inside the nucleon. There have been during the 
last 10 years a lot of 
experimental (at  MAMI, Jlab and MIT-Bates) and theoretical
activities to determine these quantities, 
a summary can be found in \cite{hose06}. 
As illustration Fig.\ref{fig:gps} compares the result of a HBCHPT calculation 
at ${\cal O}(p^3)$ of two particular combinations of spin-independent and 
spin-dependent GPs with the MAMI data. 
A very nice agreement is obtained. Clearly it would be interesting 
to perform an ${\cal O}(p^4)$ calculation to check the convergence of the
series.  HBCHPT  ${\cal O}(p^4)$ calculations have been performed \cite{kpv04}
but only for the spin dependent polarizabilities. 
The first HBCHPT calculation of the GPs 
can be found in \cite{hhkd00}. In the 
existing unpolarized as well as
polarized experiments it is not possible
to resolve the 6 GPs.  However the double polarization experiment A1/1-00 
which is
being realized at MAMI will allow to disentangle them
and its result is thus eagerly awaited. A new program at HIGS will
also allow to measure these GPs. 
These experiments will shed new light on
our understanding of the internal structure of the proton and serve
as a check on the theoretical explanations of the polarizabilities. 
In the next section I will concentrate
on spin-dependent observables obtained in V$^2$CS and discuss 
two specific combinations of the spin GPs.
\subsubsection{V$^2$CS: spin structure functions and sum rules} \label{v2cs}

There are many on-going activities concerning the spin of the nucleon,
as for example understanding how it is built from quarks and gluons. Apart
from the GPs,
other interesting quantities to test CHPT are the spin structure 
functions and their moments.
Experiments with polarized beams and polarized $p$, $d$,
$^3He$ targets are performed  
over a very wide range of $Q^2$, with $Q^2$ the negative of the photon 
virtuality (momentum transfer squared). 
Nowadays  rather low values for this momentum transfer are reached at JLAB
in the range of 
applicability of CHPT. Indeed  preliminary results 
for $Q^2$ values as low as 0.05 ~GeV$^2$ have been reported  while data 
taking have started for
very low $Q^2$ down to 0.015 GeV$^2$ \cite{chen06}.  
At high energy systematic and controlled 
theoretical calculations can also be performed. The region of
intermediate momentum transfer  is accessible using quark/resonance
models or can be investigated using dispersion relations \cite{dkt01}. 
Thus ultimately
these investigations will  lead to an
understanding of how in  QCD the transition from the
non--perturbative to the perturbative regime takes place, guided by the precise
experimental mapping of spin--dependent observables from low momentum 
transfer to the multi--GeV region. 

It is common to express the spin-dependent amplitude of V$^2$CS in terms of 
two structure 
functions, called $S_1(\nu,Q^2)$ and $S_2(\nu,Q^2)$. In the rest
frame and in the Coulomb gauge $\epsilon_0=\epsilon'_0$  the V$^2$CS 
forward matrix element is given by:
\begin{eqnarray}
\epsilon^\prime\cdot T\cdot\epsilon|_{\rm rest}
&=&\frac{1}{2m}\, \chi^\dagger\left\{i\vec{\sigma}\cdot
\left(\vec{\epsilon}^{\,\prime}\times\vec{\epsilon}\right)\left[m\nu 
S_1(\nu,Q^2)-Q^2S_2(\nu,Q^2)\right]\right.\nonumber\\
& &\phantom{\frac{1}{N}\;\chi^\dagger}\left.
+i\left[\vec{\sigma}\cdot(\vec{\epsilon}^{\,\prime}\times\hat{q})\,
\vec{\epsilon}\cdot\hat{q}-\vec{\sigma}\cdot(\vec{\epsilon}\times\hat{q})\,
\vec{\epsilon}^{\,\prime}\cdot\hat{q}\right]\left(\nu^2+Q^2\right)S_2(\nu,Q^2)
\right\}\chi\; ,\label{match}
\end{eqnarray} 
where 
$\nu=p \cdot q/ m $ the energy transfer and $Q^2 = -q^2 \ge 0$ the (negative of
the) photon virtuality. 
Note that while $S_1 (\nu,Q^2)$ is even under crossing 
$\nu \leftrightarrow -\nu$,  the structure function $S_2(\nu,Q^2)$ is odd.
Because of unitarity, there is a basic connection between spin  
structure functions in  V$^2$CS and the ones $G_{1,2}$ probed in inelastic  
electroproduction experiments. One has 
\be
{\rm Im}~{S}_i (\nu , Q^2)  = 2\pi~ G_i (\nu , Q^2) ~, \quad (i = 1,2)~.
\ee 
This is simply related to the fact that
the imaginary part of the Compton tensor is given in terms of nucleon
plus meson states, the lowest one being the pion--nucleon state.
This gives rise to sum rules as the well-known Drell-Hearn-Gerasimov 
sum rule \cite{gdh66} for real photons which relates the anomalous magnetic 
moment to the 
total photoabsorption cross sections
$\sigma_{1/2}$
and $\sigma_{3/2}$ corresponding to the excitation of intermediate states
with spin projection 3/2 and 1/2, respectively: 
\be 
-{2\;\pi^2\;\alpha\;\kappa^2 \over m^2} =\int {d\nu \over \nu}\, \left[
\sigma_{3/2} - \sigma_{1/2} \right]~.
\ee
The value of the l.h.s. of the sum rule is $-204 \,\mu$b for the proton and 
$-232 \,\mu$b  for the neutron. This sum rule is presently under active experimental
investigation at MAMI, ELSA, GRAAL, CEBAF and Spring-8. First measurements
on the proton agree with this sum rule within some assumptions from 
contributions
coming from un-measured regions. For the neutron only estimates can be given.
For an experimental review see \cite{helbing}.

The DHG sum rule has been generalized to finite $Q^2$ \cite{dkt01,jo01}. 
There are different possibilities, an interesting definition being: 
\be
I_1(Q^2)={ 2 m^2 \over Q^2} \int_0^{x_0} m \nu  G_1(x,Q^2) dx~. 
\ee
Indeed while at low $Q^2$ one recovers the DHG sum rule this definition 
leads to the Ellis-Jaffe sum-rule at very large momentum transfer \cite{ej74}.
In this equation
$x=Q^2/2m\nu$ is the standard scaling variable and $x_0$ corresponds 
to the pion production threshold. Other sum rules can be defined involving
also the structure function $G_2$ as the so-called Burkhardt-Cottingham 
sum rule \cite{bc70}. Expanding the $V^2CS$ structure functions at low energies 
$\nu$, that is around $\nu = 0$, one has for example
\be
\bar{S}_1^{(0)}(0,Q^2)  ={4e^2 \over m^2} I_1(Q^2) \, \, ,
\ee
where $\bar{S}_1^{(0)}(0,Q^2)$ is the first constant term in the expansion of 
$S_1$
and the bar means that the elastic contribution (nucleon pole term) has been 
subtracted. 
For the relations of the other structure functions see \cite{behem03}.
Two other interesting quantities as mentioned above are the following combinations of GPs, the forward spin polarizability
$\gamma_0$ and the longitudinal-transverse polarizability $\delta_0$. They
are defined and given by:
\bea
\gamma_0(Q^2)&\equiv& {1 \over 4 \pi^2} \int {d\nu\over\nu^3}
(1-x)(\sigma_{1/2}(\nu,0) - \sigma_{3/2}(\nu,0))= 
{1 \over 8\pi}\left(\bar{S}_1^{(2)}(0,Q^2)
- \frac{Q^2}{m}\bar{S}_2^{(3)}(0,Q^2)\right)~,
\label{eq:genpol} \\ \nonumber
\delta_0(Q^2)&\equiv&{1 \over 2 \pi^2} \int {d\nu\over\nu^3}
(1-x) \lim_{Q^2 \to 0}
\biggl({\nu\over Q}(\sigma_{1/2}(\nu,0)\biggr)={1 \over 8\pi}\left(\bar{S}_1^{(2)}(0,Q^2)
+\bar{S}_2^{(1)}(0,Q^2)\right)~.
\eea
They involve an extra $1/\nu^2$ weighting compared to the first moments so  
that they have the experimental advantage that the uncertainty due to the
unmeasured region at large $\nu$ is minimized.  

The two spin structure functions $S^{(1,2)}$ have been calculated within 
HBCHPT \cite{jko00,bu01,ksv02} and to fourth order (one loop) in the IR 
regularization 
\cite{behem02,behem03}.
At this order no unknown low-energy constants appear and thus 
parameter-free predictions are obtained. It is, however, well-known that the 
excitation of the $\Delta (1232)$ plays a significant role 
in the spin sector of the nucleon.  
A first attempt to include the 
$\Delta$ explicitly has thus been done in \cite{behem03}. There the 
relativistic 
Born graphs were calculated in order to get an estimate of the contribution 
of this resonance.  
In order to take into account the fact that the  $\Delta \to N \gamma$
transition occurs at finite $Q^2$  the possibility of introducing a 
transition form factor $G_{\Delta N \gamma} (Q^2)$
as extracted from pion electroproduction in the $\Delta(1232)$-resonance
region \cite{trans} was also studied in that reference. A less pronounced 
though important resonance contribution is related
to the vector mesons. These were thus also included in \cite{behem03}.
\begin{figure}[htb]
\epsfysize=9.0cm
\begin{minipage}[t]{6 cm}
\vskip -0.3cm
\hskip 1cm
\includegraphics*[width=7.5cm]{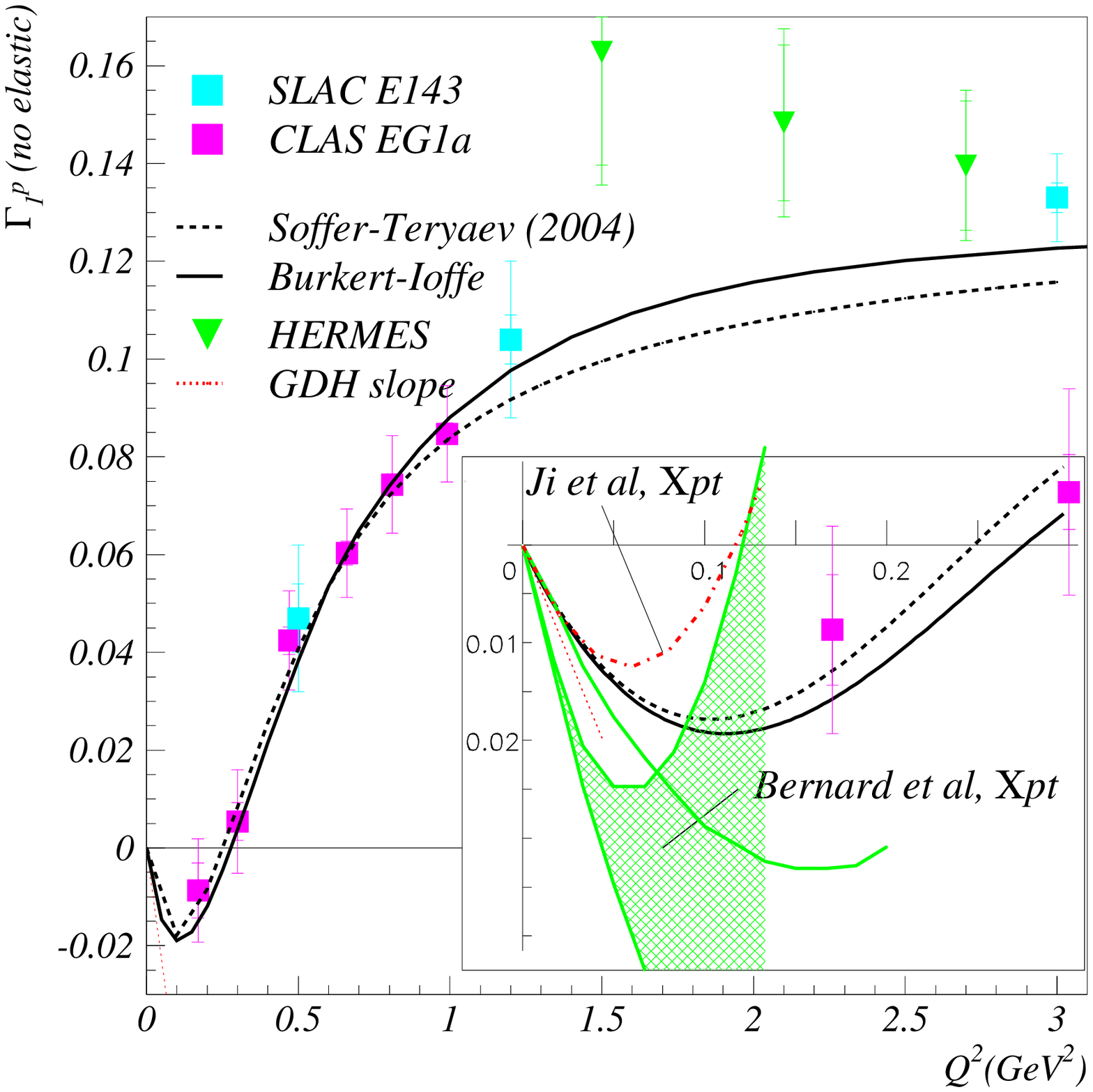}
\vskip -7cm
\hskip 10cm
\includegraphics*[width=6.cm]{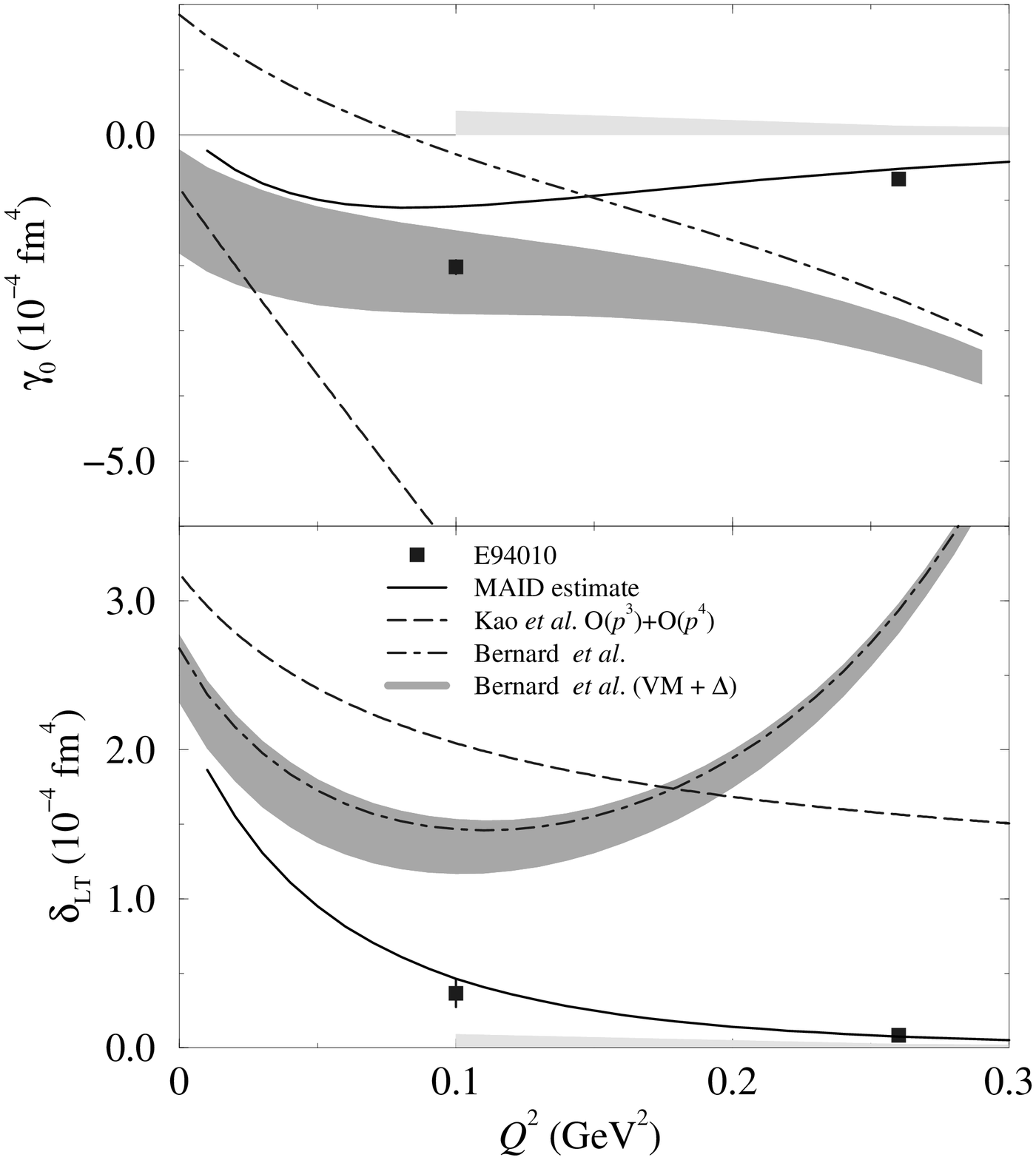}
\vskip -5.9cm
\end{minipage}
\caption{Left panel: Experimental results for $\Gamma_1^p(Q^2)$
compared to model predictions and CHPT calculations. The full lines (bands)
at low $Q^2$ are the next to leading order CHPT predictions by 
Ji et al. \cite{jko00} and Bernard et al. \cite{behem03}. Right panel: neutron
generalized spin
polarizabilities $\gamma_0$ and $\delta_{LT}$. The dashed lines  are the 
HBCHPT calculation by \cite{ksv02}. 
Figure courtesy of A. Deur. 
 \label{fig:ssf}}
\end{figure}

On the l.h.s. of Fig.\ref{fig:ssf} is shown the 
first moment $\Gamma_1^p(Q^2)$ for the proton:
$
\Gamma_1 (Q^2) = Q^2/2m^2 \, {I}_1 (Q^2)
$
over a wide range of $Q^2$. We are interested in the region $Q^2 < 0.1$ GeV
as detailed on the insert of the figure. There are 
shown 
the JLAB CLAS EG1a data \cite{fa03} together 
with the HBCHPT results
from Ji et al \cite{jko00} and the IR ones from Bernard et al \cite{behem03}.
The difference
between these two CHPT results comes from higher order terms which
improves the convergence of the series of the latter  at higher values of
$Q^2$ as already explained. The data are at somewhat too high $Q^2$ to compare
with these  results. However preliminary data at lower $Q^2$ have been 
reported at Chiral Dynamics 2006. 
As can be inferred from Fig.2 of Chen's proceedings \cite{chen06} 
there is a nice agreement between the IR calculation,
black curve in the shaded area denoted Bernard et al. in the Figure, up to 
values 
of $Q^2 \sim 0.07$ GeV$^2$ which is roughly the range of validity of the 
calculation. 
It turns out that the inclusion of the $\Delta$ and vector
mesons as just described above makes the agreement somewhat worse. However,
a full ${\cal O}(\epsilon^3)$ calculation has to be performed, the shown
results giving just a first estimate of the contribution of  these degrees 
of freedom. 
For results on other moments and the neutron, see Chen's proceedings.

On the right panel of Fig. \ref{fig:ssf} are shown the neutron generalized 
polarizabilities,
see Eq.~(\ref{eq:genpol}). $\gamma_0$ is one of the quantities known to be 
rather sensitive
to the $\Delta$ degrees of freedom as illustrated in the figure by 
comparing the two CHPT
calculations with and without $\Delta$. 
The data point at $Q^2=0.1$~ 
GeV$^2$ nicely lies in the band of the calculation of ref.\cite{behem03} 
with $\Delta$
degrees of freedom. The proton case not shown here is more problematic 
due to a very bad convergence of the chiral series \cite{behem02}. 
$\delta_{LT}$ is an interesting quantity, it can indeed 
serve as a testing ground of the chiral dynamics of QCD since contrary to 
$\gamma_0$ the $\Delta$ contributes only marginally and 
its chiral expansion at the photon point
is very well behaved, for 
a discussion see \cite{behem02}.
There is, however, a problem here
with the IR result linked to the unphysical cuts appearing in the regular
function $R$ as discussed in Section \ref{ir}. In principle these cuts lie 
far away
from the region where CHPT is valid and are of no relevance. In the particular
case of concern here they lead to a change of curvature already 
at $Q^2 \sim 0.1$
GeV$^2$ and thus to a disagreement with the lower data point. Clearly for that
particular quantity data points at rather low $Q^2$ are needed to be able to
compare the experimental result with 
the IR one. Alternatively one could use another regularization
which does not have this problem like EOMS for example. Let me point out again 
that
the same happens in the case of the nucleon mass or, as we have seen,
in the anomalous magnetic
moment, both quantities go to infinity as $M_\pi$ goes to infinity due to these
unphysical cuts. In the relativistic framework they would go to zero. This has,
however, in those cases no impact on the result in the low energy region.

As we have seen there are in fact four independent spin independent GPs.
$\gamma_0$ for example is a particular combination of three
of them.  HBCHPT predictions to ${\cal O}
(q^4)$
for all spin-flip GPs have been obtained
in \cite{kpv04}. 
 
\subsection{\it{Photo-and Electroproduction}} \label{elecpro}

I will concentrate here on Pion and Electroproduction in the threshold
region. For a discussion of the $\Delta$ region, see the very
recent review \cite{pvy06} and references therein. I will also concentrate
on neutral pions. For charged pions and its relation to the nucleon
axial radius and pion charge  (vector) radius, see \cite{bem02}. For a 
discussion of pion electroproduction and chiral Ward identities see 
\cite{fusc03}.

\subsubsection{off the nucleon} \label{nuc}

\begin{figure}[htb]
\epsfysize=9.0cm
\begin{minipage}[t]{6 cm}
\includegraphics*[width=6cm,angle=270]{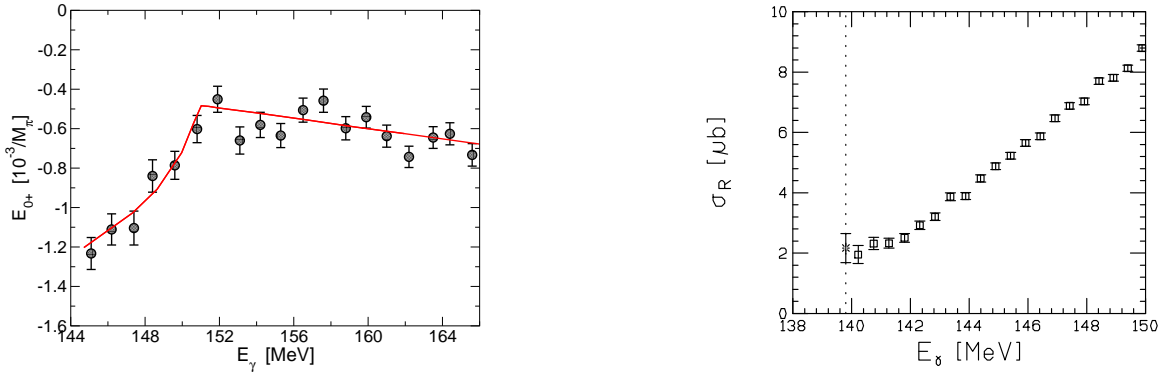}
\vskip -5.2cm
\hskip 10cm
\includegraphics*[width=6cm]{jack.epsi}
\vskip -7cm
\end{minipage}
\caption{Left panel: CHPT prediction (solid line) 
for the electric dipole amplitude
in   $\gamma p \to \pi^0 p$ in comparison to the  data  from MAMI.
Right panel: Reduced total cross section for $\gamma d \to \pi^0 d$. The
data from SAL are depicted by the boxes, the CHPT threshold prediction is the
star on the dotted line (indicating the threshold photon energy).
\label{fig:photo}} 
\end{figure}

The  electric dipole amplitude 
$E_{0+}$ is a very interesting quantity since it vanishes in
the chiral limit.  It admits an expansion in terms of the quark (pion)
mass, $
E_{0+} = a M_\pi + b M_\pi^2 + \ldots~,$
where $a, b, \ldots$ are calculable coefficients. This defines a venerable
low-energy theorem (LET) which has been the source of a lot
of activities, related to what is now known as the LEG  (low 
energy guess) of the 70's \cite{VZ,deB}, for details see \cite{BGKM, BKM1}. 
See also \cite{em95} for a definition of  a LET.
In case of the 
neutron, i.e. for $E_{0+}^{\pi^0 n}$,
the coefficient $a$ is zero because of gauge invariance. There have been
many 
experimental and theoretical developments concerning the electric
dipole amplitude for neutral pion production off protons 
\cite{BGKM,BKM1,bkmpi0}. Let me summarize what is the status today.
Even though the convergence for the electric dipole amplitude is slow
due to some strong final-state interactions, 
a CHPT calculation to order $p^4$ does allow to
understand its energy dependence in the threshold region
once three LECs are fitted to the total and differential cross section
data \cite{bkme0p}, see the left panel of Fig.\ref{fig:photo}. The threshold
value $E_{0+}^{\rm thr} (\pi^0 p) = -1.16$ (in units $10^{-3}/M_\pi$) thus obtained agrees with the data,
$E_{0+}^{\rm thr} (\pi^0 p) = -1.31\pm 0.08$ \cite{fuchs},
                                $-1.32\pm 0.05$ \cite{berg},
                                $-1.33\pm 0.08 \pm 0.03$ \cite{schmidt}.
Even more interesting is the
case of the neutron. Here, CHPT predicts a sizeably larger $E_{0+}$
than for the proton (in magnitude).
The CHPT predictions for $E_{0+} (\pi^0 p,\pi^0 n)$ in the threshold
region  clearly
exhibit the unitary cusp due to the opening of the secondary
threshold, $\gamma p \to \pi^+ n \to \pi^0 p$ and $\gamma n \to \pi^-
p \to \pi^0 n$, respectively. Its strength is directly proportional
to the isovector (charge exchange) $\pi N$ scattering length. An average value
of $3.3 \pm 0.5$ was given by the MAMI A2 collaboration in \cite{bal97}. 
However, more work with polarized
targets is required to have a more precise determination
of this strength allowing 
to get additional informations \cite{bern96} on zero-energy pion-nucleon 
scattering as well as on isospin breaking effects in the $\pi N$ system 
\cite{b98}.  
Note that while $E_{0+} (\pi^0
p)$ is almost vanishing after the secondary threshold, the neutron
electric dipole amplitude is sizeable.

Quite in contrast to what was believed for a long
time, there exists a set of LETs for the slopes of the P--waves
$P_{1,2} = 3E_{1+} \pm M_{1+} \mp M_{1-}$ at threshold, for example
\be
\frac{1}{|\vec q \,|} P_{1, {\rm thr}}^{\pi^0 p} = \frac{e g_{\pi
  N}}{8 \pi m^2} \left\lbrace 1 + \kappa_p + \mu \left[ -1 - 
\frac{\kappa_p}{2} + \frac{g_{\pi N}^2}{48 \pi}(10 -3\pi) \right]
+ {\cal O}(\mu^2) \right\rbrace \, \, , 
\ee
and similarly for the slope of $P_2$ at threshold. 
Here, $g_{\pi N}$, $m$ and $\kappa_p$ are the pion-nucleon coupling
constant, the proton mass and the proton anomalous magnetic moment,
in order.
\begin{figure}[t]
\epsfysize=9.0cm
\includegraphics*[width=8.3cm]{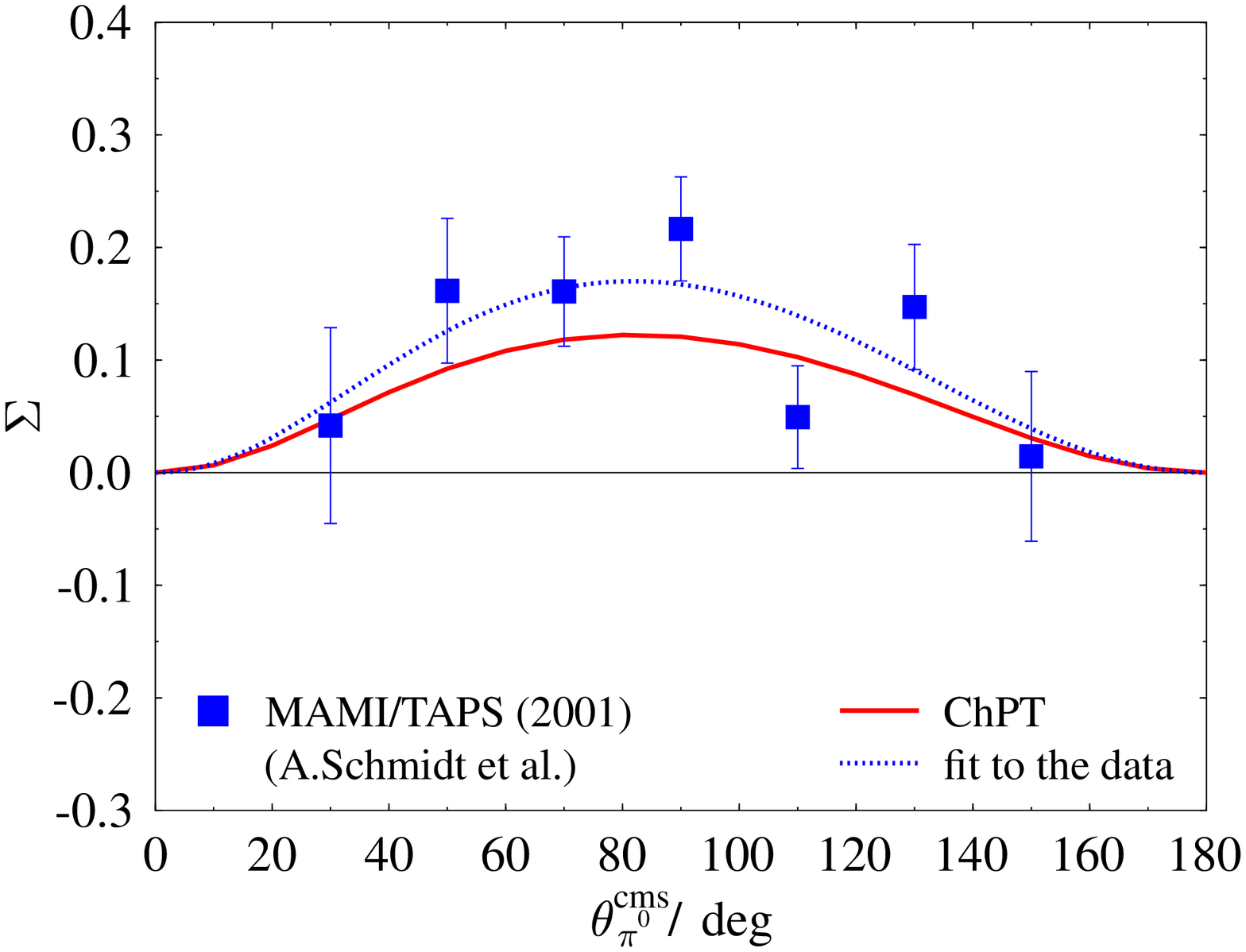}
\vskip -4.2cm
\hskip 8cm
\begin{tabular}{|l||c|c|c||}
    \hline
 & CHPT  ${\cal O}(p^3)$
 & MAMI/TAPS \cite{schmidt} & SAL \cite{berg} \\
    \hline   
$P_{1, {\rm thr}}$ &  $9.14 \pm 0.50$ & $9.47 \pm 0.08 \pm 0.29$ & $9.08 \pm
 0.14$    \\
$P_{2, {\rm thr}}$ & $-9.70 \pm 0.50$  & $-9.46 \pm 0.10 \pm 0.29$ & --  
\\
    \hline
  \end{tabular}
\vskip 2.7cm
\caption{The photon asymmetry at $E_\gamma = 159.5$~MeV.  
Solid line: CHPT prediction. The data are from \protect\cite{schmidt}.
Dashed line: Fit to the data. Figure  courtesy of  Reinhard Beck.
\label{fig:a}
}
\end{figure}

To disentangle the P-wave multipoles
$P_2$ and $P_3$, one has to perform an experiment involving
polarization. This has been achieved at  MAMI, where the photon
asymmetry $\Sigma (\theta)$ in $\vec{\gamma} p \to \pi^0
p$  has been measured at $E_\gamma = 159.5\,$MeV,
see Fig.\ref{fig:a} \cite{schmidt}. 
The analysis of these data together with the unpolarized ones leads to the
values  for the slopes of $P_1$ and $P_2$ at threshold given on the r.h.s. of 
this figure (in units  of $q \cdot 10^{-3}/ M_{\pi^+}^2$).
As can be seen the third order theoretical predictions compare
well with the MAMI data.
The fourth order corrections to these predictions have been analyzed
in detail in \cite{bblmvk}. One obtains for the sum of renormalized Born, 
third 
and fourth order loop and counterterm contributions
\bea
\frac{1}{|\vec q \,|} P_{1, {\rm thr}}^{\pi^0 p} 
&=&\,\,\,\, (0.460 + 0.017 - 0.133 + 0.0048\,\xi_1)~{\rm 
GeV}^{-2}~, \\
\frac{1}{|\vec q \,|} P_{2, {\rm thr}}^{\pi^0 p}
&=& -(0.449 + 0.058 - 0.109 + 0.0048\,\xi_2)~{\rm GeV}^{-2}~,
\eea
where the $\xi_{1,2}$ are LECs contributing to $P_1$ and $P_2$ at
fourth order,
respectively. In a resonance saturation picture, these LECs only 
depend on the $N\Delta$ transition magnetic moment $\kappa^*$ 
(for details see \cite{bblmvk}).
We note the rather sizeable (25\%) correction from the fourth order
loops which at first sight seems to destroy the agreement between the LETs
and the data. However, it is known that $\kappa^* \simeq 4\ldots 6$.
For $\kappa^* = 4$, the delta contribution
almost completely cancels the large fourth order loop effect and thus the
predictions for the P--wave slopes are within 7\% of the empirical
values, see Table~3 in ref.\cite{bblmvk}. 
Note, however, that the empirical finding $ P_1^{\rm
  exp}/ |\vec q \,| = - P_2^{\rm exp}/|\vec q \,|$ is difficult to reconcile with any theory.

\begin{figure}[hbt]
\begin{center}
\epsfysize=2.5in
\epsffile{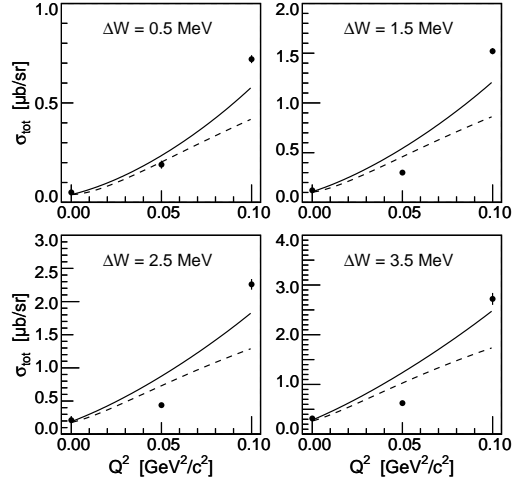}
\end{center}
\caption{
Total cross section of pion electroproduction off protons for a polarization
of the photon $\epsilon =
          0.8$. 
 The solid (dashed) line is the prediction of CHPT
          (MAID). The data points at $Q^2=0, \, 0.05$ and $0.1$~GeV$^2$ are from
          \protect\cite{schmidt}, \protect\cite{merkel} and \protect\cite{distler}. 
          \label{fig:a0}  }
\end{figure}

Producing the pion with virtual photons offers further insight
since one can extract the longitudinal S--wave multipole $L_{0+}$ and
also novel P--wave multipoles. Data have been taken at
NIKHEF~\cite{welch} \cite{benno} and MAMI~\cite{distler} for
photon virtuality of $k^2 = -0.1$~GeV$^2$.  CHPT calculations have been
performed in the relativistic framework \cite{blkm94} and then redone 
in the heavy fermion formalism \cite{bkmel}. The abovementioned data for the 
differential cross
sections, 
the only one available at the time the calculations were done, 
were used to determine the
three novel S--wave LECs. Note, however, that the photon four--momenta are 
already somewhat too 
large for
CHPT tests since the loop corrections are large \cite{bkmel}. Also an operator 
of dimension five, i.e. one
order higher than the calculation  done had to be taken into 
account  since it was shown that the two S--waves are
over-constrained by a LET valid up to order $p^4$. 
In \cite{bkmel}, 
many predictions for $k^2 \simeq -0.05$~GeV$^2$ were given. 
At MAMI, data have then been taken for this value of $k^2$ \cite{merkel},
 see also \cite{mer06}.
The measured
cross section is significantly lower than predicted by chiral
perturbation theory \cite{bkmel} or by the most sophisticated 
phenomenological model of the Mainz group \cite{MAID}. As an example,
 the total cross section is
shown  in  Fig.\ref{fig:a0}
as a function of $Q^2 = -k^2 \geq 0$ from the photoproduction point at
$Q^2 = 0$ up to the older measurement at $Q^2 = 0.1\,$GeV$^2$. In all
energy bins above threshold (with $\Delta$W the pion energy with
respect to the threshold value) one observes a ``kinky structure'', a
non-smooth momentum dependence at odds with expectations based on 
resonance excitation. This is even more strikingly observed in the
fitted value for the P-wave combination $|P_2|^2 + |P_3|^2$ given in
Table~II of ref.\cite{merkel}. It comes out 
significantly smaller than
what CHPT or models predict. This is rather surprising since $P_3$ is 
believed to be saturated by the $\Delta$ and $P_2$, as we have seen, is further 
constrained
by a LET, this combination should thus be understood 
to some precision. To resolve the problem of the $Q^2$ dependence of the
total cross-section
MAMI performed an experiment where three different $Q^2$ values were measured. 
Preliminary data seem to indicate that the  data points of the 
previous measurement at $Q^2=0.1$GeV$^2$ are somewhat high \cite{garll07}. 
Furthermore, the BigBite Collaboration at Jefferson
Lab plans to measure neutral pion electroproduction from low to intermediate 
$Q^2$ in small steps of $\Delta Q^2$, see \cite{bigbite}. On the 
theoretical side a refit will have to be done once data at lower $Q^2$ will be
available. Also up to now the P-waves have only been calculated to third order
beyond the photon point.
A fourth order calculation is  clearly needed. 

There exists an interesting sum rule, the Fubini-Furlan-Rosetti (FFR) 
derived in the sixties \cite{ffr65} which relates the nucleon anomalous 
magnetic 
moment to an integral over the invariant amplitude $A_1$ of pion 
photoproduction:
\be
\kappa^{V,S} =\frac {8 m^2}{e \pi g_{\pi N}} \int \frac{ d \nu'}{\nu'}
{\rm Im} \, A_1^{(+,0)} (\nu',t=0) \, \, ,
\label{ffrsr}
\ee
with $\kappa_V$ and $\kappa_S$ the isovector and isoscalar anomalous magnetic
moments and $A_1^{(+,0)}$ the respective combinations of $A_1$. 
It is exact in the chiral limit of QCD and thus all quantities in 
Eq.~(\ref{ffrsr})
are to be understood in the limit of vanishing quark masses. 
It allows to analyze the relation between dispersion relations 
\cite{pdt05} and the chiral
representation for neutral pion photoproduction. 
As pointed out in \cite{pdt05} this sum rule is a nice example
where use of HBCHPT is not appropriate since in this framework 
the nucleon pole positions  are slightly moved leading e.g. to an 
incorrect curvature of the discrepancy function (which measures the 
corrections to the sum-rule due to the finite pion mass) for energies below 
threshold. It has thus been considered in the framework of covariant chiral 
perturbation theory \cite{bkm05}. It was shown in that reference that one can
achieve a good description of the energy dependence of the discrepancy function
for the proton together with the one of the electric dipole amplitude at 
threshold and the P-wave slopes at threshold. Relations between the LECs and 
the subtraction constants of the dispersive analysis could thus be established
with an unprecedented accuracy \cite{bkm05,pdt06}. 

%
%
%
 
\subsubsection{off deuterium} \label{deut}

The question arises how to measure the neutron amplitude? 
This is of particular interest in order to understand isospin breaking 
violation.  
The natural neutron target is the deuteron. The electric
dipole $E_d$ amplitude  has been calculated to order $p^4$ 
in Ref.\cite{bkmpl}. 
It was shown that the next--to--leading order
three--body corrections and the possible four--fermion contact terms
do not induce any new unknown LEC and one therefore can calculate
$E_d$ in a parameter--free manner. Furthermore, the leading order
three--body terms (the well known charge exchange contribution) are dominant, 
but one finds a good convergence for
these corrections and also a sizeable sensitivity to the elementary
neutron amplitude.  The CHPT prediction in
comparison to the SAL data \cite{sald} for the reduced cross section of 
coherent
neutral pion production off deuterium is shown in the right panel of
Fig.\ref{fig:photo}. The predicted value for the deuteron electric dipole
amplitude $E_d = E_{0+}^{\pi^0 d}$ at threshold \cite{bkmpl}  is 
$E_d^{\rm thr}  = -1.8\pm0.2$ compared 
to the experimental result $-1.5 \pm 0.1$ \cite{sald}. It shows a rather 
strong dependence
in the values of the elementary amplitude $\pi^0 n$. 
Neutral pion electroproduction has been investigated in \cite{kbm02}.
As can be seen from the figures in this paper 
 the predicted 
differential 
cross sections are satisfactorily described although some systematic 
discrepancies for the higher values of the excess energy $\Delta W$  remain. 
The threshold multipoles $|E_d|$ and $|L_d|$
are consistent with the data.


\subsection {\it {Two-pion production}} \label{2piprod}

One of the nice successes of baryon CHPT is the prediction of the 
electromagnetic
two--pion production off the proton. I will briefly discuss it here. 
Due to space limitation I will not review here the reaction 
$\pi N \to \pi \pi N$. The interested reader can consult \cite{febeme00}
and references therein.

\begin{figure}[t]
\begin{center}
\psfig{file=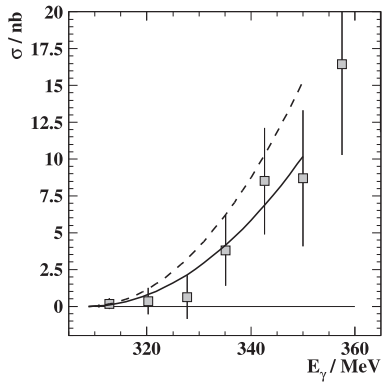,width=5cm}
\end{center}
\caption{Total cross section for the process $\gamma p \to \pi^0 \pi^0 p$
         as measured by the TAPS collaboration at MAMI \protect\cite{wolf00}.
         The solid line is the fourth order central HBCHPT prediction, 0.6nb
in Eq.~(\ref{eq:2pi0})
and the dashed line the upper limit (0.9nb). Figure taken 
from \cite{kotal}.         
          \label{fig:tpi0} }
\end{figure}

Electromagnetic production of two pions off a nucleon can be used to
study the excitation of certain nucleon resonances, in particular the
$\Delta (1232)$, the Roper $N^* (1440)$ or the $D_{15} (1520)$. However,
close to threshold one observes an interesting effect due to the chiral pion
loops of QCD. To be specific, consider the reaction $\gamma p \to \pi\pi N$,
where the two pions in the final state can both be charged, both neutral or
one charged and one neutral. To leading order in the chiral expansion, the
production of two neutral pions is strictly suppressed. However, at
next--to--leading order, due to finite chiral loops the production cross
section for final states with two neutral pions is considerably
enhanced \cite{bekames94}.  Also, in a small window above threshold, the
potentially large contribution from double--delta excitation is strongly
suppressed, leaving a window in which one can detect much more neutrals than
expected. This prediction was further sharpened in \cite{bekame96},
where all fourth order corrections including the excitation of the Roper and its
successive decay into two neutral pions were considered. The predicted near
threshold cross section is
\be
\sigma_{\rm tot} (E_\gamma ) \leq 0.91~{\rm nb}~\biggl({ E_\gamma -
  E_\gamma^{\rm thr} \over 10~{\rm MeV}} \biggr)^2~,
\label{eq:2pi0}
\ee
with $E_\gamma^{\rm thr} = 308.8\,$MeV the threshold energy for
$\gamma p \to \pi^0 \pi^0 p$ (in the lab system). 
This prediction can only be applied
for the first 30 MeV above threshold. A measurement by the TAPS 
collaboration~\cite{kotal} has shown that such an enhancement of the
the 2$\pi^0$ production indeed happens, see Fig.\ref{fig:tpi0}, proving once
again the importance of pion loop effects, which can lead to rather unexpected 
predictions and results. New data with much higher precision measured with 
the Crystal Ball and TAPS at MAMI will soon be published \cite{kotze07}.

\vskip -0.3cm
\section{Axial properties} \label{axial}
I will be very brief here and only report on very recent determinations  of
the induced pseudoscalar coupling constant $g_P$. For more details on axial 
properties see \cite{bem02}. 
A very precise prediction of the induced pseudoscalar form factor
 based on chiral Ward identities has been given in \cite{bkma92}:
\be
G_P(t) = \frac{4 m g_{\pi N} F_\pi} {M_\pi^2 -t} -\frac{2}{3} g_A m^2 r_A^2
\label{eq:gp}
\ee
leading to
\be 
g_P=(M_\mu/2m) G_P(t=-0.88 M_\mu^2)=8.26 \pm 0.23.
\label{eq:gp0}
\ee
\begin{figure}[t]
\epsfysize=18.0cm
\begin{center}
\vskip -0.3cm
\includegraphics*[width=5cm,angle=90]{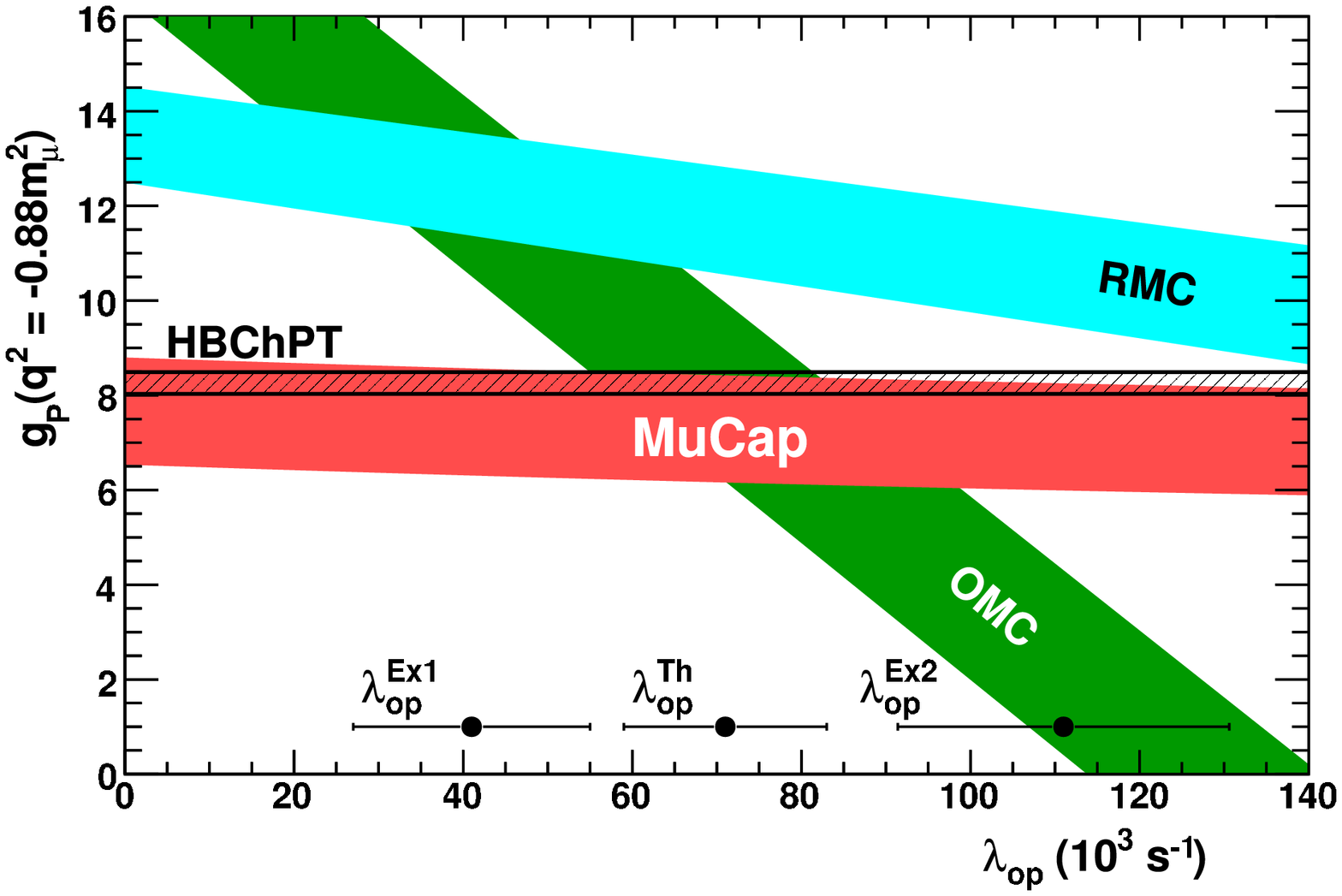}
\end{center}
\caption{Comparison of a HBCHPT (hatched line) calculation of $g_P$ 
with experimental measurements as a function of the ortho-para transition
rate $\lambda_{op}$ in the $p \mu p$ molecule. Figure courtesy of C. Petitjean
for the MuCap collaboration. 
 \label{fig:gp}}
\end{figure}
\noindent In  Eq.~(\ref{eq:gp}) $M_\mu$ is the muon mass and
$r_A$  the axial radius which can be determined from 
charged pion electroproduction or neutrino-proton scattering. Note
that the difference between the two extracted  experimental values of $r_A$ 
can be understood
by a computable and unique loop correction within CHPT \cite{bkmb92}.
The effect on Eq.~(\ref{eq:gp0})
of the three-pion continuum determined in a two-loop HBCHPT
calculation  is negligibly small \cite{ka03}.
The value, Eq.~(\ref{eq:gp0}) where the error is almost entirely due to 
the uncertainty on the 
value of the pion-nucleon coupling constant, agreed with 
the much less precise experimental 
determination extracted from ordinary muon capture (OMC) \cite{ba81}. 
This was however challenged by the TRIUMF result from radiative muon capture:
$g_P=12.35 \pm 0.88 \pm 0.38$
\cite{jo96}.  
This has lead to some sizeable theoretical activity 
\cite{flms97,mmk98,am98,bhm00,amk02} and it
was conjectured that in fact a sum of various small effects could make 
up for the  discrepancy. 
However, both the RMC and OMC experimental  determination of $g_P$
depend on the ortho-para transition rate $\lambda_{op}$ in the 
$p \mu p$ molecule as shown on Fig.\ref{fig:gp}. Its value is poorly 
known due to mutually inconsistent experimental and theoretical results.  
Very recently, a new experimental result on $g_P$ has been reported
by the MuCap collaboration 
\cite{mucap}. It is also
shown on the figure. It is nearly independent of molecular effects and  is 
nicely consistent with the CHPT prediction. The experimental MuCap
result $g_P=7.3 \pm 1.1$, takes into account 
recently calculated radiative corrections 
\cite{cms07}.        

\section{Three flavors} \label{3flav}

\subsection{\it {Chiral dynamics with strange
 quarks: General remarks}} \label{cdy}

I will now turn to the case of 3 flavors. 
As we have briefly seen in the introduction matters are more delicate
in SU(3) since the strange quark plays a special role. Indeed its mass
is of the order of 100~MeV. The question is whether one can consider
it as light compared to the QCD scale $\Lambda_{QCD} \sim 200$~MeV or should 
it be
treated as heavy. Related to that is the question 
whether standard CHPT is a well 
converging series, the relevant expansion parameter being in that
case $m_K/\Lambda_\chi \sim 0.4$ and whether  $\bar s s$ sea quark 
pairs may play a
significant role in chiral dynamics leading to different patterns
of chiral symmetry breaking in the $N_f=2$ and $N_f=3$ chiral limits.
These are all very interesting questions which have been mostly 
touched upon in the meson sector and which are still rather open
questions. 
  
\begin{figure}[htb]
\centerline{
\epsfysize=2.in
\epsffile{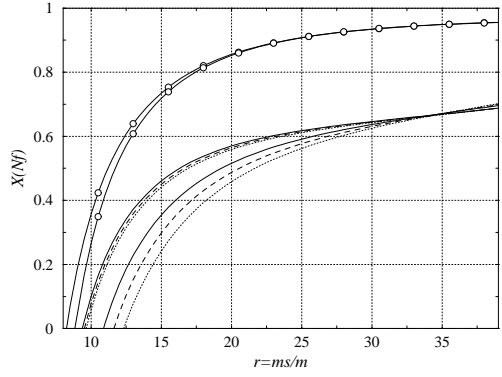}
}
\vskip 0.1cm
\caption{Upper and lower bounds for $X(3)$ as a functions of $r=m_s/m$ for 
$F=85$ MeV with the T-matrix model of Ref.\cite{oop99}. The allowed 
range for X(2) is also shown (line with an open circle). Figure taken from
\cite{ds00}
\label{fig:x(3)} }
\end{figure}

Let me briefly summarize what is
known on the dependence of the chiral order parameters on $N_f$.
The order parameters $F^2$, the decay constant, and $\Sigma$,
the quark condensate, 
can be obtained in terms of eigenvalues of the Euclidean Dirac 
operator $\gamma_\mu D_\mu$ defined in a box $L \times L \times L \times L$
with periodic boundary conditions via the well-known Banks-Casher  
formula \cite{bc80}. Being dominated by 
the IR end of the Dirac spectrum one expects a paramagnetic 
effect such that 
\be
\Sigma(N_{f}) < \Sigma(N_f-1) \sim 1/L^4 \, , \quad F^2(N_{f}) <  F^2(N_f-1)
\sim 1/L^2 \, \, .
\ee
Let us concentrate on $N_f=2,3$ and define what is exactly meant by 
$\Sigma(3)$ and $\Sigma(2)$. One considers $N_f$ quarks as massless,
keeping the remaining masses at their physical mass so that: 
\be
\Sigma(3)=\lim_{m_s \to 0} \Sigma(2,m_s) \,\, ,
\ee
where, in the limit $N_f=2$, $\Sigma$ depends on the physical mass $m_s$ 
(I don't consider here the heavier quarks).  
It can be shown that $\Sigma(3)$ and $\Sigma(2)$ are related via:
\be
\Sigma(3)=\Sigma(2) + m_s Z^s + {\cal O}(m_s^2 \log m_s^2),  
\ee
with $ Z^s$ related to the correlator $\Pi(0)$
\be 
m m_s \partial \Sigma(2) / \partial m_s =M_\pi^2 M_K^2 \Pi(0) \, \, , \quad 
\Pi(0)=i \frac{m m_s}{M_\pi^2 M_K^2} \lim_{m \to 0}
\int dx \langle T \{\bar u u(x) \bar s s(0) \} \rangle_{\rm {conn}} \, \, .
\ee
$\Pi(0)$ measures the violation of the OZI rule in the isoscalar-scalar 
(vacuum) channel.  In the limit of large $N_c$  the OZI rule is exact 
thus  the correlator $Z^s$ vanishes and the difference between $\Sigma(3)$
and $\Sigma(2)$ cancels. This is the standard scenario.
The real world seems, however, to be different from the large $N_c$ 
scenario in the scalar sector. For example the  scalar meson $f_0 (980)$
is found to be rather light and narrow and it couples strongly to both
$K \bar K$ and $\pi \pi$ channels in violation to the large $N_c$
expectation.  
At present one does not really know the size of the violation. 
Investigating a superconvergent sum rule for the OZI violating correlator
$\Pi$ with the strong constraint that its imaginary part satisfies a 
Weinberg type sum rule, 
Moussallam \cite{mou00} gave an estimate $2 \lesssim 16 \pi^2 \Pi \lesssim 6$. 
The 
$f_0 (980)$ contributes in an important way to these numbers.   
Within CHPT this OZI violation 
is encoded in certain LECs whose values will thus depend on its strength.
 These are the SU(3) ${\cal O}(p^4)$ meson LECs $L_6, L_8$ which 
enter the meson masses and $L_4$ which merely shows up in the decay 
constants.
For example the previous estimate for $ \Pi$ leads to $ 
0.4 \lesssim  10^{3} L_6(M_\eta) \lesssim 0.8$ to be compared with the 
standard scenario value $L_6(M_\eta)=(0.0 \pm 0.3)
10^{-3}$. This 
translates into
\be
\Sigma(3)=\Sigma(2) [ 1 -0.54 \pm 0.27 ] \,\, ,
\ee
where the central value indicates that the condensate decreases by a  factor 
of two when one decreases the strange mass from its physical value   
down to zero  but 
the uncertainties are large enough to give marginal consistency with the 
standard scenario. The correlator $\Pi$ has been recalculated on more
general grounds in \cite{ds00}.  
It is particularly interesting to look at the
$m_s$ dependence of $\Sigma(N_f)$ obtained in this analysis. 
This is shown on Fig.~\ref{fig:x(3)}
where the so-called Gell-Mann-Oakes-Renner ratio 
\be
X(N_f)=\frac {2 \hat m \Sigma (N_f)}{F_\pi^2 M_\pi^2}
\ee
is displayed as a function of $m_s/m$. For more details see ref.~\cite{ds00}. 

More work on the theory side, for example
three dynamical flavors lattice  calculations with 
quark masses as close as possible to their physical values, as well as more 
experimental informations are
needed in order to precisely determine the condensates. 

In the baryon sector the SU(3) case
has been much less studied. Here I will first
present the status of calculations done in standard CHPT. I will then 
discuss shortly proposal to modify the expansion considering  
the perturbative expansion as not justified.   
There are also interesting questions  
like for example what is the 
structure of some strange states like the $\Lambda(1405)$
or the $S_{11}(1535)$ 
where a unitary extension of CHPT
has to be used. 
This is a nice tool which has been developped
by \cite{kswa95,ksw95,or98,om01,lk02,room03}.  
I won't have space to discuss all this here,
I refer the interested reader 
to the early review \cite{oor00}.

Before going to the applications let me briefly sketch the straightforward 
modifications which have to be done when
going from $SU(2)$ to $SU(3)$.
The effective meson-baryon Lagrangian is now expressed in terms of the baryon
octet 
\begin{eqnarray}
B  =  \left(
\matrix  { {1\over \sqrt 2} \Sigma^0 + {1 \over \sqrt 6} \Lambda
&\Sigma^+ &  p \nonumber \\
\Sigma^-
    & -{1\over \sqrt 2} \Sigma^0 + {1 \over \sqrt 6} \Lambda & n
    \nonumber \\
\Xi^-
        &       \Xi^0 &- {2 \over \sqrt 6} \Lambda \nonumber \\} 
\!\!\!\!\!\!\!\!\!\!\!\!\!\!\!\!\! \right)  \, \, \, ,
\end{eqnarray}
and pseudoscalar Goldstone fields $(\phi =\pi, K, \eta)$ collected in 
a now $3 \times 3$ unimodular, unitary matrix $U(\phi)$,
\be
U(\phi) =u^2(\phi)=\exp \{ i \phi /F\}  \,\, .
\ee
In the case of isospin breaking the neutral pion and the $\eta$ mix,
so that the physical fields are related to the pure SU(3) components
via a mixing angle $\varepsilon$ with
\be
\tan 2 \varepsilon = \frac{\sqrt 3}{2R} 
\,\,,\,\,\,\,\,\,\,\,\, R=\frac {m_s-\hat m}{m_d-m_u}
\label{eq:mixang}
\ee
with $ \hat m =(m_u+m_d)/2$. The value of $R$
has been determined in \cite{leut96}, leading to the usually quoted result 
$R=40.8 \pm 3.2$. Note that in the determination of $R$ enters a quantity 
denoted $Q^2$ which is taken in \cite{leut96} to be equal to
$22.7 \pm 0.8$ using the difference between the charged and neutral kaon
 masses and the partial width  $
\eta \to \pi^+ \pi^- \pi^0$. However a smaller value of this 
quantity was obtained in \cite{bij}
leading to smaller values of $R$.
The value of $R$ has never been tested experimentally. In fact $R$ could be 
obtained by looking at the ratio of charged and neutral $K_{l3}$ decay, see 
\cite{fras07}. 
The complete and minimal chiral effective meson-baryon
Lagrangian at third order can be found 
in \cite{fm06}, see also \cite{ka90}.
As compared to the $SU(2)$ case one has more fields and more operator 
structures
in the EFT and consequently more LECs.  For example 
the dimension one Lagrangian reads:
\be
{\cal L}_{MB}^{(1)}  =  \langle\bar B \, [ \, i \nabla\!\!\!\!/ \, ,
\, B] \,\rangle \,
- \, \tilde m_0 \langle \bar B  \, B \rangle 
+ {D \over 2} \, \langle\bar B \, \{u\!\!\!/ \gamma_5 , B \}\, \rangle
+ {F \over 2} \, \langle \bar B \, [ u\!\!\!/ \gamma_5 , B ] \,
\rangle  \,\,\, ,
\label{LMB1}
\ee
with $\tilde m_0$ the average octet mass in the chiral limit and   
$D \simeq 0.81$ and $F \simeq 0.46$, two 
axial-vector coupling constants which can be determined from hyperon beta
decays.  
The leading symmetry breakers within $SU(2)$ and $SU(3)$ are:
\be
c_1  \bar \psi_N  \langle \chi_+ \rangle \psi_N  \to b_0 \langle \bar B B \rangle +b_D  \langle \bar B \{ \chi_+,
B \} \rangle
+b_F \langle \bar B [ \chi_+,
B ] \rangle \, \, .
\ee 
A certain amount of LECs could be determined using lattice data. Also,
the various operators in $SU(2)$ and $SU(3)$ are related by matching
conditions which are important constraints that should be implemented in 
any SU(3) analysis. The full matching conditions to fourth order in the
chiral expansion have been derived in \cite{fm04}. To leading order,
one has for example:
\be
\tilde m_0 =m_0[1+{\cal O}(m_s)], \quad g_A= D+F +{\cal O}(m_s), \quad 
c_1 =b_0+ \frac{1}{2} (b_D+b_F) + {\cal O}(\sqrt m_s)    \, \, .
\ee

\subsection{\it {Baryon masses and Chiral Extrapolation}} \label{mb} 

\vskip 0.25cm
\begin{figure}[t]
\epsfysize=8cm
\begin{center}
\begin{minipage}[t]{7cm}
\epsfig{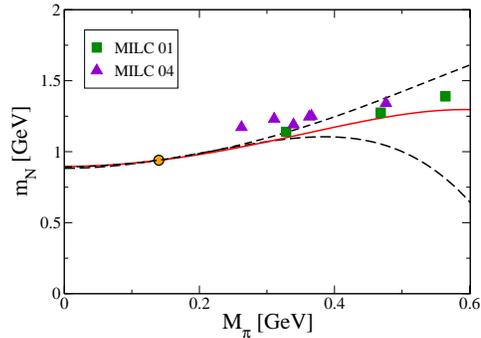}
\end{minipage}
\begin{minipage}[t]{18cm}
\caption{SU(3) analysis of the nucleon mass. The solid line gives the 
best fit and the dashed line the theoretical uncertainty under
the constraint that the nucleon mass takes its physical value for the
physical quark masses. The square/triangle are the MILC 2001 \cite{ber01} /
 2004 \cite{aamilc04} data.  \label{fig:MN3}}
\end{minipage}
\end{center}
\end{figure}

Explicit expressions for the baryon masses in terms of the quark masses
to one loop order with isospin breaking
can be found in \cite{fm04}. They are very lengthy 
and contain a large number of LECs. For illustration I will just give the
expression up to third order $\sim m_q^{3/2}$:
\begin{eqnarray}
m_B &=& \tilde m_0 + \left(\gamma_{1,B} + \tilde{\gamma}_{1,B} \frac{\varepsilon}{\sqrt{3}} \right) \, B\, \hat m 
          + \left(\gamma_{2,B} + \tilde{\gamma}_{2,B} \frac{\varepsilon}{\sqrt{3}} \right) \, B\, m_s  \nonumber \\
& & +\left(\delta_{1,B} + \tilde{\delta}_{1,B} \frac{\varepsilon}{\sqrt{3}} \right) 
              \, \frac{\sqrt{2}\, B^{3/2}}{4\pi F_\pi^2} \, \hat m^{3/2}
            +  \delta_{2,B} \, \frac{B^{3/2}}{4\pi F_\pi^2} \, (\hat m+m_s)^{3/2} \\
          &+& \left(\delta_{3,B} + \tilde{\delta}_{3,B} \frac{\varepsilon}{\sqrt{3}} \right) 
              \, \frac{\sqrt{2}\, B^{3/2}}{4 \sqrt{3}\pi F_\pi^2} \, (\hat m + 2m_s)^{3/2}
           +  \tilde{\delta}_{4,B}\, \frac{\varepsilon}{\sqrt{3}}
              \, \frac{B^{3/2}}{4\pi F_\pi^2} \, (\hat m-m_s)(\hat m+m_s)^{1/2}~, \nonumber 
\end{eqnarray}
where $\gamma_{I,B}$ are functions of the symmetry breakers 
$b_0, b_D$ and $b_F$, $\delta_{I,B}$ are combinations of 
$F$ and $D$ and $\varepsilon$ is the mixing angle defined in 
Eq.~(\ref{eq:mixang}). 
This expression is clearly of the same form as its SU(2)
counterpart, Eq.~(\ref{eq:mextra}) in the isospin limit.

With these expressions fits to the MILC 2001 data have been performed  
\cite{fms05} based on two different regularization schemes (cut-off and 
dimensional regularization) and with the constraint that the 
physical value of the nucleon mass is obtained for the physical quark mass. 
In these fits the dimension two LECs where taken from \cite{bm97} and 
the dimension four LECs were determined from a best description of the 
lattice data,
see the solid line on Fig.~\ref{fig:MN3}. It turns out that the later MILC 
2004 data at lower quark masses are too high compared to this best fit due to
the constraint at the physical value.  Taking into account
also the kaon mass dependence of the nucleon mass and the uncertainty
due to the MILC 2004 data, the following ranges for various
(isoscalar) quantities have been obtained:
\be 
\quad 710~{\rm MeV} \lesssim
  \tilde{m_0} \lesssim 1070~{\rm MeV}~,  \quad 39.5~{\rm MeV} 
\lesssim \sigma(0) \lesssim  46.7~{\rm MeV}~,
\quad 0.07 \lesssim y \lesssim 0.22~,
\ee
where the $\pi N$ $\sigma$ term, $\sigma(0)$, and the strangeness fraction of 
the proton, $y$, have been defined in Section \ref{nucm}.

\begin{figure}[htb]
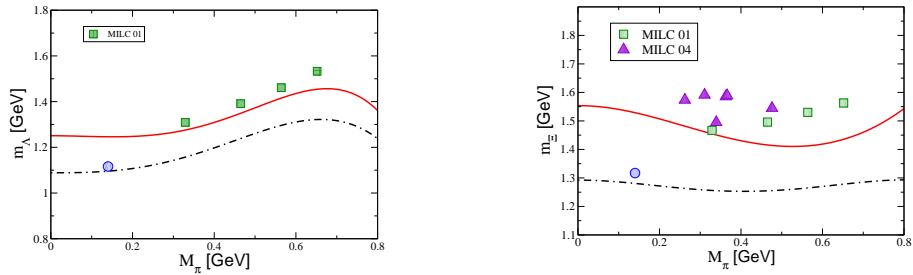

\epsfysize=9.0cm
\begin{minipage}[t]{6 cm}
\hskip 3cm
\includegraphics*[width=5cm]{mL4u.eps}
\vskip -3.6cm
\hskip 10cm
\includegraphics*[width=5cm]{mX4u.eps}
\vskip -5.9cm
\end{minipage}
\caption{Pion mass dependance of the $\Lambda$ mass (left panel) and the 
$\Xi$ mass (right panel). 
 \label{fig:MXL}}
\end{figure}
Once the nucleon has been fitted one can predict the masses of the other octet
members. While the $\Sigma$ mass is well reproduced the $\Lambda$ and $\Xi$
masses come about $10-15 \%$ too high.  Predictions for the pion mass 
dependence
of the  $\Lambda$ are shown in Fig.\ref{fig:MXL} in comparison to the MILC 
data, see the upper solid line. For comparison the  lower dot-dashed line is 
the 
result when the dimension four LECs are taken from  \cite{bm97}, where in 
that case the masses at the physical pion mass 
are well reproduced. 
Interestingly the $\Xi$ pion mass dependence is much flatter that what one 
would expect from the MILC data. One in fact expects the $\Xi$ to be less 
sensitive to variations in the pion mass since it contains only one valence 
light quark and the MILC result is a mystery to be resolved.

\subsection{\it{Baryon Electromagnetic Properties }} \label{elecb}

I will restrict myself here due to lack of space to the discussion of the 
nucleon form factors.
Calculations of the magnetic moments of the whole octet can be found in
\cite{mest97}.
There are three independent diagonal vector currents in SU(3):
\be
J_\mu^{(i)} =\bar q \frac{\lambda^i}{2} \gamma_\mu q \, \quad i=3,8,0
\label{eq:jmu}
\ee
which are proportional to the isovector and isoscalar electromagnetic
currents and the baryon number current, respectively. In Eq.~(\ref{eq:jmu}),
$\lambda^{3,8}$ are the usual Gell-Mann matrices and  $\lambda^{0}=
\sqrt{2/3}$ I.  
The electromagnetic $J_\mu^{{\rm EM}}$ and the strangeness $J_\mu^s$ currents are linear combinations
of these. I will discuss the first current here and the second in the following
section. One has: 
\be
J_\mu^{{\rm EM}}= J_\mu^{(3)} +\frac {1} {\sqrt{3}}  J_\mu^{(8)} \, \, .
\ee
Extension of the SU(2) calculations of the em form factors presented in
Section \ref{ff} has been done in \cite{km01}. I will mainly discuss here the 
difference between the HBCHPT and the IR results specially for the convergence 
of the
chiral series in the case of the electric radii $\langle r_E^2 \rangle$ 
as shown on Table 
\ref{tab:elrad}. For more details on the 
$Q^2$ dependence of the form factors and the magnetic radii, 
see this reference.
The table shows the predictions for  $\langle r_E^2 \rangle$ for the third
and fourth order in both formalisms. The IR scheme yields overall considerable
improvement over the HBCHPT scheme, especially for $\Lambda$, $\Sigma^+$, and 
\begin{table}[t] 
\begin{center}
\begin{minipage}[htb]{18. cm} 
\caption{
Predictions for the electric radii $\langle r_E^2 \rangle$ [fm$^2$].
The errors for the relativistic fourth order predictions display
the uncertainty due to two order 2 LECs taken from the best fit to the
magnetic moments at fourth order.
The errors for the experimental $\Sigma^-$ radius
values refer to statistical (first) and systematic (second) errors.}
\label{tab:elrad}
\end{minipage}
\vskip 0.4truecm
\begin{tabular}{|c||c|c||c|c||c|} 
\hline
\multicolumn{1}{|c||}{}&
\multicolumn{2}{|c||}{HB}
 & \multicolumn{2}{|c||}{IR} &  \multicolumn{1}{|c|}{}\\
\hline
& ${\cal O}(p^3)$ & ${\cal O}(p^4)$ &
  ${\cal O}(p^3)$ & ${\cal O}(p^4)$ & exp.\\ 
\hline 
$\Sigma^+$        &    0.59  &    0.72  &    0.63  &    0.60$\pm$0.02 & --- \\
$\Sigma^0$        & $-$0.14  & $-$0.08  & $-$0.05  & $-$0.03$\pm$0.01 & --- \\
$\Sigma^-$        &    0.87  &    0.88  &    0.72  &    0.67$\pm$0.03 & 0.61$\pm$0.12$\pm$0.09 \cite{selex}\\
$\Xi^0$           &    0.36  &    0.08  &    0.15  &    0.13$\pm$0.03 & --- \\
$\Xi^-$           &    0.67  &    0.75  &    0.56  &    0.49$\pm$0.05 & --- \\
\hline
\end{tabular} 
\end{center} 
\end{table}
\noindent $\Xi^0$. Also it shows the hierarchy in the size of the electric radii expected
from naive quark model considerations,  $\langle (r_E^p)^2 \rangle >
\langle (r_E^{\Sigma^{\pm}})^2 \rangle > \langle (r_E^{\Xi^-})^2 \rangle$. 
Furthermore the experimental $\Sigma^-$ radius given by the SELEX 
collaboration 
\cite{selex} is in very good agreement with the calculated IR value.

\subsection{\it{Strange form factors of the nucleon}}  \label{strange}

One particularly interesting question which we have already touched upon
in the context of the $\pi N$ $\sigma$ term and which is connected to the
violation of the OZI rule is: what is the  strangeness content in the nucleon?
There has been a lot of work done on this subject both from theory 
and experiment. The question of the spin content of the proton for example
has been widely studied. 
Also it was pointed out a long time ago \cite{gh76}
that the dispersion theoretical analysis of the nucleon's electromagnetic
form factors allows one to get bounds on the violation of the OZI rule.
It was found there that the ratio of the vector coupling constants to the
nucleon of the $\phi$ and of the $\omega$ which is almost 
entirely
due to the coupling of $\bar s s$ to the nucleon is fairly large $\sim -0.4$
indicating a strong violation of the Zweig's rule. 
Interesting quantities in this context are
the strange form-factors. 
Jaffe showed \cite{ja89}, see also \cite{hmd96}, that under certain assumptions the information 
encoded in the 
isoscalar nucleon form factors could be used to extract strange matrix 
elements and that the strange form factor $F_1^{(s)}$ thus obtained was rather 
large in magnitude due to the
strong coupling of the $\phi$ to the nucleon in dispersion theoretical
analysis. However a realistic isoscalar spectral function including
the $\rho \pi$ exchange led to sizeably reduced values for the strange
magnetic moment and radius \cite{mmsv97}.   
To determine the strange form factor experimentally one measures parity 
violating (PV)  
asymmetries in PV
electron scattering from nucleons and nuclei.
There is a whole experimental program dedicated to such measurements
with on going 
experiments from the 
G0 \cite{g0}, HAPPEX \cite{ha} (for the nucleon) and \cite{hanuc}
(for $^4He$), Mainz A4 \cite{a4} and SAMPLE \cite{sa} 
collaborations with, at present, no strong evidence for the presence of
strange-quark effects in the nucleon. $^4He$  is particularly interesting 
since due to the spin-parity of this nucleus it can lead to a direct 
determination of the strangeness electric form factor provided that other
effects like isospin symmetry breaking or meson exchange
contributions to the charge operators are negligible.
Recently values for the strange magnetic moment $\mu_s$ and 
the strange electric radius $\langle r^2_{E,s} \rangle$ have been 
extracted from a complete world set of the data obtained by these
collaborations up to $Q^2 \sim 0.3$ GeV$^2$
leading to \cite{yrct06}
\be
\langle r^2_{E,s} \rangle = (0.014 \pm 0.096 \mp 0.00)  \, {\rm {fm}}^{2} \,\, ,
\quad \quad  \mu_s =0.12 \pm 0.55 \pm 0.07 \, . \label{dstrangeff}
\ee

Let us look in more detail at these strange form factors. 
They are defined in terms of the strangeness current:
\be
J_\mu^s=\sqrt {\frac{2}{3}} J_\mu^{(0)} -\frac {2} {\sqrt{3}}  J_\mu^{(8)}
\, \, ,\label{scurrent}
\ee
in complete 
analogy to what is done in the electromagnetic case.
They have been calculated in lattice simulations  as well as 
hadron models. 
I will concentrate here on CHPT. 
As can be seen the baryon number current enter Eq.~(\ref{scurrent}). 
Thus one expects
that LECs will contribute which do not appear in any 
other known processes. It was thus claimed \cite{ri97} that CHPT would not
be able to say anything about these form factors. However with the advent 
of experimental
results which could be used to fix some of these LECs, one hoped 
to be able to  predict for example the anomalous magnetic moment. At present
an  ${\cal {O}}(q^4)$ calculation of the radius and the magnetic moment
has been done \cite{hprz03} in HBCHPT (note that here the definition of the
radius does not involve the normalization factor as it is
usually the case for the em radii). It leads to 
\bea
\mu_s&=& 1.2 + 2.5(b_s(\lambda=1GeV) +0.6 b_8 (\lambda=1GeV)) \, \, ,\nonumber \\
\langle r^2_{M,s} \rangle &=& -0.16 + [ 0.12  +0.3 b_s^r(\lambda=1GeV)] 
 \, {\rm {fm}}^2  \, \, .
\label{strangff}
\eea
In this formula $b_s$, $b_s^r$ and $b_8$ are undetermined singlet LECs.
The contributions $\sim 1.2$ and $\sim 0.12$ arise from loop graphs and 
LECs which were obtained from
measured octet baryon magnetic moments \cite{pr00}.
The first number on the r.h.s. of the expression for 
$\langle r^2_{M,s} \rangle$
corresponds to the ${\cal {O}}(q^3)$ result. It has been first obtained
in \cite{hms98} and shown to  
satisfies a Low Energy Theorem: 
\be 
\langle r^2_{M,s} \rangle =-\frac{5 D^2 -6DF +9F^2} {48 \pi F_K^2}
\frac{m}{M_K} + {\cal {O}}(M_K^0) \, \, . 
\ee
Using this ${\cal {O}}(q^3)$  parameter free prediction together with an 
extrapolation of 
the SAMPLE result at $Q^2=0.1$ GeV$^2$ down to zero $Q^2$ a determination of
the strange magnetic moment 
has been made. Unfortunately as can be seen from Eq.~(\ref{strangff}) there is 
a large 
cancellation between the ${\cal {O}}(q^3)$ and some of the  ${\cal {O}}(q^4)$
terms making the strange radius very sensitive to the unkwown  LEC $b_1^r$.
Thus neither the magnitude nor the size of the magnetic radius can be
determined in a model independent manner. The value of $b_1^r$ and $ 
b_s(\lambda=1GeV) +0.6 b_8 (\lambda=1GeV)$ have been 
obtained in \cite{hprz03} by a matching with dispersion relations and found 
to be $-1.1$ and $-0.6$ respectively in nice agreement with the expectation 
from dimensional analysis.  
This leads to $\langle r^2_{M,s} \rangle \sim 0.34 \, {\rm {fm}}^2$ 
and $\mu_s=-0.36$. This last quantity is within the error bars of the best 
fit Eq.~(\ref{dstrangeff}).
A quenched lattice calculation of
$G_M^{(s)}(q^2)$ at five different kinematic points \cite{dlw98} 
is in agreement 
with the value of  $\mu_s$
but in disagreement with the one of $ \langle r^2_{M,s} \rangle$. 

While the SAMPLE collaboration measures the strange magnetic form factor,
the other collaborations which have chosen different kinematics measure
a combination of the strange electric and magnetic form factor. Combining
the SAMPLE and the HAPPEX results in order to fix two unknown LECs the 
$Q^2$ dependence of the strange electric form factor has been determined  to 
${\cal {O}}(q^3)$
\cite{hbm99}, leading to a strange electric radius fairly small and 
positive:
\be
\langle r^2_{E,s} \rangle = (0.05 \pm 0.09)  \, {\rm {fm}}^2 
\ee
to be compared with Eq.~(\ref{dstrangeff}).
Few things have to be stressed with respect to this number. First
this calculation uses the strange magnetic moment to 
$O(q^3)$. Its central value is found to be $\mu_s = 0.18$ not in agreement
with the central value discussed above. Second the experimental values
have been updated since the calculation has been done. 
Third, a $O(q^4)$ calculation should of course be performed to evaluate 
the $1/m$ corrections which can, in some cases, be non negligible. 

Up to now the experimental results concerning the strange form factors 
seem to give rather small values. It was pointed out \cite{kl06} that 
the inequality
of up and down quarks produces in fact effects that mimic the strange
form factors and thus could be of importance for a precise determination 
of these form factors. Isospin violation in the vector form factors of the
nucleon have thus been recently calculated in two flavor CHPT with extraction 
of some LECs from resonance saturation \cite{kl06}. Some upper limits have
been obtained $G_M^{u,d}(t) < 0.05$ and $ G_E^{u,d}(t) < 0.01$, for
more details see that reference. As we have previously said, the knowledge of 
isospin symmetry breaking 
effects are also of utmost importance for the measurements on $^4 He$.
They have been determined in a very recent paper \cite{vsklgkmr07} 
and found to be of comparable magnitude to those associated with strangeness 
components in the nucleon electric form factor at the low momentum transfers
of interest.    

More low-$t$ data on the nucleon's strange vector form factors will clearly be
extremely useful in getting a better understanding  of these quantities.

\subsection{\it{Kaon-Nucleon scattering}} \label{knscatt}

Since the review \cite{BKM95r} has been written not much work has been 
done on the subject of kaon-nucleon scattering
in pure CHPT. In fact there is a big difference between the 
strangeness S=1 and the S=-1 
channels. While the first one is purely elastic at low energies, the second
one involves inelastic channels. Especially the isospin 0 
$\bar K N$ channel is completely dominated by the nearby subthreshold 
$\Lambda$(1405) resonance. It thus seems difficult to explain this 
channel in pure CHPT.

I will briefly report
on a calculation by Kaiser \cite{ka01} of the threshold 
T-matrix of kaon-nucleon and antikaon-nucleon scattering to one loop 
order in HBCHPT. The following empirical values have been obtained
\cite{ma81} from a combined dispersion relation analysis of $\bar K N$ and 
$K N$ scattering data. 
\bea
&T_{K N}^{(0)} =0.4 \, {\rm {fm}} \, , \quad \quad \quad \quad \quad \quad
 \quad \quad \quad \quad \, \,
&T_{K N}^{(1)} =-6.3 \, {\rm {fm}} \, \, ,
\\ \nonumber
&T_{\bar K N}^{(0)} =(-32.6 + 13.0 i) \, {\rm {fm}} \, , \quad \quad \quad
\quad \quad
&T_{\bar K N}^{(1)} =(7.1 + 11.5 i) \, {\rm {fm}} \, \, .
\eea
Using the $K N$ empirical values to fix the two LECs entering the calculation,
Kaiser predicted the $\bar K N$ threshold T-matrices:
\be
T_{\bar K N}^{(0)} =(30.4 + 6.2 i) \, {\rm {fm}} \, , \quad \quad \quad
\quad \quad \quad  \quad  \, \,
T_{\bar K N}^{(1)} =(7.1 + 10.4 i) \, {\rm {fm}} \, \, .
\ee
The complex valued isospin-1 amplitude is in very good agreement with
the corresponding empirical value, whether, as expected, CHPT fails
completely in the case of the isospin 0. It is interesting to look at the
chiral expansion of these quantities. One has: 
$ T_{\bar K N}^{(0)} =(0 + 2.29 - 1.87) \, {\rm {fm}}$,
$ T_{ K N}^{(1)} =(-7.63 + 7.83 - 6.54) \, {\rm {fm}}$ and
$ T_{\bar K N}^{(1)} =(3.81 + 5.06 - 1.74 +10.39 i) \, {\rm {fm}}$.
It shows a pattern which seems generic for the SU(3) baryon CHPT calculations
namely the cancellations of large contributions at second and third chiral 
order, see for example the baryon masses and their magnetic moments 
\cite{bm97,mest97}.

In order to be able to describe the S=-1 channel one has to employ 
non-perturbative methods
\cite{kswa95,ksw95,or98,om01,lk02,room03} which will allow for example to 
generate the $\Lambda$(1405) as a quasi-bound $\bar K N$ state 
from the 
lowest order attractive chiral meson-baryon interaction in this channel.  
Recent calculations have been made within unitarized CHPT in 
\cite{bnw05,opv05}.  
These  
studies are very important in view of the kaonic atoms measurement 
at DEAR \cite{baal95} (analog to the pionic atoms measurement described in
Section \ref{isoviol}) 
which allow to measure the $\bar K N$ threshold
amplitude to a very good accuracy. There is also a foreseen even 
better determination by the DEAR/SIDDHARTA Collaboration \cite{ss}. 
The pertinent formula for the energy
shift and decay width in terms of the scattering amplitudes has been 
obtained in \cite{mrr04} in the framework of effective field theory, that
accounts for a systematic expansion in isospin breaking effects.   
Let me just briefly summarize what is the present status. 
  It was pointed out in \cite{mrr04} 
that the 
scattering lengths obtained within unitarized CHPT 
are in disagreement with the DEAR measurement \cite{bal05}, statement
which was confirmed in \cite{bnw05}. However some recent papers 
\cite{opv05,o06}
show that it is in fact possible to obtain fits compatible both with DEAR and 
$K^- p$ scattering data. For a thorough discussion of these issues, see 
\cite{bnm06}.

\subsection{\it {Kaon-Photoproduction}} \label{kpho}

We have been discussing in section \ref{elecpro} pion photo- and electroproduction.
In fact some kaon photoproduction
data have also been taken at the electron stretcher ring ELSA (Bonn) 
over a wide energy range. An exploratory study of the 
reactions $\gamma p \to
\Sigma^+ K^0, \Lambda K^0$ and $\Sigma^0 K^+$ has thus been done within
HBCHPT in \cite{sm97} to third order. Clearly as discussed in the case of the 
pion a full scale $q^4$ calculation would be necessary, however as stressed 
in  \cite{sm97}
it is first necessary  to see whether the method is applicable in view of the 
not that small expansion parameter in SU(3) as discussed previously.
 
The calculation goes as in the SU(2) case except
that now one has to take into account kaon and $\eta$ loops. Also
more LECs, 13,  contribute. One combination is
fixed from the nucleon axial radius while the others are determined within
resonance saturation. 
One immediate outcome  is the investigation of the
effect of such loops on the $SU(2)$ predictions. As expected, they are
small, e.g. for neutral pion photoproduction off protons, $E_{0+, {\rm {thr}}}
^K =(e F M_\pi^3)/(96 \pi^2 F_\pi^3 M_K)=0.14 \cdot 10^{-3}/M_{\pi^+}$
which is just 1/10th of the empirical value and considerably smaller than the
pion contribution.  


For the reaction  $\gamma p \to K^0 \Sigma^+$ no data
points exists in the first 100 MeV above threshold. In \cite{sm97} a 
prediction is made for the electric dipole amplitude $ E_{0+, {\rm {thr}}}(
K^0 \Sigma^+)=1.07 \cdot 10^{-3}/M_\pi$. An interesting observation is
that in that case the leading $P$-wave multipoles are very sensitive to the 
yet unmeasured
magnetic moment of the $\Sigma^0$ because it is enhanced by the 
coupling constant ratio $g_{pK\Lambda}/g_{pK\Sigma^0} =(D+3F)/\sqrt3(F-D) 
\simeq -5$.
Results of the calculation
for the reactions  $\gamma p \to K^+ \Lambda$, 
and  $\gamma p \to K^+ \Sigma^0$ \cite{sm97} have been 
compared to the available data \cite{bal94}. While the total cross
section for the second reaction agrees with the two data points, the
lowest bin from ELSA is slightly bigger than the calculation. The predicted
recoil polarization comes out generally too small except for the 
forward angles for the reaction $\gamma p \to K^+ \Lambda$, however the
shape is well described. Also the important sign difference between the
recoil polarization of the two reactions is well reproduced. It stems from 
an intricate interference of the complex $S$- and $P$- wave multipoles.   

Results are encouraging but clearly more work has to be done on the 
theory side, e.g. inclusion of higher order effects and higher partial waves
as well as a better handle on the various coupling constants. On the 
experimental side one would need data closer to threshold together with 
finer energy binning. Note that here also unitarized CHPT calculations have 
been done \cite{kww97}.

\subsection{\it Proposals for improved chiral expansion} \label{impch}

As we have seen the strange mass is not that small and in 
some processes one can 
question the validity of the standard perturbative expansion,
see for example \cite{dhb99} for a discussion of such processes and also some
of the one discussed here. 
Let me discuss some proposals to improve on the chiral expansion.
There are two different ways to proceed. 
Either 
the strange quark is treated
on equal footing with  the up and down quarks following the standard 
scenario or not.  
In the first case  
different methods of regularizations can be proposed to improve on 
the convergence (one should keep in mind the remark already done
in Section \ref{or} that the resulting physics should be independent of the
choice of the regularization scheme. However at a given order
one choice might be more efficient than another one). 
We have already seen in the previous sections that the IR
regularization can lead to improved convergence compared to HBCHPT. 
Another regularization named long distance regularization has been proposed 
first by Donoghue 
and Holstein \cite{dh98}. 
On the other hand, A. Roessl \cite{ro99}
proposed a different way of 
treating the kaon making use of the fact that $M_\pi/M_K$ is not that
large. This is the so-called Heavy Kaon CHPT (HKCHPT). Note that the approach 
\cite{dsg02} is somewhat different. There only the two first orders 
in the chiral expansion in the meson sector
are questioned due to the possible small value of the  condensate in SU(3), 
the series is then supposed to be well converging. I will here briefly  
summarize HKCHPT and the long distance regularization.

\subsubsection{Heavy kaon CHPT} \label{hkchpt}

Roessl's idea \cite{ro99} is to  consider the kaon as a heavy source 
in much the
same way as the nucleon in HBCHPT (a closely related work applying 
reparameterization invariance instead of the reduction of relativistic
amplitude is presented in \cite{ouss01}). Since the kaons appear now as matter 
fields, the chiral Lagrangian for pion-kaon interactions decomposes
into a string of terms with a fixed number of kaons field. One has
\be
\cal{L}_{HKCHPT} = \cal{L}_\pi + \cal{L}_{\pi K K} + \cal{L}_{\pi K K K K}
+ \cdots
\ee
where the first term is the conventional pion effective Lagrangian. Then
each term of the string is chirally expanded. Similarly to the baryons
the Lagrangian for processes with at most one kaon in the in/our states  
contains terms with an odd 
number of derivatives and the power counting has to be modified due to the 
new large mass scale $M_K$. Matching conditions allow to fix the LECs
of HKCHPT from the ones based on the standard chiral 
expansion with light kaons. This framework has been applied for example
to the calculation of the pion-kaon sigma terms \cite{fkm02}. 
 
\subsubsection{long distance regularization} \label{ldreg}

Donoghue and Holstein \cite{dh98} observed that the large
corrections from one-loop graphs  found in some processes arised
to a large extent from propagation at short distances--smaller than the 
physical size of baryons--where the EFT cannot represent the correct
physics. Using a cut-off regularization instead of dimensional regularization
enables to  
remove this short-distance physics. The method thus introduces an
additional parameter $\Lambda$ and a cutoff function. $\Lambda$ has to be
chosen not too low as to remove any truly long distance physics and not too
large as not to include spurious short distance physics. It is assumed
to be of the order of $\Lambda \gg 1/ r_B  \sim 300-600$ MeV,
where $r_B$ is the size of the baryon. The loop integrals will usually
give rise to strong dependence in the cutoff in the form $\Lambda^3$, $\Lambda
^2$, $\Lambda$ and $\ln \Lambda$. These are absorbed into  renormalized values
of the LECs.  

This regularization has been used in the calculation of different processes
without \cite{dhb99,dh04} and with decuplets degrees of freedom
\cite{bhlo02,dh04}.
It was shown to greatly 
improve on the convergence of the chiral series. 
It amounts to 
multiplying the dimensional 
regularized quantity by a cut-off dependent function.
In the case of the strange magnetic radius \cite{dh04}, for example
one has using a dipole regulator:
\be
\langle r^2_{M,s} \rangle_{{\rm {cutoff}}} = \langle r^2_{M,s} 
\rangle_{{\rm {dim. reg}}}
X(\Lambda/M_K)  = -0.162 \, {{\rm fm^2}} X(\Lambda/M_K)  \, , \quad 
X(x) =\frac {x^3 (x^2 +14 x/5 +1)}{(x+1)^5} \, \, .
\ee
The function $X(x)$ is an increasing function of $x$ such that
$X(\infty)=0$.  
For $\Lambda=300 \cdots 600$ MeV the strange magnetic moment is sizeably
reduced compared to the dimensionally regularized result. One has
$\langle r^2_{M,s} \rangle_{{\rm {cutoff}}} =-0.01 \cdots -0.032 
{{\rm fm^2}}$. This calculation is formally ${\cal O}(p^3)$ though
the long distance portion of the one-loop integral contains higher
order pieces. 

There are, however, some drawbacks with this regularization. It involves
one additional parameter, the cut-off $\Lambda$ which 
might not be so easily determined. Also chiral symmetry is 
by no means guaranteed. Furthermore,
the analytic structure can be screwed up, additional unphysical
poles and/or cuts can be produced. Note also that contrary to IR regularization
it does not solve the 
problems of possible large $1/m$ corrections. Also up to now only ${\cal O}(p^3)$ calculations have been performed. It would be interesting to have 
full one loop results within this scheme.
  
\section{Conclusion} \label{conc}
Understanding the implications of spontaneous symmetry breaking, the symmetry
which governs the strong interactions at low energies,
 on baryon properties is a difficult
task. It has become possible on the theory side with the development
of effective theories and particularly the model-independent 
framework, CHPT, as well as with the progress of lattice QCD. One of the
big advantages of CHPT is that it allows for a precise determination of
the theoretical uncertainties.
On the experimental side much more precise data have become available due
to the advent of CW machines and the possibility of having polarized targets 
as well as polarized beams. 
Some successes have been obtained as we have seen in this 
review. Also Baryon CHPT has been refined in the last 
decade with the development of new 
regularization schemes. Chiral extrapolations have been performed 
and studies of generalized parton distribution are actually
developed, I unfortunately did not have the space to report on it here.     
However, work remains to be done to further sharpen our understanding
of QCD at low energies. Particularly the $\Delta$ degree of freedom
has still to be better understood for some processes. Also calculations
in the resonance region have to be generalized. Chiral extrapolations
have to be pursued, specially the study of unstable particles. 
Isospin breaking effects have also to be well under control considering
the degree of accuracy reached in many processes. $SU(3)$ 
calculations have to be continued so as to understand better the 
mechanism of chiral symmetry breaking.  

\vskip 0.6truecm
{\Large {\bf {Acknowledgments}}}
\vskip 0.3truecm
I would like to thank 
Ulf-G. Mei{\ss}ner, S. R. Beane, N. Fettes,
J. Gasser,
T. Hemmert, N. Kaiser, B. Kubis, J. Kambor,  
H. Krebs, T.-S. H. Lee, 
A. Rusetsky and J. Stern
with whom I had enjoyable and enriching collaboration. 
I am particularly grateful to 
Ulf-G. Mei{\ss}ner
for many useful comments and careful reading of the manuscript.
I would also like to thank O. P\`ene and H. Wittig for sharing their insights 
on lattice
QCD and reading carefully the part of the review related to it.
This work was supported in part by the EU Integrated Infrastructure Initiative
Hadron Physics Project (contract number RII3-CT-2004-506078) and the 
EU Contract No. MRTN-CT-2006-035482, ``FLAVIAnet''.


\end{document}